\documentclass[11pt,a4paper,nofootinbib,showpacs,preprint,utf8]{article}

\usepackage{jheppub}
\makeatletter
\renewcommand{\@fpheader}{}
\makeatother
\usepackage[english]{babel}
\usepackage{amsmath}
\usepackage[table]{xcolor}
\usepackage{colortbl}
\usepackage[utf8]{inputenc}
\usepackage[utf8]{inputenc}
\DeclareUnicodeCharacter{2212}{-}
\usepackage[english]{babel}
\usepackage{amsmath, amssymb, amsfonts, bm, bbm, slashed, subdepth}
\usepackage{orcidlink}
\usepackage{pifont}
\usepackage{graphicx}
\usepackage{mathrsfs}
\usepackage[toc,page]{appendix}
\usepackage{enumerate}
\usepackage{setspace}
\usepackage{booktabs, tabularx}
\usepackage{units}
\usepackage{color}
\usepackage{float}
\usepackage{multirow}
\usepackage{array}
\usepackage{makecell}
\newcolumntype{C}[1]{>{\centering\arraybackslash}m{#1}}
\usepackage[pscoord]{eso-pic}
\usepackage[normalem]{ulem}
\usepackage{import}
\usepackage{url}
\usepackage{subcaption}
\usepackage{tikz}
\usetikzlibrary{snakes}
\usepackage{cancel}
\allowdisplaybreaks[1]
\usepackage{tikz}
\usepackage{tikz-feynman}
\usetikzlibrary{positioning}
\usepackage{scalerel}
\usepackage{soul}
\usepackage{comment}

\usepackage{tabularx}
\usepackage{array} 

\newcolumntype{Y}{>{\centering\arraybackslash}X}

\usepackage{hyperref}
\hypersetup{
	setpagesize=false,
	bookmarksnumbered=true,
	colorlinks=true,
	linkcolor=blue,
	citecolor=red,
	hypertexnames=true
}

\newcommand{\be}{\begin{equation}}
	\newcommand{\ee}{\end{equation}}
\newcommand{\ba}{\begin{array}}
	\newcommand{\ea}{\end{array}}
\newcommand{\bea}{\begin{eqnarray}}
	\newcommand{\eea}{\end{eqnarray}}
\newcommand{\balg}{\begin{align}}
	\newcommand{\ealg}{\end{align}}
\newcommand{\bit}{\begin{itemize}}
	\newcommand{\eit}{\end{itemize}}

\definecolor{bostonuniversityred}{rgb}{0.8, 0.0, 0.0}

\makeindex

\begin{document}

\title{Probing mixed-state dark matter and $b \to s \mu^+\mu^-$ anomalies in a scalar-assisted baryonic gauge theory}
\author[a]{{Taramati}}
\author[b]{{,~Manas Kumar Mohapatra}}
\author[c]{,~Utkarsh Patel}
\author[b]{{,~Rukmani Mohanta}}
\author[a,d]{{,~Sudhanwa Patra}}
\affiliation[a]{Department of Physics, Indian Institute of Technology Bhilai, Durg-491002, Chhattisgarh, India}
\affiliation[b]{School of Physics, University of Hyderabad, Hyderabad-500046, India}
\affiliation[c]{Saha Institute of Nuclear Physics 1/AF, Salt Lake, Kolkata, 700064, West Bengal, India}
\affiliation[d]{Institute of Physics, Bhubaneswar, Sachivalaya Marg, Bhubaneswar-751005, India}

\emailAdd{taramati@iitbhilai.ac.in}
\emailAdd{manasmohapatra12@gmail.com}
\emailAdd{utkarsh.patel@saha.ac.in}
\emailAdd{rmsp@uohyd.ac.in}
\emailAdd{sudhanwa@iitbhilai.ac.in}
\abstract{We explore a Standard Model extension based on a local $U(1)_B$ symmetry, where a baryon-charged scalar mediates interactions between a fermionic dark matter candidate and Standard Model quarks. In this setup, the dark matter relic abundance is shaped not only by standard annihilation channels but also by additional coannihilation processes induced by a new scalar. The presence of this mediator provides a unified link between dark sector and flavor physics, yielding distinctive phenomenological consequences. We conduct a detailed study of dark matter phenomenology, emphasizing the role of the mass splitting between the dark matter particles, and the scalar mediator in determining the efficiency of coannihilation. The parameter space is examined in light of existing constraints from cosmological observations, direct and indirect detection experiments, as well as the collider searches at the \texttt{LHC}. Our analysis shows that the extended scalar sector opens up viable regions of parameter space beyond those accessible in minimal \(U(1)_B\) realizations, many of which are expected to be tested by forthcoming searches at \texttt{XENONnT} and \texttt{CTA}. Moreover, the model induces correlated signatures from flavor observables associated with the $b \to s \mu^+ \mu^-$ transitions as well, serving as complementary tests of the underlying framework.}

	\maketitle
\newpage
\section{Introduction}
\label{sec:intro}
The Standard Model (SM) of particle physics, while extremely successful in explaining a wide range of phenomena, fails to account for several key observations, such as the nature of dark matter (DM) \cite{Cirelli:2024ssz,Abercrombie:2015wmb,FileviezPerez:2019jju,MurguiGalvez:2020mcc,Taramati:2024kkn,Taramati:2025ygs, Bertone:2004pz,Bertone:2016nfn}, the observed baryon asymmetry of the Universe, and the possible hints of anomalies in $B$ meson decays. These open questions strongly motivate extensions of the SM based on additional gauge symmetries. In particular, a class of well-motivated models extends the Standard Model gauge group by gauging baryon number $(B)$, lepton number $(L)$, or their certain combinations, giving rise to additional local symmetries such as $U(1)_B$, $U(1)_L$, and others. After a systematic and careful charge assignment to exotic particles, these SM extended gauge groups provide natural, anomaly-free frameworks~\cite{FileviezPerez:2019jju,MurguiGalvez:2020mcc,Taramati:2024kkn,Taramati:2025ygs}. The lightest mass eigenstate among the exotic fermions, stabilized by a remnant discrete symmetry after $U(1)$ breaking, emerges as a viable DM candidate~\cite{Das:2022oyx,Patra:2016ofq,Patel:2024zsu,Das:2019pua,Alves:2015mua,OKADA2020135845,Neves:2021rlb,PhysRevD.101.095031,Biswas_2020,Das2022TwocomponentSA,PhysRevD.92.053007,Mishra:2020fhy,Ma:2026tyk,Mahapatra:2025vzu,PhysRevLett.129.021801}.

A particularly compelling set of models involves gauging only the baryon number, leading to an additional $U(1)_{B}$ gauge symmetry~\cite{FileviezPerez:2011pt}. Such models have been extensively studied in the literature for a wide range of phenomenological motivations~\cite{Ma:2020quj,Michaels:2020fzj,Boos:2022pyq,FileviezPerez:2018jmr,Butterworth:2024eyr,Butterworth:2025asm}. In Ref.~\cite{Ma:2020quj}, authors consider a gauged $U(1)_B$ framework with vectorlike fermions and a diquark scalar, in which a residual global symmetry stabilizes a singlet scalar DM candidate with viable relic density and observable direct-detection signatures. In Ref.~\cite{Michaels:2020fzj}, the authors demonstrate that the exotic decay $Z \to Z'\gamma$ provides a clean probe of anomalous $U(1)$ gauge symmetries. They show that anomaly-canceling fermions can induce observable, non-decoupling effects that may be accessible at future \texttt{TeraZ} experiments. As a result, electroweak precision observables and low-mass dijet searches emerge as complementary probes of baryon number gauge symmetry. Ref.~\cite{Boos:2022pyq} considers a fermionic DM in an asymptotically safe $U(1)_B$ model, where ultraviolet fixed points constrain the gauge coupling and kinetic mixing, yielding a predictive framework compatible with relic density and direct-detection bounds. In Ref.~\cite{FileviezPerez:2018jmr}, the authors investigate a Dirac DM candidate in a gauged $U(1)_B$ framework, showing that relic density and unitarity constraints impose an upper bound of order $200$~TeV on the baryon-number breaking scale. This tight connection between DM and the gauge sector renders the model highly predictive and testable, with implications for baryogenesis. More recently, Refs.~\cite{Butterworth:2024eyr,Butterworth:2025asm} studied minimal gauged $U(1)_B$ frameworks in which gauge anomaly cancellation predicts new fermions, including a fermionic DM candidate and a leptophobic gauge boson $Z_B$. These works emphasized the rich collider phenomenology of such models, highlighting signatures from long-lived charged fermions, multi-lepton final states, while demonstrating consistency with relic density, direct-detection, and \texttt{LHC} constraints. While these studies establish this framework as rich \& well-motivated for DM and collider phenomenology, a common study for DM and flavor observables has remained largely unexplored in this context.

With this motivation, the current work extends the minimal $U(1)_B$ framework by introducing an additional singlet scalar field, $S_1$, which simultaneously mediates DM interactions and couples to the quark sector. This enables a unified analysis of DM and flavor observables within a consistent gauged $U(1)_B$ theory. Such a combined analysis of DM and flavor physics is particularly timely, as flavor observables can probe energy scales complementary to those accessed by the next runs of collider experiments, thereby enhancing the discovery potential of gauged $U(1)_B$ models. Nevertheless, our framework can be stringently tested and potentially ruled out by null results from upcoming DM direct detection experiments~\cite{Macolino:2020uqq, DarkSide-20k:2017zyg} and/or by precision flavor measurements at upcoming Belle II~\cite{Belle-II:2025wpi} and \texttt{LHCb}~\cite{LHCb:2023hlw} runs. 
{\large
\begin{table}[htb]
\centering
\setlength{\tabcolsep}{1pt}
\renewcommand{\arraystretch}{2.0}
\arrayrulewidth=0.3mm
\resizebox{\textwidth}{!}{
\begin{tabular}{|c|c|c|c|c|}
\hline
\rowcolor{gray!30} \multicolumn{5}{|c|}{\textbf{\Large Lepton Universality Ratios in $b \to s \ell^+ \ell^-$ Observables}} \\
\hline
\Large \textbf{Decay Observable} & \Large \textbf{ $q^2$ Interval (GeV$^2$)} &\Large \textbf{SM Prediction} &\Large \textbf{Measured Value} &\Large \textbf{Significance with SM} \\
\hline
\multirow{2}{*}{\Large $R_K$} & \Large [0.1, 1.1] & \Large $1 \pm 0.01$~\cite{Bordone:2016gaq,Hiller:2003js} & \Large $0.994^{+0.090+0.027}_{-0.082-0.029}$~\cite{LHCb:2022vje} & \Large - \\
\cline{2-5}
 & \Large [1.1, 6.0] & \Large $1 \pm 0.01$~\cite{Bordone:2016gaq,Hiller:2003js} &\Large $0.949^{+0.042+0.023}_{-0.041-0.023}$~\cite{LHCb:2022vje} & \large- \\
\hline
\Large $R_{K_S^0}$ &\Large [1.1, 6.0] & \Large $1 \pm 0.01$~\cite{Bordone:2016gaq,Hiller:2003js} & \Large $0.66^{+0.20}_{-0.14}$ (stat) $^{+0.02}_{-0.04}$ (syst)~\cite{LHCb:2021lvy} & \Large $1.4\sigma$ \\
\hline
\multirow{2}{*}{\Large $R_{K^*}$} & \Large [0.045, 1.1] & \Large $1 \pm 0.01$~\cite{Bordone:2016gaq,Hiller:2003js} &\Large $0.927^{+0.093+0.034}_{-0.087-0.033}$~\cite{LHCb:2022vje} &\Large - \\
\cline{2-5}
 &\Large [1.1, 6.0] &\Large $1 \pm 0.01$~\cite{Bordone:2016gaq,Hiller:2003js} &\Large $1.027^{+0.072+0.027}_{-0.068-0.027}$ (stat) $\pm 0.047$ (syst)~\cite{LHCb:2022vje} &\Large - \\
\hline
\large $R_{K^{*+}}$ &\Large [0.045, 6.0] &\Large $1 \pm 0.01$~\cite{Bordone:2016gaq,Hiller:2003js} &\Large $0.70^{+0.18}_{-0.13}$ (stat) $^{+0.03}_{-0.04}$ (syst)~\cite{LHCb:2021lvy} &\Large $1.5\sigma$ \\
\hline
\multirow{3}{*}{\Large $R_\phi^{-1}$} &\Large [0.1, 1.1] & \Large $1.016$~\cite{Straub:2018kue} &\Large $1.57^{+0.28}_{-0.25} \pm 0.05$~\cite{LHCb:2024rto} &\large - \\
\cline{2-5}
 &\Large [1.1, 6.0] &\Large $1.003$~\cite{Straub:2018kue} &\Large $0.91^{+0.20}_{-0.19} \pm 0.05$~\cite{LHCb:2024rto} &\large - \\
 &\Large [15, 19] &\Large $1.002$~\cite{Straub:2018kue} &\Large $0.85^{+0.24}_{-0.23} \pm 0.10$~\cite{LHCb:2024rto} &\Large - \\
\hline
\rowcolor{gray!30}\multicolumn{5}{|c|}{\Large \textbf{Exclusive $b \to s \mu^+ \mu^-$ Observables}} \\
\hline
\Large $P_{5}^{\prime}$ &\Large [4.0, 6.0] &\Large $-0.759 \pm 0.071$~\cite{Descotes-Genon:2013vna} &\Large $-0.439 \pm 0.111 \pm 0.036$~\cite{LHCb:2020lmf} &\Large $3.3\sigma$ \\
\cline{2-5}
 &\Large [4.3, 6.0] &\Large $-0.795 \pm 0.065$~\cite{Descotes-Genon:2012isb} &\Large $-0.96^{+0.22}_{-0.21}$ (stat) $\pm 0.25$ (syst)~\cite{CMS:2017rzx} &\Large $1.0\sigma$ \\
\cline{2-5}
 &\Large [4.0, 8.0] &\Large $-0.795 \pm 0.054$~\cite{Descotes-Genon:2014uoa} &\Large $-0.267^{+0.275}_{-0.269}$ (stat) $\pm 0.049$ (syst)~\cite{Belle:2016xuo} & \Large$2.1\sigma$ \\
\hline
\Large $\mathcal{B}(B^+ \to K^+ \mu^+ \mu^-)$ &\Large [1.1, 6.0] & \Large $(1.708 \pm 0.283) \times 10^{-7}$~\cite{Parrott:2022zte} & \Large $(1.186 \pm 0.034 \pm 0.059) \times 10^{-7}$~\cite{LHCb:2022vje} & \Large $4.2\sigma$ \\
\hline
\Large $\mathcal{B}(B^0 \to K^{0*} \mu^+ \mu^-)$ & \Large [1.1, 6.0] & \Large $(2.323 \pm 0.381)\times 10^{-7}$~\cite{Bharucha:2015bzk} & \Large $(2.018 \pm 0.100 \pm 0.053) \times 10^{-7}$~\cite{LHCb:2016ykl} & \Large - \\
\hline
\Large $\mathcal{B}(B_s \to \phi \mu^+ \mu^-)$ & \Large [1.1, 6.0] & \Large $(2.647 \pm 0.319)\times 10^{-7}$~\cite{Aebischer:2018iyb,Bharucha:2015bzk} & \Large $(1.41 \pm 0.073 \pm 0.024 \pm 0.068)\times 10^{-7}$~\cite{LHCb:2021zwz,LHCb:2013tgx,LHCb:2015wdu} &\Large $3.6\sigma$ \\
\hline
\Large $\mathcal{B}(B_s \to \mu^+ \mu^-)$ &\Large - & \Large $(3.672 \pm 0.154)\times 10^{-9}$~\cite{Bobeth:2013uxa,Beneke:2019slt} & \Large $(3.361 \pm 0.028)\times 10^{-9}$~\cite{Greljo:2022jac} & \Large - \\
\hline
\end{tabular}
}
\caption{Status of  $b \to s \ell^+ \ell^-$ decay observables with SM predictions and experimental results.}
\label{Expt_Obs}
\end{table}
}
In this study, the presence of $S_1$ in the scalar sector has a significant impact on the phenomenology of DM. Depending on the charge assignment and coupling strengths, the new scalar can itself be a viable mediator in interactions between the dark sector and the SM particles. This influences the DM relic abundance through modified annihilation or co-annihilation channels, and alters predictions for direct and indirect detection experiments.

In addition to its implications for DM phenomenology, the framework considered in this work naturally gives rise to nontrivial effects in flavor physics. In particular, new particles and interactions that couple the dark sector to SM fermions can generate flavor-changing neutral-current (FCNC) processes at the loop level. Such transitions are suppressed in the SM and therefore provide a sensitive indirect probe of physics beyond it. Among the FCNC processes, the quark-level transition $b \to s \ell^+ \ell^-$ plays a central role due to its contribution to a wide range of experimentally accessible observables, including branching ratios, angular distributions, and lepton flavor universality (LFU) tests. In recent years, measurements of these observables by the \texttt{LHCb} and Belle collaborations have revealed a pattern of deviations from the SM predictions in several channels. While the latest measurements of the LFU ratios $R_K$ and $R_{K^*}$ are consistent with the SM expectations, the discrepancies still persist in the celebrated angular observable $P_5'$, and in the differential branching fractions of several decay modes such as, $B_s \to \phi \mu^+ \mu^-$ and $B ^+ \to K^+ \mu^+ \mu^- $. A concise summary of the current experimental status of the relevant $b \to s \ell^+ \ell^-$ observables, together with their Standard Model predictions, is presented in Table~\ref{Expt_Obs}. These anomalies can be explained by new interactions that contribute to semileptonic $b \to s \ell^+ \ell^-$ ($\ell = e, \mu$) transitions, which may also play a role in governing DM interactions. This observation motivates a combined analysis of DM and flavor observables within a specific and unified theoretical framework. In contrast to the explicitly leptophilic scenarios explored in Refs.~\cite{Singirala:2018mio, Singirala:2021gok, Mohapatra:2024aqd,Chao:2021qxq, Ko:2021lpx}, which rely on leptoquarks, $Z'$ bosons, or inert scalar doublets with direct couplings to charged leptons, our framework is based on a predominantly leptophobic mixed-state DM setup. The new mediator couples primarily to quarks and the dark sector, while interactions with charged leptons arise only indirectly, for example, through mixing effects. As a result, the $b \to s \mu^+ \mu^-$ transitions are generated in a controlled and predictive manner, allowing flavor observables to probe the same underlying dynamics responsible for DM interactions. Although the authors in Ref.~\cite{Sahoo:2021vug} have addressed the $b \to s \mu^+ \mu^-$ anomalies, focusing on mixed-state frameworks with vector-like particles and leptoquarks, this mainly affects their contributions to the semileptonic operators. In contrast, our analysis is performed within a concrete, well-defined model in which the mediator structure is explicitly connected to the dark sector. This enables a detailed study of exclusive decay modes and angular observables in $b \to s \mu^+ \mu^-$ transitions, while simultaneously correlating flavor constraints with DM phenomenology. Consequently, our framework provides complementary signatures beyond those accessible in purely leptophilic constructions.

The outline of the paper is as follows. In Section~\ref{sec:model}, we introduce the theoretical framework of the model and discuss its particle content, relevant Lagrangian, and the scalar and gauge sectors. DM phenomenology, including relic density computation, direct detection constraints, and indirect signatures, is presented in Section~\ref{sec:DMP}. In Section~\ref{sec:flavor}, we perform a comprehensive flavor physics analysis focusing on $b \to s \mu^+\mu^-$ transitions. The combined numerical results exploring correlations between flavor physics analyses and DM observables, and their phenomenological implications, are discussed in Section~\ref{sec:cd}. 
Finally, we present the conclusion of the work in Section~\ref {sec:conc} while an appendix is also provided at the end.

\section{Model Framework}
\label{sec:model}
The baryonic gauge extension of the SM, in which the baryon number is promoted to a local $U(1)_B$ symmetry, has been thoroughly explored in our earlier works~\cite{Taramati:2024kkn,Taramati:2025ygs}. To ensure the cancellation of gauge anomalies, the model introduces exotic fermions: a pair of $SU(2)_L$ doublets  
\begin{equation}
\Psi_{L}\,[1_c,2_L,\tfrac{1}{2}_Y,-1_B]\sim \begin{pmatrix}
    \Psi_L^+\\\Psi_L^0
\end{pmatrix}, 
\qquad 
\Psi_{R}\,[1_c,2_L,\tfrac{1}{2}_Y,2_B]\sim \begin{pmatrix}
    \Psi_R^+\\\Psi_R^0
\end{pmatrix},
\end{equation}
two pairs of singlets  
\begin{equation}
\xi_{L}\,[1_c,1_L,1_Y,2_B], 
\qquad 
\xi_{R}\,[1_c,1_L,1_Y,-1_B],
\end{equation}
and  
\begin{equation}
\chi_{L}^0\,[1_c,1_L,0_Y,2_B], 
\qquad 
\chi_{R}^0\,[1_c,1_L,0_Y,-1_B],
\end{equation}
where the  numbers in the bracket represent the charges under $SU(3)_c \times SU(2)_L\times U(1)_Y \times U(1)_B$ gauge group. The baryonic charges are assigned such that $B_1 - B_2 = -3$. In addition, one extra scalar field  
\begin{equation}
S\,[1_c,1_L,0_Y,-3_B]
\end{equation}
is required to break the extended gauge symmetry spontaneously. After $U(1)_B$ breaking, the neutral components of $\chi^0_{L/R}$ and $\Psi^0_{L/R}$ mix to form two Dirac fermions $(\Psi_1, \Psi_2)$, where the lighter state $\Psi_1$ naturally emerges as a viable DM candidate. A comprehensive discussion of anomaly cancellation, fermion charge assignments, and the diagonalization of the resulting mass matrices is provided in~\cite{Taramati:2024kkn,Taramati:2025ygs}.

In continuation of our previous work on the singlet-doublet fermionic DM (SDFDM) framework~\cite{Taramati:2024kkn}, we consider an extension of the model by introducing an additional scalar field, denoted as $S_1$. This scalar transforms as $(\mathbf{3_c}, \mathbf{1_L},\mathbf{ {-1/3}_Y}, \mathbf{{-5/3}_B})$ under the $SU(3)_c \times SU(2)_L \times U(1)_Y \times U(1)_B$ gauge group. While being a singlet under the electroweak $SU(2)_L$ symmetry, it carries non-trivial charges under $SU(3)_C$ and $U(1)_B$, making it a colored scalar state.

The motivation for introducing $S_1$ is twofold:
\begin{itemize}
 \item To explore the possible interplay between DM and flavor physics observables, by allowing new interactions that can mediate the rare FCNC decays.
 \item To explore the phenomenological implications of the additional scalar $S_1$ in the context of DM, particularly its influence on the relic abundance, direct detection cross-sections, and potential indirect detection signatures.
\end{itemize}
The field content of exotic fermions and their gauge charge assignments remain identical to those defined in Ref.~\cite{Taramati:2024kkn}, ensuring the anomaly cancellation conditions are satisfied. The extended model introduces the following Yukawa interaction term involving the new scalar $S_1$:
\begin{equation}
\label{eq:s1}
\mathcal{L}_{\text{Yukawa}} \supset Y_{S_1} \, \overline{q} \, S_1 \, \Psi_R + \text{h.c.},
\end{equation}
where $q$ represents the SM quark left-handed doublet, $\Psi_R$ is the exotic right-handed fermion from the SDFDM sector, and $Y_{S_1}$ denotes the corresponding Yukawa coupling. This Lagrangian term not only induces new interactions relevant to flavor-physics processes but also facilitates interactions between the visible and dark sectors, thereby influencing DM phenomenology. Additionally, the scalar $S_1$ contributes to new processes that affect DM production and detection mechanisms, thereby enriching the phenomenology of the SDFDM framework. Including the $S_1$ interaction given above in Eq.~(\ref{eq:s1}), the relevant kinetic terms and Yukawa Lagrangian involving all the exotic fermions of the framework is given by:
\begin{eqnarray}
 \mathcal{L}&=&\overline{\Psi_{L}}i \slashed{D}\Psi_{L}+\overline{\Psi_{R}}i \slashed{D}\Psi_{R}+\overline{\chi_{L}}i \slashed{D}\chi_{L}+\overline{\chi_{R}}i \slashed{D}\chi_{R} +\overline{\xi_{L}}i \slashed{D}\xi_{L}+\overline{\xi_{R}}i \slashed{D}\xi_{R}\nonumber \\
&&-h_1 \overline{\Psi_{L}} \tilde{H}\xi_R-h_2 \overline{\Psi_{R}} \tilde{H}\xi_L-h_3 \overline{\Psi_{L}} {H}\chi_R -h_4 \overline{\Psi_{R}} {H}\chi_L\nonumber \\ &&-\lambda_{\Psi} \overline{\Psi_{L}} S \Psi_{R} -\lambda_{\xi} \overline{\xi_{L}} \tilde{S} \xi_{R}-\lambda_{\chi} \overline{\chi_{L}} \tilde{S} \chi_{R}+Y_{S_1} \, \overline{q} \, S_1 \, \Psi_R  .
\label{eq:Lag}
\end{eqnarray}

\subsection{Scalar \& gauge sector}
Within the current framework, we extend the scalar sector by introducing an additional scalar field, $S_1$. This scalar $S_1$ can interact with scalar $S$ and the SM Higgs doublet $H$ at tree level. Accordingly, the most general renormalizable scalar potential consistent with the symmetries of the model can be written as:
\begin{eqnarray}
\label{eq:VHS}
V(H,S,S_1) &=&
   -\mu^2_{H}\, H^\dagger H  + \lambda_{H}\,(H^\dagger H)^2
   -\mu^2_{S}\, S^\dagger S  + \lambda_{S}\,(S^\dagger S)^2 
   -\mu^2_{S_1}\, S_1^\dagger S_1  + \lambda_{S_1}\,(S_1^\dagger S_1)^2 \nonumber \\
&& +\, \lambda_{HS}\,(S^\dagger S)(H^\dagger H) 
   + \lambda_{HS_1}\,(S_1^\dagger S_1)(H^\dagger H) 
   + \lambda_{SS_1}\,(S_1^\dagger S_1)(S^\dagger S),
\end{eqnarray}
 where $\lambda_{HS}$, $\lambda_{SS_1}$ and $\lambda_{HS_1}$ are portal coupling strengths. Upon the spontaneous breaking of the associated symmetries, the scalar fields $H$ and $S$ acquire nonzero vacuum expectation values $v$ and $v_B$, respectively. In contrast, the scalar $S_1$ is inert and does not acquire a VEV. The mass matrix for these scalars in the eigenbasis~$(H, S, S_1)$ can be written as:
\begin{eqnarray}
M^2_{HSS_1} =
\begin{pmatrix}
2\lambda_{H}\, v^2 & \lambda_{HS}\, v v_B & 0 \\[6pt]
\lambda_{HS}\, v v_B & 2\lambda_{S}\, v_B^2 & 0 \\[6pt]
0 & 0 & -\mu_{S_1}^2 \;+\; \tfrac{1}{2}\,\lambda_{HS_1} v^2 \;+\; \tfrac{1}{2}\,\lambda_{SS_1} v_B^2
\end{pmatrix},
\label{eq:mateigen1}
\end{eqnarray}
After matrix diagonalization, the mass eigenvalues in the basis~$(h,s,S_1)$ are given as:
\begin{eqnarray}
&&m^2_{h} = v^2 \lambda_H +v_B^2 \lambda  -\sqrt{ (v^2 \lambda_H -v_B^2 \lambda)^2+( \lambda_{HS} v v_B)^2}, \nonumber \\
&&m^2_{s} = v^2 \lambda_H +v_B^2 \lambda  +\sqrt{ (v^2 \lambda_H -v_B^2 \lambda)^2+( \lambda_{HS} v v_B)^2},\nonumber \\
&&m_{s_1}^2 =  -\mu_{S_1}^2 \;+\; \tfrac{1}{2}\,\lambda_{HS_1} v^2 \;+\; \tfrac{1}{2}\,\lambda_{SS_1} v_B^2.
\label{eq:exofer11}
\end{eqnarray}
Here, $m_h\simeq125$~GeV is the mass of the SM-like Higgs particle. The mixing between physical states $h$ and $s$ can be parameterized by a $\theta$ mixing angle defined by
\begin{equation}
\tan\!\left(2\theta\right) \;=\;
\frac{\lambda_{HS}\, v v_B}{\lambda_{H} v^2 - \lambda_{S} v_B^2},
\end{equation}
which arises from diagonalizing the $(H,S)$ sub-matrix from Eq.~(\ref{eq:mateigen1}). The scalar masses can be recast in terms of the $\theta$ angle as,
\begin{eqnarray}
&&m^2_{h} = 2v^2\lambda_H \cos^2\theta + 2v_B^2\lambda \sin^2\theta + \lambda_{HS} v\, v_B \sin 2\theta, \nonumber \\
&&m^2_{s} =2v^2\lambda_H \sin^2\theta + 2v_B^2\lambda \cos^2\theta - \lambda_{HS} v\, v_B \sin 2\theta.
\label{eq:exofer12}
\end{eqnarray}
From Eq.~(\ref{eq:mateigen1}), it can be seen that $S_1$ does not mix with the other scalars; thus, the mass eigenstate $s_1$ is purely derived from the $S_1$ field. Hence, for the remainder of this paper, we may interchangeably use $S_1$ and $s_1$, as both refer to the same physical state. In the subsequent sections, we present a detailed analysis of the implications of the new scalar $S_1$ on flavor physics observables, including its potential contributions to flavor-changing processes and rare decays. Furthermore, we investigate its impact on DM phenomenology, focusing on the relic abundance, direct detection prospects, and possible indirect detection signatures arising from its interactions.

In the gauge sector, the presence of the additional scalar $S_1$ does not modify the mass spectrum of the electroweak gauge bosons $Z$ and $W^{\pm}$, nor the mass of the extra gauge
boson $Z'$ associated with the $U(1)_B$ symmetry (see Ref.~\cite{Taramati:2024kkn,Taramati:2025ygs} for more  details). In our analysis, we neglect the kinetic mixing between the $U(1)_Y$ and $U(1)_B$ gauge groups. This approximation is justified by current experimental bounds, which tightly constrain the kinetic mixing parameter and render its effects on the gauge boson masses and couplings phenomenologically sub-leading within the parameter space considered~\cite{Hook:2010tw,PhysRevD.100.095001,10.1093/ptep/ptac117}. Under the zero kinetic mixing assumption, the mass of $Z'$ is given by,
\begin{equation}
M_{Z'}=3g_Bv_B
\label{eq:zpmass}   
\end{equation}

\section{DM Phenomenology}
\label{sec:DMP}
With the extended model setup introduced in the previous section, we further explore the framework's phenomenological relevance. In particular, we focus on two crucial aspects: the impact of the new scalar field $S_1$ on DM observables, and its role in flavor physics, including possible contributions to the rare FCNC decays. The $S_1$ Yukawa interaction connects the dark matter and flavor sectors via the loop-level flavor observables.

Throughout our analysis, we also emphasize the impact of scalar $S_1$ and of various mass splittings in the dark sector on the DM-related phenomenology. The Feynman diagrams relevant to DM annihilation involving scalar $S_1$ are provided in Figure~\ref{fig:annihilation_diagrams}, and the relevant co-annihilations involving $S_1$ and other dark sector particles are provided in Figure~\ref{fig:coann_diagrams} in the Appendix.  The fermionic mass eigenstate $\Psi_1$ is taken to be the stable dark matter candidate, with properties determined by singlet–doublet mixing (see Ref.~\cite{Taramati:2024kkn}). The presence of the extra scalar $S_1$ potentially has a significant impact, since it alters DM annihilation channels and modifies DM scattering rates with nucleons. Thus, the relevant input parameters are:  
\begin{equation}
\, M_{\Psi_1},\;~ M_{Z'},\;~ \Delta M(\Psi_2,\Psi_1),\;~ \sin\theta_{\rm DM},\;~ v_B,\;~ m_{S_1}, \;~ \Delta M(S_1,\Psi_1),\;~ Y_{S_1}\,,
\label{eq:parameters}
\end{equation}
where $M_{\Psi_1}$ is the DM mass, $M_{Z'}$ is the mass of the leptophobic gauge mediator, $\Delta M(\Psi_2,\Psi_1)$ is the splitting between exotic neutral fermions, $\theta_{\rm DM}$ is the DM mixing angle, $v_B$ is the $U(1)_B$ breaking scale, $m_{S_1}$ is the scalar mass, $\Delta M(S_1,\Psi_1)$ is the mass gap between $S_1$ and DM, and $Y_{S_1}$ is the Yukawa coupling strength for the Lagrangian term in Eq.~(\ref{eq:s1}).

To study DM phenomenology, we implement the model in \texttt{SARAH}~\cite{Staub:2013tta} package, generate the spectrum with \texttt{SPheno}~\cite{Porod:2011nf}, and perform numerical computations of relic abundance and cross-sections using \texttt{micrOMEGAs}~\cite{Belanger:2018ccd} code. This methodology enables us to systematically track the impact of the new scalar sector on both DM and flavor phenomenology. In the following discussion, we perform three key analyses: the relic density calculation, constraints from direct searches, and possible indirect detection signals, with an emphasis on the roles of $S_1$ and dark-sector mass splittings in shaping the allowed parameter space.

\begin{figure}[htbp]
\centering
 \centering
    \begin{subfigure}[b]{0.8\textwidth}
        \centering
        \includegraphics[width=\textwidth]{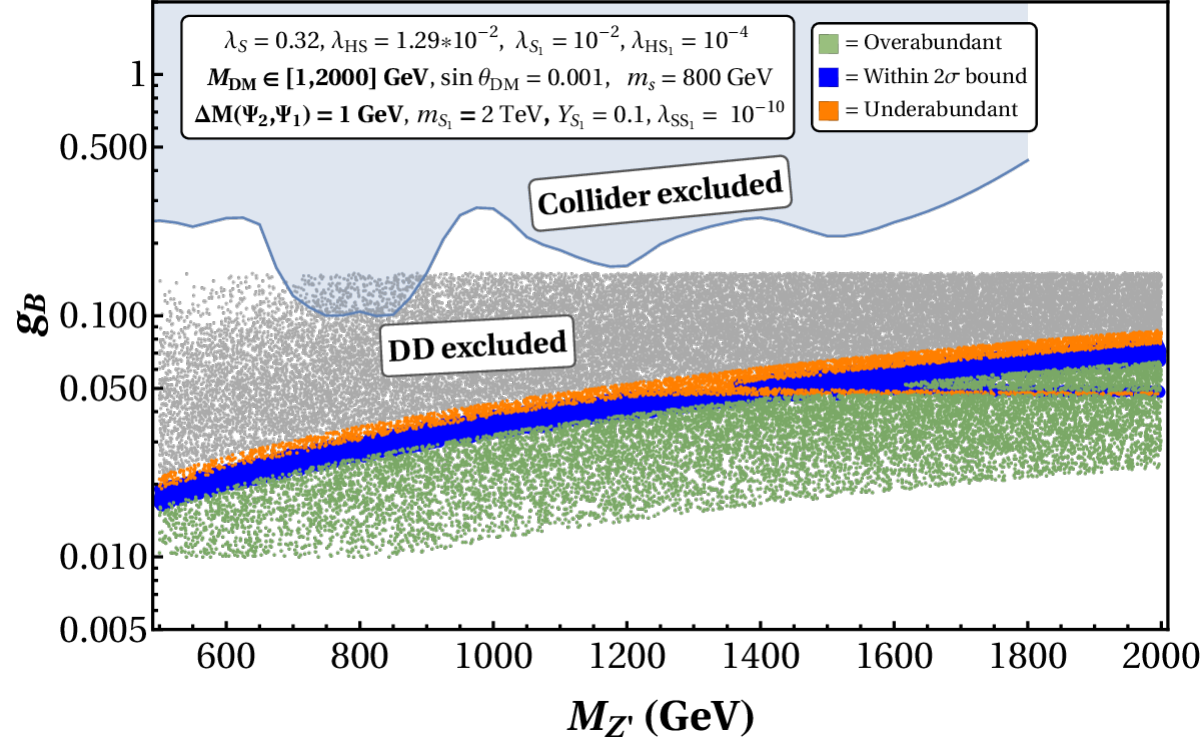}
        \caption{\small }
        \label{fig:relic1a}
    \end{subfigure}
    
    \vspace{0.5cm} 
    
    \begin{subfigure}[b]{0.8\textwidth}
        \centering
        \includegraphics[width=\textwidth]{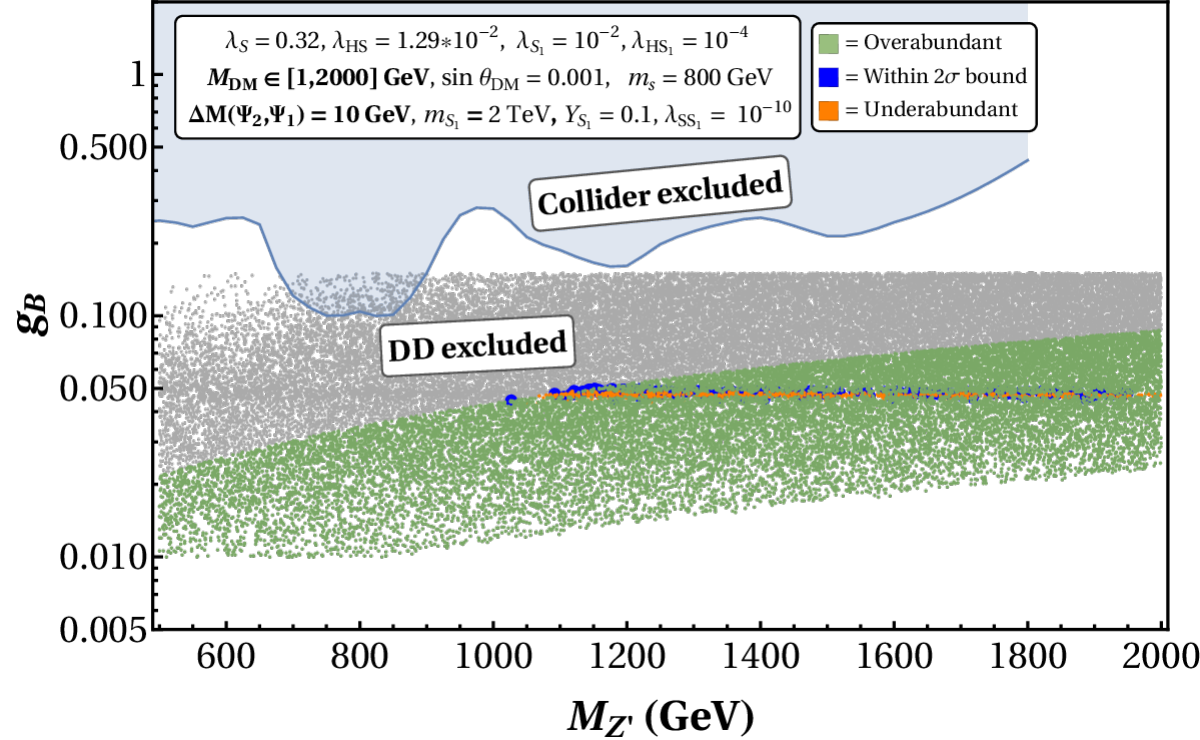}
        \caption{\small }
        \label{fig:relic1b}
    \end{subfigure}
\caption{\small Parameter space in the $(M_{Z'}, g_B)$ plane for two benchmark scenarios. The relic density constraint from Planck is indicated by the color coding: green points correspond to overabundant DM, orange points lie within the $2\sigma$ allowed range, and blue points give underabundant relic density. The gray shaded area is excluded by direct-detection (DD) bounds from \texttt{LZ (2022}, while the light-blue shaded region is ruled out by collider searches for a heavy $Z'$ boson. \textbf{Top panel (a):} mass splitting $\Delta M({\Psi_2,\Psi_1})=1~\text{GeV}$. \textbf{Bottom panel (b):} mass splitting $\Delta M({\Psi_2,\Psi_1}) = 10~\text{GeV}$. Other relevant parameters are fixed as $\lambda_S=0.32$, $\lambda_{HS}=1.29\times 10^{-2}$, $\lambda_{S_1}=10^{-2}$, $\lambda_{HS_1}=10^{-4}$, $\lambda_{SS_1}=10^{-10}$, $\sin\theta_{\text{DM}} = 0.001$, $m_s=800~\text{GeV}$, $m_{S_1}=2~\text{TeV}$, and $Y_{S_1} = 0.1$. The DM mass is varied within the range of 1 GeV to 2 TeV.}
\label{fig:relic1}
\end{figure}

\subsection{Relic Density and Direct Detection Analysis}
\label{subsec:relic}
In this sub-section, we study the impact of variations of model input parameters on the DM final relic, the obtained DD cross section, and the associated phenomenology. Specifically, we examine the effects of changing the mass splittings~$\Delta M(\Psi_2,\Psi_1)$,~$\Delta M(S_1,\Psi_1)$, and coupling~$Y_{S_1}$. The parameter scans and their associated plots presented ahead are generated by fixing certain input parameters: $\lambda_{S} = 0.32$, $\lambda_{HS} = 1.29 \times 10^{-2}$, $\sin \theta_{\text{DM}} = 0.001$, $m_{s} = 800~\text{GeV}$, $m_{S_1} = 2~\text{TeV}$, $\lambda_{S} = 1\times 10^{-2}$ and $\lambda_{HS_1} = 1\times 10^{-4}$. The choice of these input parameter values is motivated partly by earlier studies. To enable a direct comparison, we keep few benchmark values identical to those employed in Refs.~\cite{Taramati:2024kkn,Taramati:2025ygs}.
For some other parameters like $\lambda_{HS_1}$, $m_{S_1}$, representative values are chosen within their allowed ranges, since the impact of their variation is not the focus of this study and may be taken up in a future work. The coupling $\lambda_{SS_1}$ is set at a small value of $1 \times 10^{-10}$ to ensure a negligible contribution of scalar $S$'s VEV~$(v_B)$ on the mass of the $S_1$.

\subsubsection{Relic Density}
\label{subsec:relic}
Firstly, we depict two scatter plots in Figure~\ref {fig:relic1}, highlighting the allowed parameter space in the $(M_{Z'}, g_B)$ plane from DM relic, direct detection, and $Z'$ collider searches over a range of DM mass. In the top panel (Figure~\ref {fig:relic1a}) where $\Delta M({\Psi_2,\Psi_1})=1~\text{GeV}$, for small values of $g_B$, we see that although this region is allowed by both direct detection and collider searches, a significant portion of the parameter points here result in DM overabundance (green shaded region) and is thus excluded by the DM relic requirements. For intermediate $g_B$ values, a narrow band of blue points~(within $2\sigma$ DM relic) is also present in the plot for the entire range of $Z'$ mass, which depicts the $Z'$ resonance region for DM candidate, i.e., where $M_{\rm DM}\simeq M_{Z'}/2$. It corresponds to a significantly enhanced DM annihilation via the $s$-channel $Z'$ pole, yielding the correct DM relic density with observations. This blue region is particularly interesting since it shows the portion of parameter space where DM survives all existing constraints from relic requirements~(\texttt{Planck 2018}~\cite{Planck:2018vyg}), direct detection~(\texttt{LZ 2022}~\cite{LZ:2022lsv}), and collider searches~(\texttt{ATLAS}), and thus constitutes a phenomenologically viable region in the upcoming experiments like~\texttt{XENONnT}.  At larger $g_B$ values, the annihilation cross section becomes 
too efficient, and for most of the parameter points here, we obtain an underabundant DM relic (orange and gray shaded regions). The orange points correspond to plot points that remain consistent with the \texttt{LZ} and collider constraints but produce underabundant DM. In the gray region, in addition to being underabundant, the DM is also ruled out by the \texttt{LZ} data.

The obtained plot behavior changes noticeably when $\Delta M({\Psi_2,\Psi_1})$ is increased to $10~\text{GeV}$ (bottom panel, Figure~\ref{fig:relic1b}) while keeping other input parameters the same as those in Figure~\ref{fig:relic1a}. In this case, with the increased mass splitting, the strength of DM co-annihilations with the next-to-lightest state, $\Psi_2$, becomes less significant, leading to a larger number of plot points where the DM relic is overabundant (green shaded region). Consequently, most of the points of interest~(in comparison with Figure~\ref{fig:relic1a}) drop out from the $2\sigma$ relic region here, and only a handful of points appear that are viable with all the phenomenological requirements~(blue colored points). Again, for larger values of $g_B$, most of the points are excluded by the \texttt{LZ 2022}, as indicated by the gray-shaded region. Overall, the figure demonstrates that the $Z'$ resonance region provides the most interesting region of parameter space in the $M_{Z'}-g_B$ plane, with a clear dependence on the dark-sector mass splitting.
\begin{figure}[htbp]
\centering
 \begin{subfigure}[b]{0.490\textwidth}
        \centering
        \includegraphics[width=\textwidth]{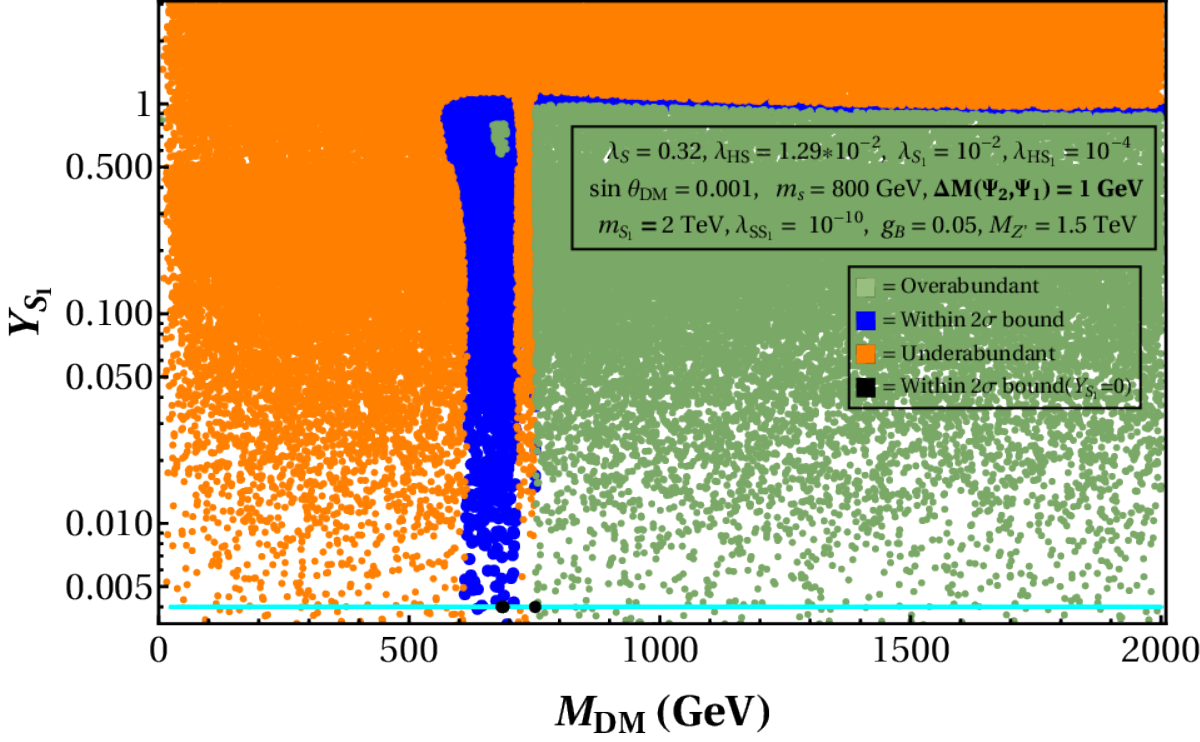}
        \caption{\small }
        \label{fig:relic2a}
    \end{subfigure}
    \hfill
 \vspace{0.5cm}
    \begin{subfigure}[b]{0.490\textwidth}
        \centering
        \includegraphics[width=\textwidth]{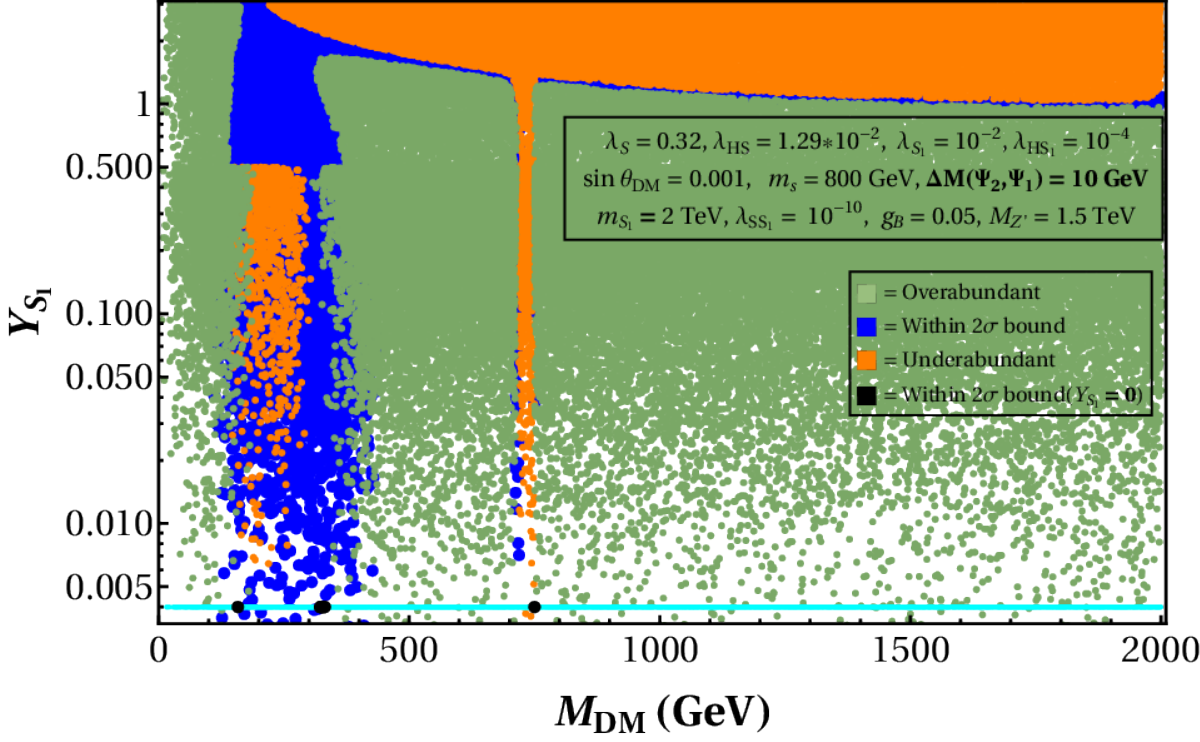}
        \caption{\small }
        \label{fig:relic2b}
    \end{subfigure}
\caption{\small The relic abundance $\Omega h^2$ in the $M_{\rm DM}-\,Y_{S_1}$ plane 
for two different choices of mass splitting $\Delta M(\Psi_2,\Psi_1)$. 
\textbf{Left} (a) $\Delta M(\Psi_2,\Psi_1) = 1$ GeV,
\textbf{Right} (b) $\Delta M(\Psi_2,\Psi_1) = 10$ GeV. 
The color code indicates the relic density: green points are over-abundant, blue points lie within the $2\sigma$ bound of the observed value ($\Omega h^2 \simeq 0.110 \pm 0.012$), and orange points are under-abundant. The analysis is performed with fixed parameters: $\lambda_S=0.32$, $\lambda_{HS}=1.29\times 10^{-2}$, $\lambda_{S_1}=10^{-2}$, $\lambda_{HS_1}=10^{-4}$, $\lambda_{SS_1}=10^{-10}$, $\sin\theta_{\text{DM}} = 0.001$, $m_s=800~\text{GeV}$, $m_{S_1}=2~\text{TeV}$, $M_{Z'} = 1.5$ TeV, and $g_{B} = 0.05$. All displayed points are consistent with direct-detection limits from the \texttt{LZ 2022} data.}
\label{fig:relic2}
\end{figure}
Next, in Figure~\ref{fig:relic2} we show the relic abundance $\Omega h^{2}$ in the $M_{\mathrm{DM}}-Y_{S_1}$ plane for two benchmark mass splittings, $\Delta M(\Psi_{2},\Psi_{1})=1\ \mathrm{GeV}$ (left panel plot~\ref{fig:relic2a}) and $\Delta M(\Psi_{2},\Psi_{1})=10\ \mathrm{GeV}$ (right panel plot~\ref{fig:relic2b}). The fixed input parameters  for these plots are: $\lambda_S=0.32$, $\lambda_{HS}=1.29\times 10^{-2}$, $\lambda_{S_1}=10^{-2}$, $\lambda_{HS_1}=10^{-4}$, $\lambda_{SS_1}=10^{-10}$, $\sin\theta_{\text{DM}} = 0.001$, $m_s=800~\text{GeV}$, $m_{S_1}=2~\text{TeV}$, $g_B = 0.05$ and $M_{Z'} = 1.5 $ TeV. All points shown satisfy the projected direct-detection sensitivities from the most stringent \texttt{LZ} data. The color coding classifies the obtained relic abundance as: green color reflects DM over-abundance, blue points are within the \texttt{Planck} $2\sigma$ band, and orange color corresponds to DM under-abundance. For both plots, the vertical features (sharp, nearly vertical blue bands) reflect the regions of resonant enhancements of the DM annihilation cross section. In our parameter space of interest, there are two evident resonance regions: the $Z'$ pole at $M_{\mathrm{DM}}\simeq M_{Z'}/2$  (visible near $M_{\mathrm{DM}}\simeq 0.75\ \mathrm{TeV}$ for $M_{Z'}=1.5\ \mathrm{TeV}$ in both plots) and a pole emerging from exotic scalar $S$ mediated interactions at $M_{\mathrm{DM}}\simeq m_{s}/2$ (visible near $M_{\mathrm{DM}}\simeq 400\ \mathrm{GeV}$ for $m_{s}=800\ \mathrm{GeV}$ in the right panel plot). In the left panel (Figure~\ref{fig:relic2a}), where the splitting is fixed to $\Delta M(\Psi_{2},\Psi_{1})= 1~{\rm GeV}$, the coannihilation channels remain highly efficient in the low DM mass region up to about $800~{\rm GeV}$, independent of the value of the Yukawa coupling $Y_{S_1}$. As a result, the scalar resonance pole is absent, and the strong DM co-annihilations via $\Psi_2$ drive the final DM relic to be underabundant here. Beyond this scale, the coannihilation channels are suppressed, leading to overabundant regions at higher DM masses (indicated by green points). In this regime, the coannihilation rate is significant, and the relic density requirement is satisfied only for relatively larger values of the Yukawa coupling, typically $Y_{S_1} \sim 1$ (thin blue horizontal curve around $Y_{S_1} \sim 1$ in Figure~\ref{fig:relic2a}). 

Increasing the splitting to $\Delta M(\Psi_{2},\Psi_{1}) = 10~{\rm GeV}$, shown in the right panel (Figure~\ref{fig:relic2b}), suppresses the strength of $\Psi_2$ assisted co-annihilations. Consequently, apart from the $Z'$ and scalar $S$ resonance regions, the parameter space mostly produces an over-abundance of final DM relic~(green regions) up to a value of 1 for $Y_{S_1}$. However, when $Y_{S_1}$ exceeds unity, the interaction term $Y_{S_1}\,\overline{q} S_1 \Psi_R$ becomes significant, reactivating coannihilation channels and opening up allowed regions consistent with relic density bounds. For even larger values of $Y_{S_1}$, the annihilation cross section is strongly enhanced, driving the relic abundance into the underabundant regime, as reflected in the extended orange region towards the top side of the plot. A robust feature across both panels is the vertical band near $M_{\rm DM} \simeq M_{Z'}/2$, originating from the $Z'$-mediated resonance. The plots also depict the role of the Yukawa coupling $Y_{S_1}$ on the final DM relic abundance. This coupling controls the strength (and width) of $S_1$ scalar-mediated DM annihilations. It is evident that non-zero values of $Y_{S_1}$ allow a larger region of parameter space to produce a final DM relic within the Planck $2\sigma$ limit (blue colored regions). For a direct comparison, we have also plotted the obtained DM relic for the case where $Y_{S_1}$ is set to zero (horizontal cyan lines), and the regions of parameter space where the obtained final relic falls within a $2\sigma$ bound are marked by black dots. Thus, the mass splitting $\Delta M(\Psi_{2},\Psi_{1})$, together with the Yukawa coupling $Y_{S_1}$, plays a crucial role in shaping the viable parameter space.

\begin{figure}[htbp]
   \centering
    \begin{subfigure}[b]{0.49\textwidth}
        \centering
        \includegraphics[width=\textwidth]{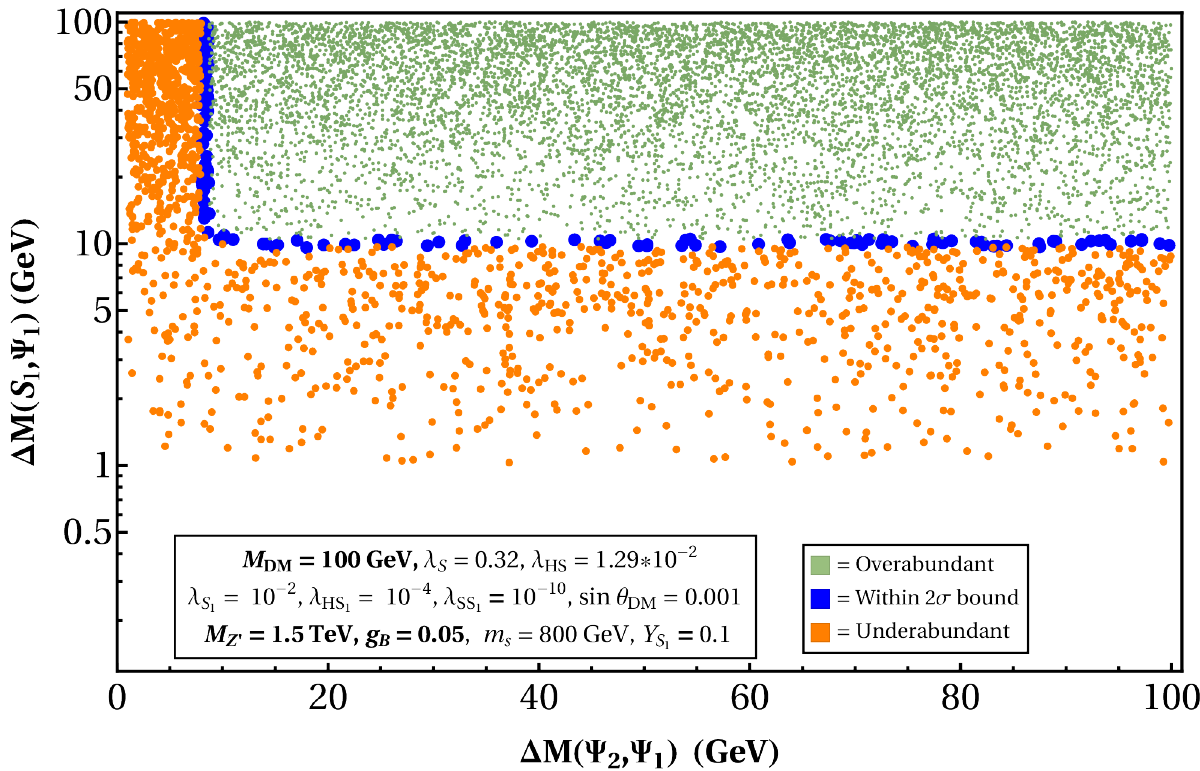}
        \caption{\small $M_\text{DM} = 100$ GeV}
        \label{fig:relic3a}
    \end{subfigure}
    \hfill
    \begin{subfigure}[b]{0.49\textwidth}
        \centering
        \includegraphics[width=\textwidth]{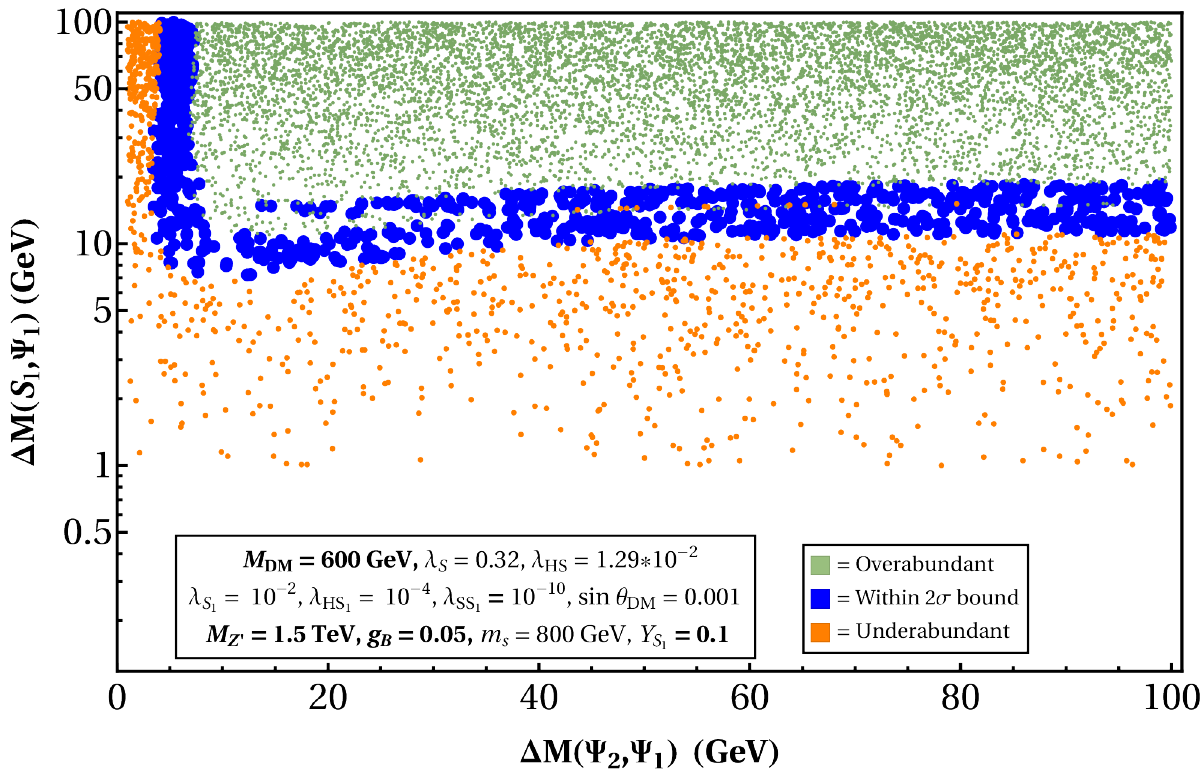}
        \caption{\small $M_\text{DM} = 500$ GeV}
        \label{fig:relic3b}
    \end{subfigure}
    
    \vspace{0.5cm} 
    
    \begin{subfigure}[b]{0.49\textwidth}
        \centering
        \includegraphics[width=\textwidth]{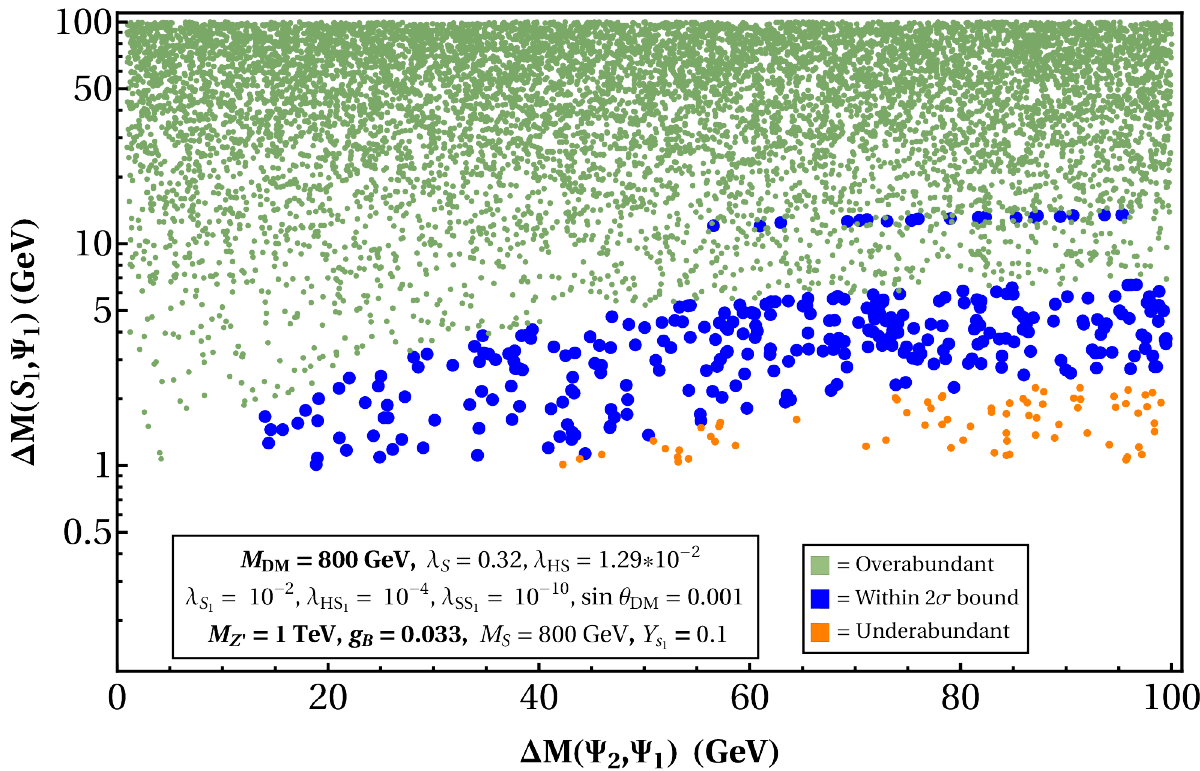}
        \caption{\small $M_\text{DM} = 800$ GeV}
        \label{fig:relic3c}
    \end{subfigure}
    \hfill
    \begin{subfigure}[b]{0.49\textwidth}
        \centering
        \includegraphics[width=\textwidth]{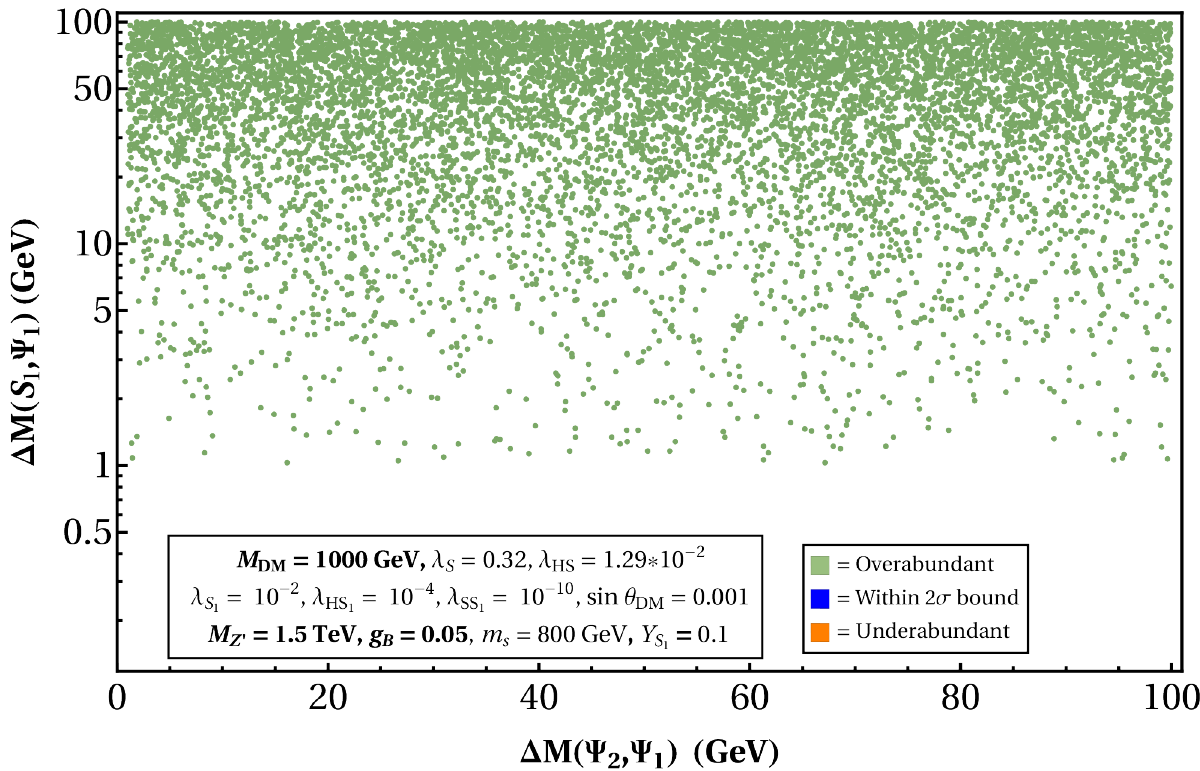}
        \caption{\small $M_\text{DM} = 1000$ GeV}
        \label{fig:relic3d}
    \end{subfigure}
    \caption{\small Relic density allowed parameter space in the $\Delta M(\Psi_2,\Psi_1)$–$\Delta M(S_1,\Psi_1)$ plane for different DM masses: (a) $M_{\rm DM}=100~\text{GeV}$, (b) $M_{\rm DM}=500~\text{GeV}$, (c) $M_{\rm DM}=800~\text{GeV}$, and (d) $M_{\rm DM}=1000~\text{GeV}$ with fixed gauge boson mass $M_{Z'} = 1.5 $ TeV. The color coding denotes overabundant (green), underabundant (orange), and points consistent with the observed relic density within $2\sigma$ (blue). All displayed points are consistent with direct detection limits from experiments such as \texttt{LZ 2022}.}
    \label{fig:relic3}
\end{figure}

The scatter plots in Figures~\ref{fig:relic3} and \ref{fig:relic4} illustrate the dependence of the relic abundance on the mass splittings $\Delta M(\Psi_{2},\Psi_{1})$ and $\Delta M(S_1,\Psi_{1})$ for two different choices of the $Z'$ mass: $M_{Z'}=1.5~\text{TeV}$ (Figure~\ref{fig:relic3}) and $M_{Z'}=1~\text{TeV}$ (Figure~\ref{fig:relic4}). Each subplot corresponds to a different value of DM mass, with $M_{\rm DM}=100,\ 500,\ 800,\ 1000~\text{GeV}$ for subplots \ref{fig:relic3a}, \ref{fig:relic3b}, \ref{fig:relic3c}, and \ref{fig:relic3d}, respectively, and $M_{\rm DM}=100,\ 400,\ 600,\ 800~\text{GeV}$ for subplots \ref{fig:relic4a}, \ref{fig:relic4b}, \ref{fig:relic4c}, and \ref{fig:relic4d}, respectively. The remaining parameters are kept fixed at $\lambda_{S}=0.32$,
$\lambda_{HS}=1.29\times 10^{-2}$, $\lambda_{S_1}=10^{-2}$, $\lambda_{HS_1}=10^{-4}$, $\lambda_{SS_1}=10^{-10}$, $\sin\theta_{\text{DM}}=0.001$, $m_{s}=800~\text{GeV}$, $m_{S_1}=2~\text{TeV}$, and $Y_{S_1}=0.1$ for all the plots. Blue points correspond to Planck $2\sigma$ allowed region, while orange (green) points denote underabundant (overabundant) DM. Direct detection and collider constraints are always satisfied in the shown parameter space due to the small gauge coupling $g_{B}=0.05, 0.033$ and the suppressed scalar mixing, $\sin{\theta}=0.001$. Below, we discuss the characteristic features obtained in these plots for representative benchmark values of $M_{\rm DM}$ as different cases.
\subsubsection*{Case I:  $\mathbf{M_{\rm \textbf{DM}}=100}~\text{GeV}$ (Figure~\ref{fig:relic3a} and Figure~\ref{fig:relic4a})} 
   In this case, the DM mass lies well below both the scalar $S$ resonance ($m_s/2= 400~\text{GeV}$) and the $Z'$ resonance ($M_{Z'}/2=750~\text{GeV}$). Thus, DM annihilations alone are inefficient to provide the correct relic abundance at freeze-out. However, when one of the mass splittings is very small, co-annihilation channels assisted via $\Psi_{2}$ or $S_1$ become sufficiently efficient to produce the observed relic 
   abundance, resulting in a narrow 'L' shaped blue strip visible in the plots. For even smaller mass splittings, the DM depletion becomes too efficient, and the final relic falls into the under-abundant region (orange points). When both the mass splittings become large enough, then, as expected, most of the parameter space is overabundant (green points). No other interesting plot features appear here.

\begin{figure}[htbp]
    \begin{subfigure}[b]{0.49\textwidth}
        \centering
        \includegraphics[width=\textwidth]{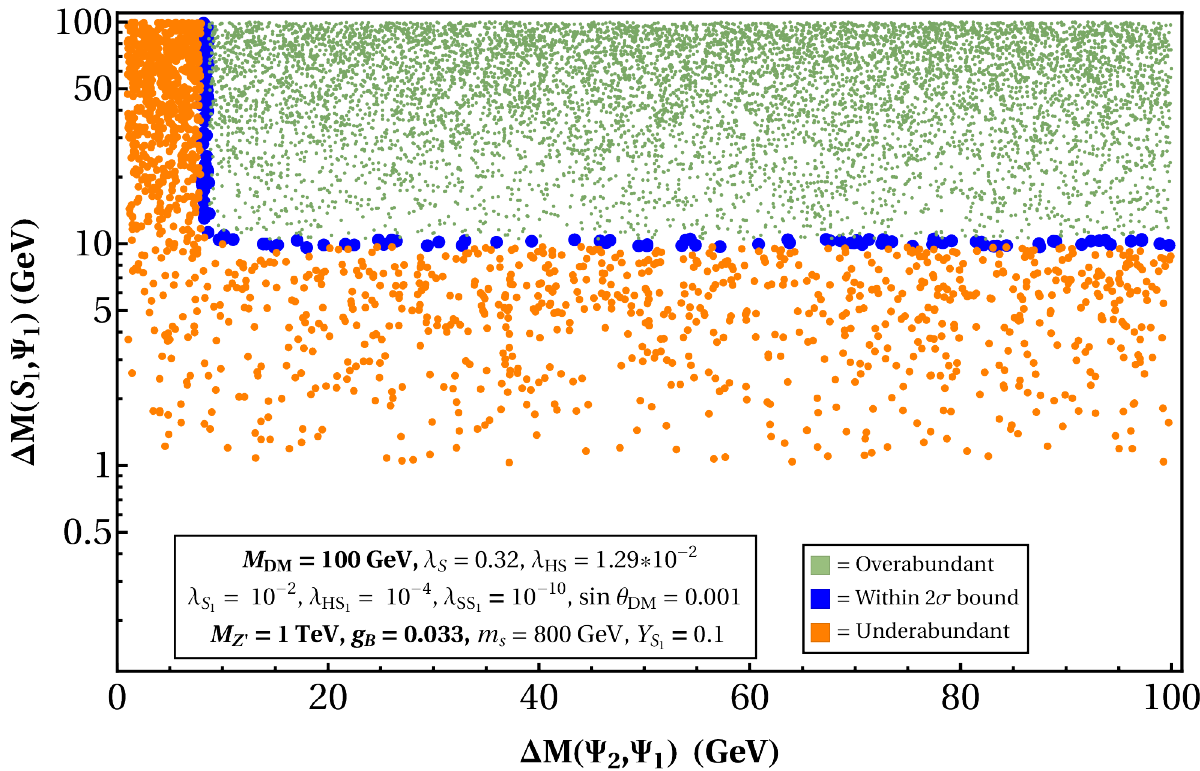}
        \caption{\small $M_\text{DM} = 100$ GeV}
        \label{fig:relic4a}
    \end{subfigure}
    \hfill
    \begin{subfigure}[b]{0.49\textwidth}
        \centering
        \includegraphics[width=\textwidth]{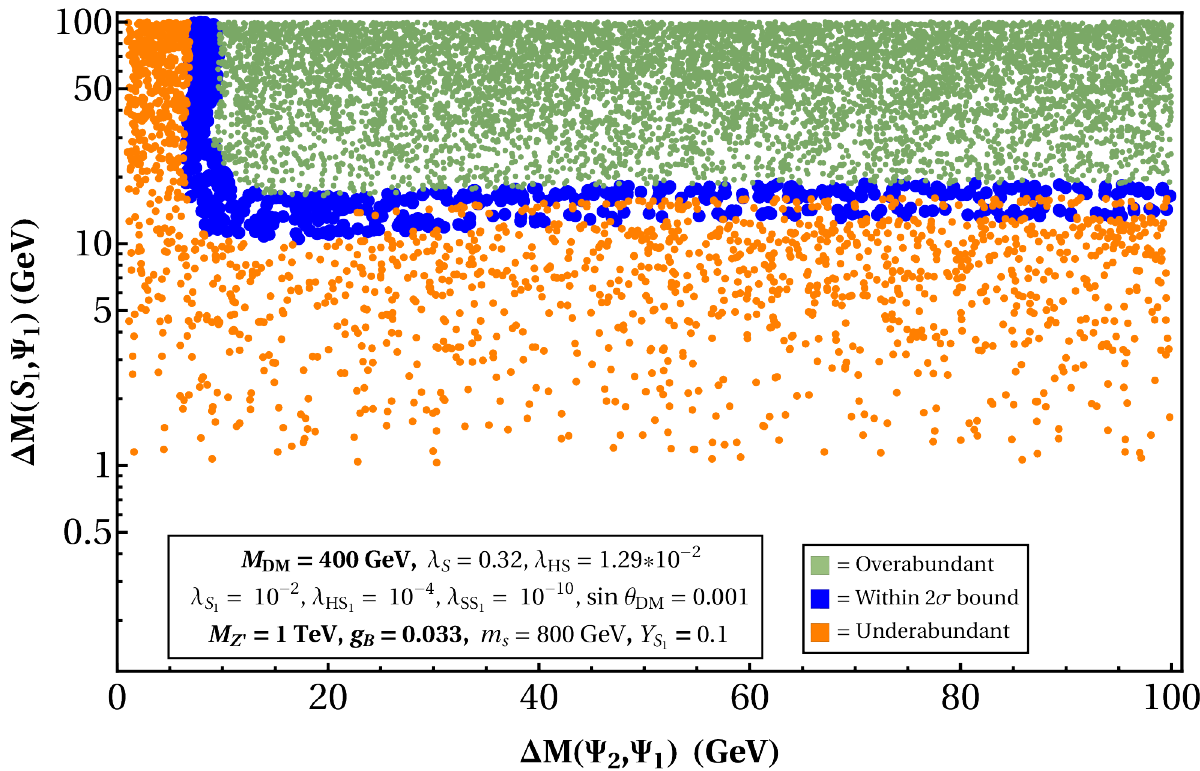}
        \caption{\small $M_\text{DM} = 400$ GeV}
        \label{fig:relic4b}
    \end{subfigure}
    
    \vspace{0.5cm} 
    
    \begin{subfigure}[b]{0.49\textwidth}
        \centering
        \includegraphics[width=\textwidth]{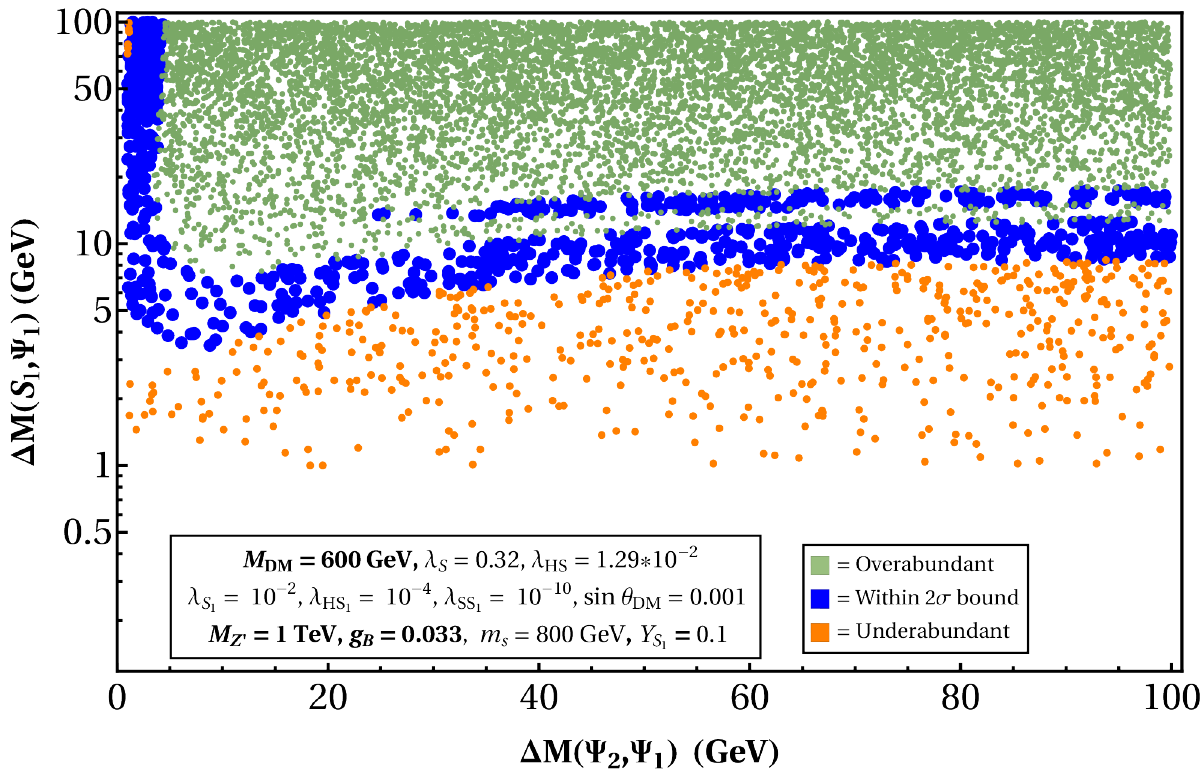}
        \caption{\small $M_\text{DM} = 600$ GeV}
        \label{fig:relic4c}
    \end{subfigure}
    \hfill
    \begin{subfigure}[b]{0.49\textwidth}
        \centering
        \includegraphics[width=\textwidth]{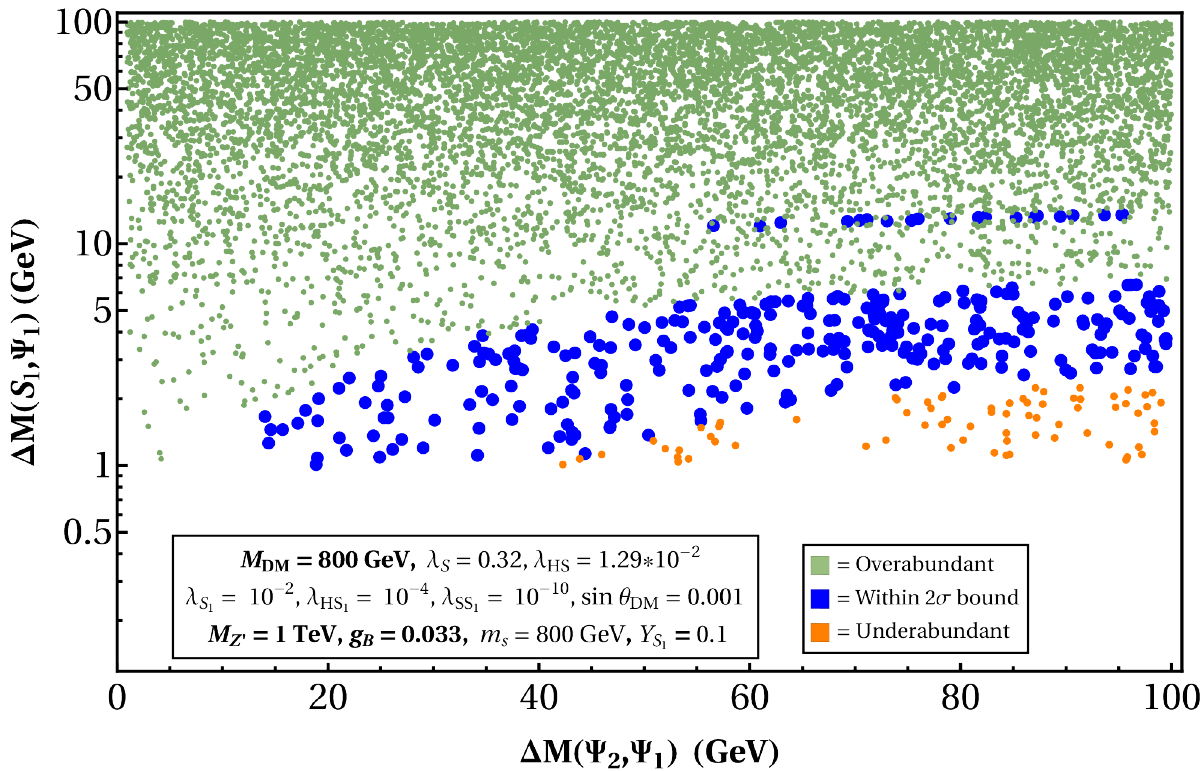}
        \caption{\small $M_\text{DM} = 800$ GeV}
        \label{fig:relic4d}
    \end{subfigure}

    \caption{\small Relic density allowed parameter space in the $\Delta M(\Psi_2,\Psi_1)$–$\Delta M(S_1,\Psi_1)$ plane for different DM masses: (a) $M_{\rm DM}=100~\text{GeV}$, (b) $M_{\rm DM}=400~\text{GeV}$, (c) $M_{\rm DM}=600~\text{GeV}$, and (d) $M_{\rm DM}=800~\text{GeV}$ with fixed gauge boson mass $M_{Z'} = 1$ TeV. The color coding denotes overabundant (green), underabundant (orange), and points consistent with the observed relic density within $2\sigma$ (blue). All displayed points are consistent with direct detection limits from experiments such as \texttt{LZ 2022}.}
    \label{fig:relic4}
\end{figure}

\subsubsection*{Case II:  $\mathbf{M_{\rm \textbf{DM}}=400, 500, 600}~\text{GeV}$ (Figures~\ref{fig:relic3b}, \ref{fig:relic4b} and \ref{fig:relic4c}, respectively)} 
    In this case, the DM mass lies close to both the scalar resonance at $m_s/2 = 400~\text{GeV}$ and the $Z'$ resonance at $M_{Z'}/2 = 750~\text{GeV}$. As a result, resonant annihilation channels become active and significantly enhance the annihilation cross section at freeze-out. In addition, for small mass splittings, co-annihilation processes further enhance DM depletion, allowing the correct relic abundance to be achieved over a larger region of parameter space. This results in an extended blue band of relic-density-allowed points spanning a substantial portion of the plot. We further find that, in this regime, the relic density remains mostly insensitive to variations in $M_{Z'}$~(compare Figures~\ref{fig:relic3b} and \ref{fig:relic4c}).
\subsubsection*{Case III: $\mathbf{M_{\rm \textbf{DM}}=800}~\text{GeV}$ (Figure~\ref{fig:relic3c} and Figure~\ref{fig:relic4d})}     
     In this case, the DM mass is nearly degenerate with the scalar partner ($m_s=800~\text{GeV}$) and close to the $Z'$ pole, $(M_{Z'}/2=750~\text{GeV})$. Thus, the scalar coannihilations majorly set the obtained DM relic. This explains the diagonal distribution of blue points, reflecting the condition $\Delta M(S_1,\Psi_1)\simeq \Delta M(\Psi_2,\Psi_1)$, where multiple states contribute simultaneously to the freeze-out DM abundance. For larger mass splittings, the strength of co-annihilation channels weakens, and the relic density becomes too large (green region). In this scenario, the results again show no significant dependence on $Z'$ mass (compare Figures~\ref{fig:relic3b} and \ref{fig:relic4c}).

\subsubsection*{Case IV: $\mathbf{M_{\rm \textbf{DM}}=1000}~\text{GeV}$ (Figure~\ref{fig:relic3a} and Figure~\ref{fig:relic3a})}     
    In this regime, the DM mass lies above the scalar mass and far from the $Z'$ resonance. Both annihilation and coannihilation channels are inefficient, and as a result, the relic density is always too large. Hence, no viable parameter points remain (i.e., all are overabundant).

Overall from Figures~\ref{fig:relic3} and \ref{fig:relic4}, we observe that at low $M_{\rm DM}$~($<400$ GeV), co-annihilations via $\Psi_2$ and $S_1$ can play a crucial role in deciding the obtained relic abundance, while at intermediate masses~($400-800$ GeV), resonance-assisted DM annihilations provide large allowed regions from the DM relic. For high DM masses~($>800$ GeV), both annihilation and co-annihilation become inefficient, leading to predominantly overabundant final relics.
 
\subsubsection{Direct Detection Constraints}
\label{subsec:DD}
Direct detection searches are among the most concrete approaches for detecting particle DM candidates. They aim to observe possible nuclear recoils produced through elastic DM-nucleon interactions. Within our model, such interactions can arise predominantly through the exchange of scalar states $h$ and $s$, as well as interactions mediated by neutral gauge bosons $Z$ and $Z'$. Moreover, the presence of the additional scalar $S_1$ provides a distinctive channel for such scatterings, thereby significantly affecting the phenomenology of direct detection constraints.
\begin{figure}[htbp]
\centering
\begin{subfigure}[b]{0.48\textwidth}
    \centering
    \includegraphics[width=\textwidth]{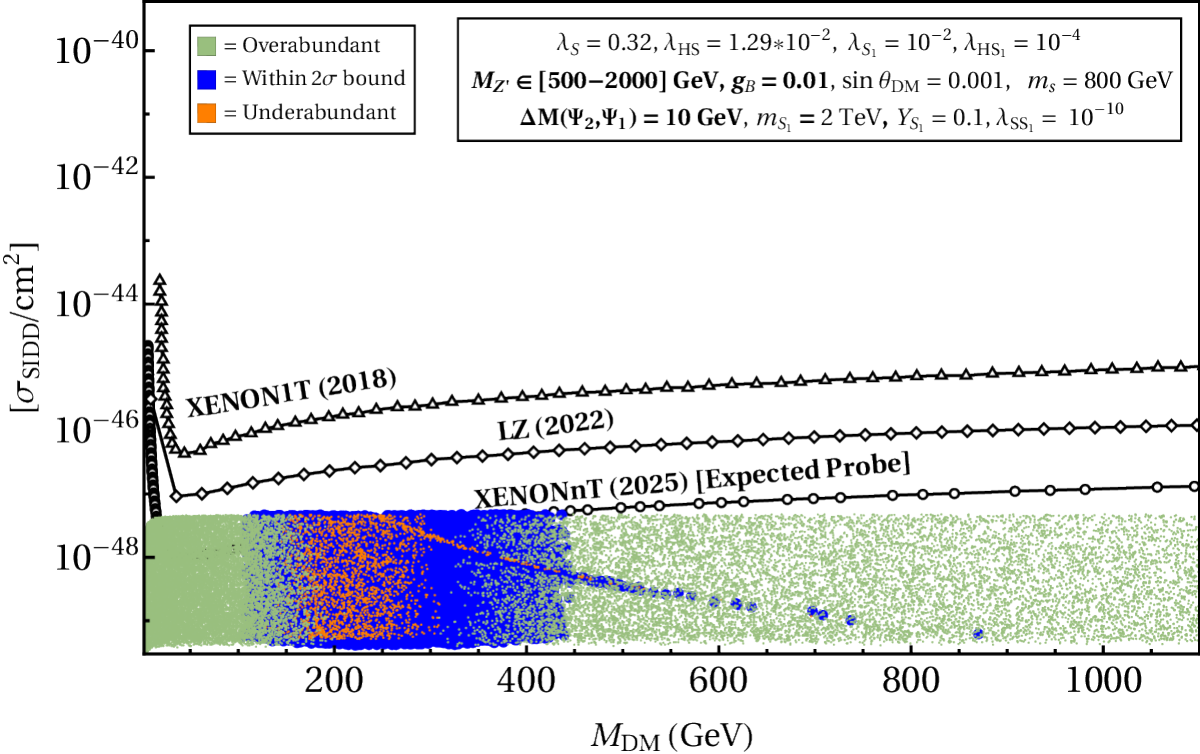}
    \caption{\small $\Delta M(\Psi_2,\Psi_1)=10~\text{GeV}, g_B = 0.01$}
    \label{fig:dd_a}
\end{subfigure}
\hfill
\begin{subfigure}[b]{0.48\textwidth}
    \centering
    \includegraphics[width=\textwidth]{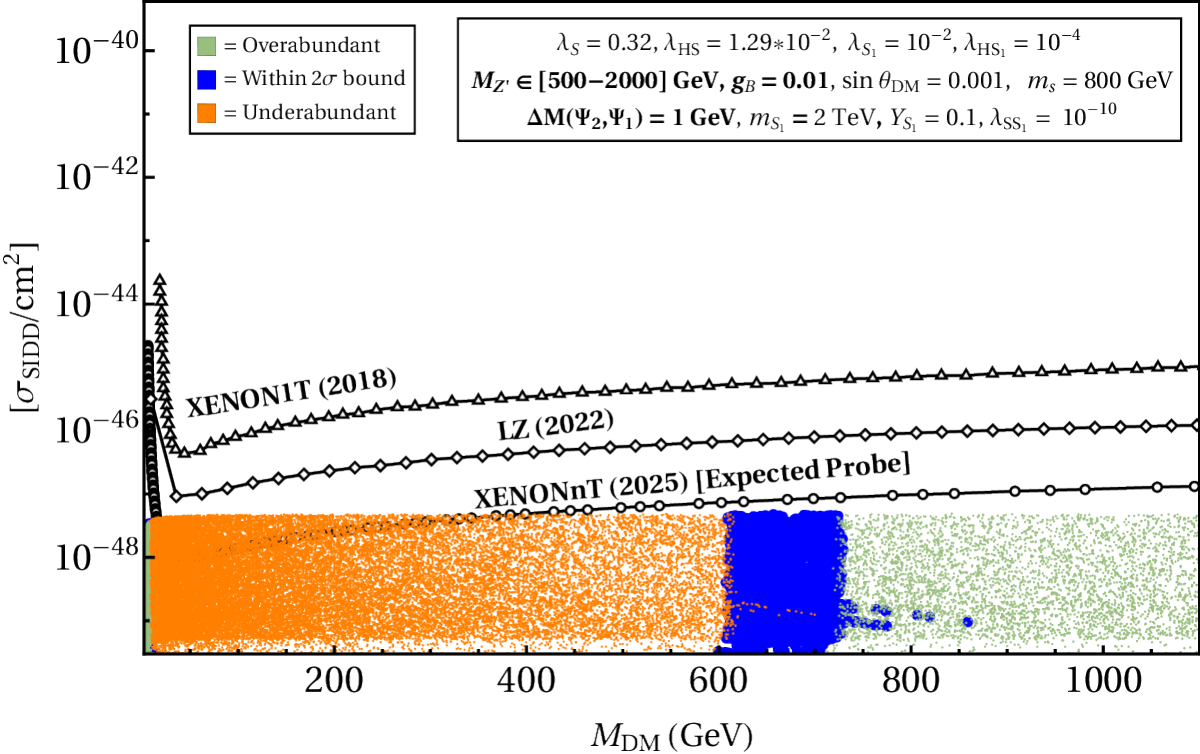}
    \caption{\small $\Delta M(\Psi_2,\Psi_1)=1~\text{GeV}, g_B = 0.01$}
    \label{fig:dd_b}
\end{subfigure}

\vspace{0.5cm}

\begin{subfigure}[b]{0.48\textwidth}
    \centering
    \includegraphics[width=\textwidth]{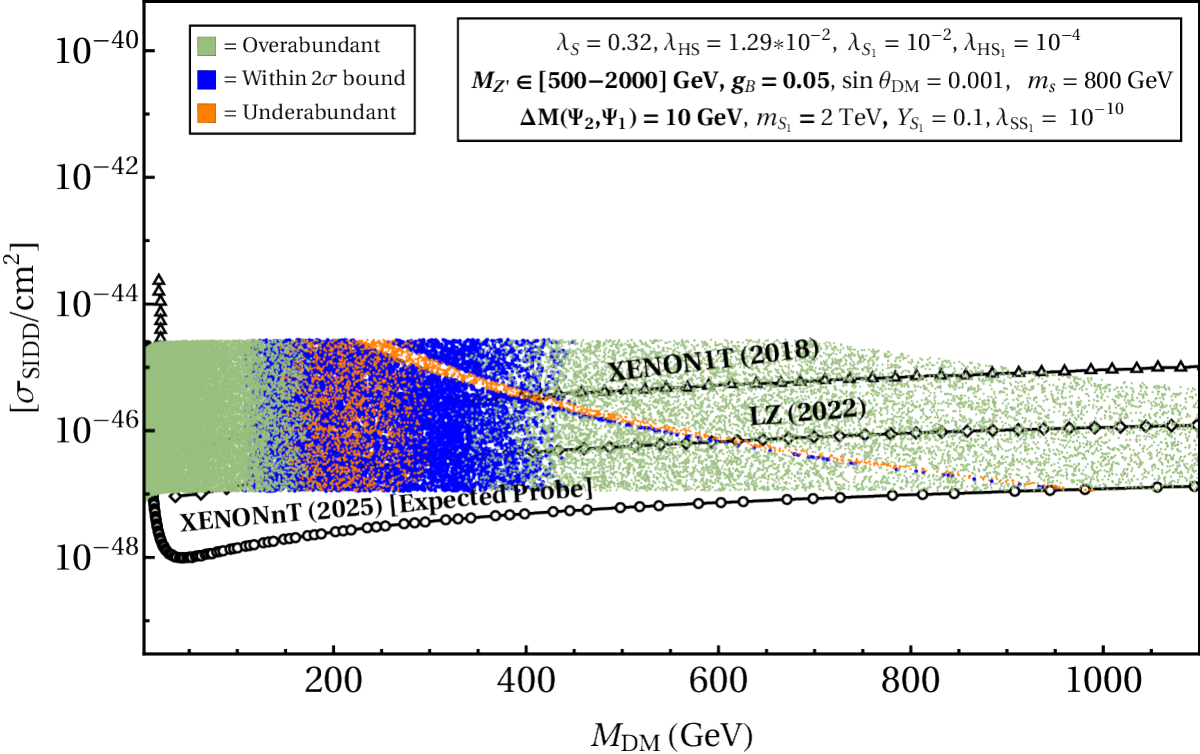}
    \caption{\small $\Delta M(\Psi_2,\Psi_1)=10~\text{GeV}, g_B = 0.05$}
    \label{fig:dd_c}
\end{subfigure}
\hfill
\begin{subfigure}[b]{0.48\textwidth}
    \centering
    \includegraphics[width=\textwidth]{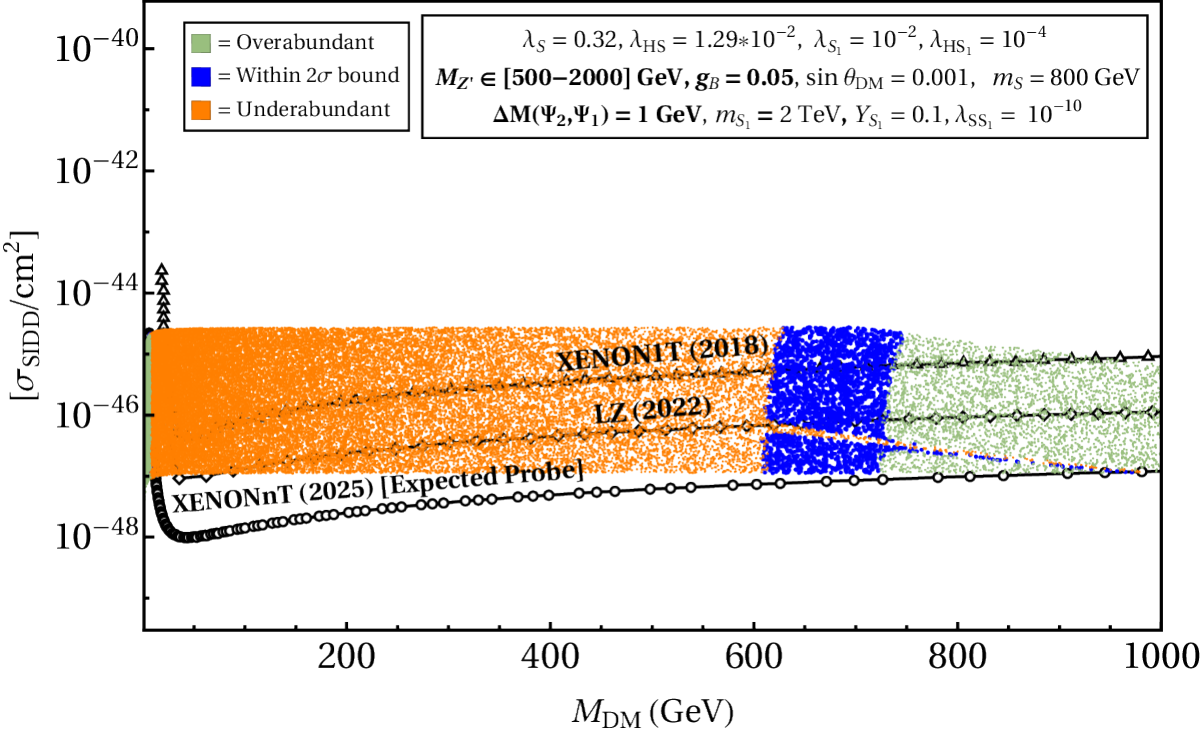}
    \caption{\small $\Delta M(\Psi_2,\Psi_1)=1~\text{GeV}, g_B = 0.05$}
    \label{fig:dd_d}
\end{subfigure}

\caption{\small Spin-independent dark matter - nucleon scattering cross section $\sigma_{\rm SIDD}$ as a function of the dark matter mass $M_{\rm DM}$ for different benchmark scenarios. The top row corresponds to $M_{\rm DM}=10~\text{GeV}$ 
with (a) $\Delta M(\Psi_2,\Psi_1)=1~\text{GeV}$ and (b) $\Delta M(\Psi_2,\Psi_1)=10~\text{GeV}$, while the bottom row corresponds to $M_{\rm DM}=1~\text{GeV}$ with (c) $\Delta M(\Psi_2,\Psi_1)=1~\text{GeV}$ and (d) $\Delta M(\Psi_2,\Psi_1)=10~\text{GeV}$. In all cases, the fixed parameters are 
$\lambda_{S}=0.32$, $\lambda_{HS}=1.29\times 10^{-2}$, $\lambda_{S_1}=10^{-2}$, $\lambda_{HS_1}=10^{-4}$, $\lambda_{SS_1}=10^{-10}$, $\sin\theta_{\rm DM}=0.001$, $m_{s}=800~\text{GeV}$, $m_{S_1}=2~\text{TeV}$, and $Y_{S_1}=0.1$. The $Z'$ mass $M_{Z'}$ are varied within 
$M_{Z'}\in[500,2000]~\text{GeV}$ with two fixed gauge coupling $g_B = 0.01,0.05$, corresponding 
to different $U(1)_B$ breaking scales $v_B$. Current limits from \texttt{XENON1T (2018)}, 
\texttt{LZ (2022)}, and projected \texttt{XENONnT (2025)} limits are also shown.}
\label{fig:dd1}
\end{figure}
The spin-independent DM nucleon scattering cross section, $\sigma_{\rm SIDD}$, is predominantly controlled by the $U(1)_B$ symmetry-breaking scale $v_B$. Other model parameters, such as the mass splittings and Yukawa couplings, have subdominant effects. In our analysis, we focus on the effects of two input parameters on the obtained SIDD cross-section: $v_B$ and $Y_{S_1}$. The parameter space for an obtained cross-section value is also used simultaneously to determine the DM relic abundance at that point, thereby identifying viable regions that meet both SIDD and relic requirements. As seen in the previous discussion, the obtained DM relic can be significantly affected by the strength of coannihilations via $\Psi_2$, and thus the mass splitting $\Delta M(\Psi_2,\Psi_1)$ is also allowed to vary for the scans presented below. The fixed input parameters for these scans are chosen as $\lambda_{S}=0.32$, $\lambda_{HS}=1.29\times 10^{-2}$, 
$\lambda_{S_1}=10^{-2}$, $\lambda_{HS_1}=10^{-4}$, 
$\lambda_{SS_1}=10^{-10}$, $\sin\theta_{\text{DM}}=0.001$, $m_{s}=800~\text{GeV}$, and $m_{S_1}=2~\text{TeV}$.  

In Figure~\ref{fig:dd1}, we show the obtained results for the spin-independent DM-nucleon scattering cross section $\sigma_{\text{SIDD}}$ as a function of the DM mass $M_{\rm DM}$ for two benchmark values of the fermion mass splitting, $\Delta M(\Psi_2,\Psi_1)=10~\text{GeV}$ (left column plots) and $\Delta M(\Psi_2,\Psi_1)=1~\text{GeV}$ (right column plots). The upper and lower rows correspond to two values for gauge coupling, $g_B=0.01$ and $g_B=0.05$, respectively. In each case, the $Z'$ mass is varied within the range $M_{Z'} \in [500,2000]~\text{GeV}$. 
Since $M_{Z'}$ is related to the $U(1)_B$ breaking scale by the relation given in Eq.~(\ref{eq:zpmass}), the scan over $M_{Z'}$ effectively corresponds to scanning over $v_B$ for a given $g_B$. Therefore, the plot features in this figure arise from variations in $v_B$.
\begin{figure}[htbp]
\centering
 \begin{subfigure}[b]{0.490\textwidth}
        \centering
        \includegraphics[width=\textwidth]{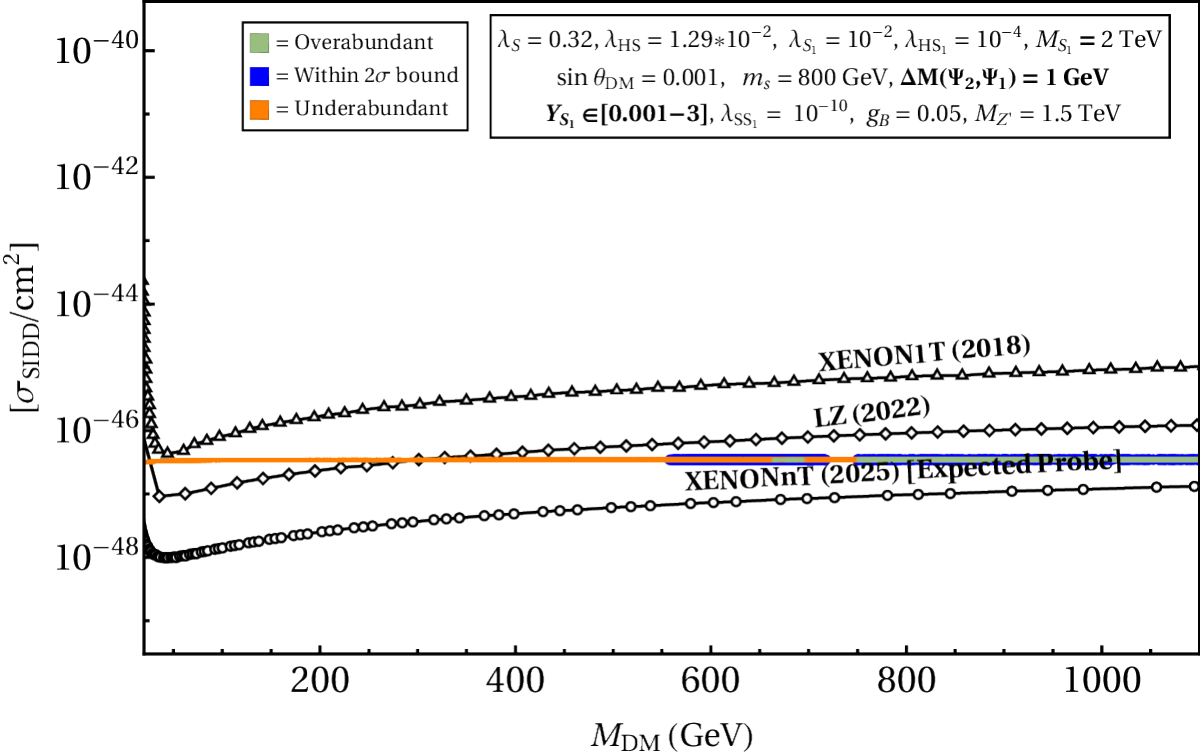}
        \caption{\small }
        \label{fig:relic6a}
    \end{subfigure}
    \hfill
 \vspace{0.5cm}
    \begin{subfigure}[b]{0.490\textwidth}
        \centering
        \includegraphics[width=\textwidth]{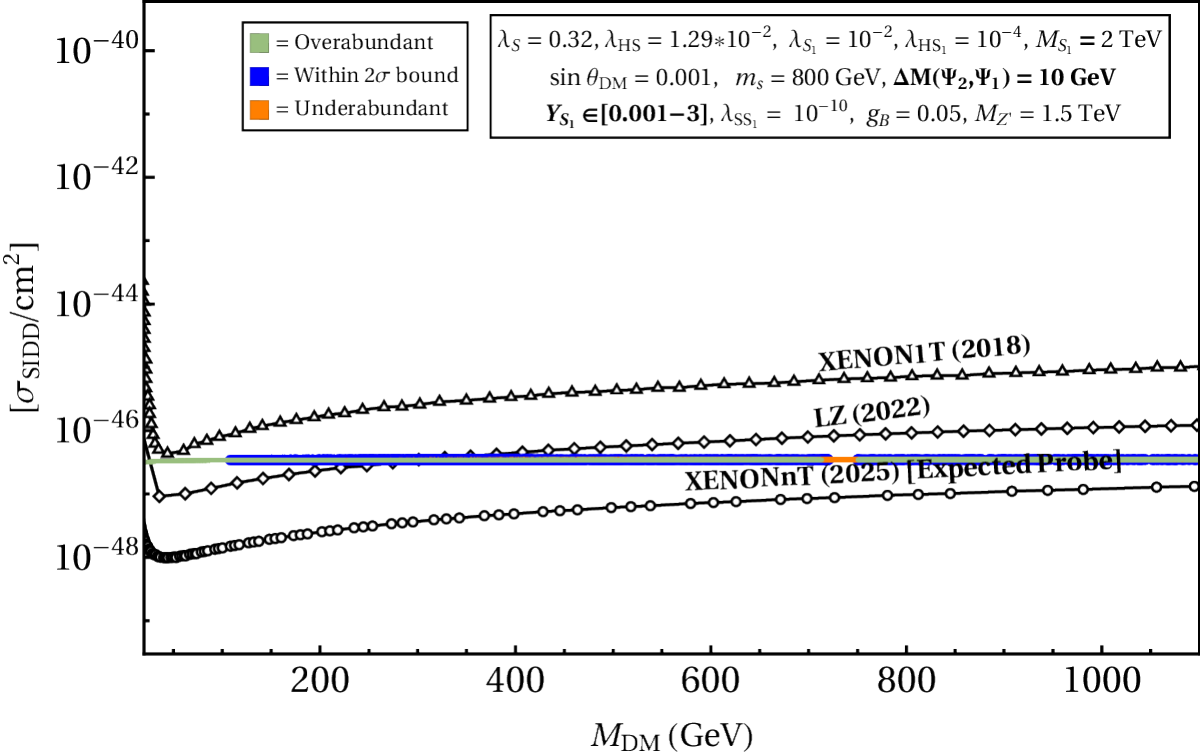}
        \caption{\small }
        \label{fig:relic6b}
    \end{subfigure}
\caption{\small Spin-independent dark-matter-nucleon scattering cross section 
$\sigma_{\rm SIDD}$ as a function of the dark-matter mass $M_{\rm DM}$. The results are shown for two benchmark choices of the input parameters: 
(a) $Y_{S_1} \in [0.001,3]$ with $\Delta M(\Psi_2,\Psi_1)=1~\text{GeV}$, and 
(b) $Y_{S_1} \in [0.001,3]$ with $\Delta M(\Psi_2,\Psi_1)=10~\text{GeV}$. 
The fixed parameters are $\lambda_{S}=0.32$, $\lambda_{HS}=1.29\times 10^{-2}$, 
$\lambda_{S_1}=10^{-2}$, $\lambda_{HS_1}=10^{-4}$, $\lambda_{SS_1}=10^{-10}$, $\sin\theta_{\rm DM}=0.001$, $m_{s}=800~\text{GeV}$, $m_{S_1}=2~\text{TeV}$, $g_B=0.05$, and $M_{Z'}=1.5~\text{TeV}$. Current bounds from \texttt{XENON1T (2018)}, \texttt{LZ (2022)}, and the projected sensitivity of \texttt{XENONnT (2025)} are overlaid for comparison.}
\label{fig:dd2}
\end{figure}
From the plots, it can be observed that the obtained value of $\sigma_{\rm SIDD}$ depends on the value of $M_{Z'}$ and hence on $v_B$. We find that increasing $v_B$ suppresses the effective DM-nucleon coupling, thereby reducing the DM-nucleon cross-section strength,~$\sigma_{\rm SIDD}$. This dependence explains the band-like regions obtained in all the plots in Figure~\ref{fig:dd1} when $M_{Z'}$ is scanned over a range of $500-2000$ GeV. Additionally, the choice of mass splitting value plays a significant role in shaping the relic-density-allowed plot regions. For smaller splitting ($\Delta M(\Psi_2,\Psi_1)=1~\text{GeV}$), efficient co-annihilation reduces the obtained relic density, leading to large under-abundant regions (orange colored). Conversely, for $\Delta M(\Psi_2,\Psi_1)=10~\text{GeV}$, the co-annihilation strength is relatively weaker, resulting in more plot area under relic-favorable (blue colored) region. The apparent difference between the $g_B=0.01$ (top row) and $g_B=0.05$ (bottom row) cases comes through the dependence of $v_B$ on $M_{Z'}$ values for a given $g_B$. For larger $g_B$~(bottom row), a given $M_{Z'}$ corresponds to a smaller $v_B$, which enhances the cross section, while for smaller $g_B$~(top row) the corresponding $v_B$ is larger, yielding smaller cross sections. In summary, the dominant effect in these plots originates from the variation of the breaking scale $v_B$ (through the scan of $M_{Z'}$), while the mass splitting $\Delta M(\Psi_2,\Psi_1)$ determines the role of co-annihilations on the obtained DM relic values. The combined impact of $v_B$ and $\Delta M(\Psi_2,\Psi_1)$ dictates which regions of the parameter space remain viable against the direct-detection limits from \texttt{XENON1T(2018)}~\cite{XENON:2018voc}, \texttt{LZ(2022)}~\cite{LZ:2022lsv} and under scrutiny from the expected data of \texttt{XENONnT(2025)}~\cite{XENON:2020kmp}.

Similarly, Figure~\ref{fig:dd2} illustrates the dependence of $\sigma_{\rm SIDD}$ on the Yukawa coupling, $Y_{S_1}$ over a range $Y_{S_1} \in [0.001,3]$. In this case, the obtained $\sigma_{\rm SIDD}$ remains essentially independent of the value of $Y_{S_1}$ and thus thin horizontal plot lines color coded with the values of obtained DM relic as a function of $\Delta M(\Psi_2,\Psi_1)$ are obtained for both Figures~\ref{fig:relic6a}~$(\Delta M(\Psi_2,\Psi_1)=1~\text{GeV})$ and~\ref{fig:relic6b}~($\Delta M(\Psi_2,\Psi_1)=10~\text{GeV})$. For given parameter space and input values, the obtained $\sigma_{\rm SIDD}$ is allowed from the LZ data for DM masses greater than $300$ GeV, for both the plots. Additionally, for~\ref{fig:relic6b}, almost all the plot points allowed by the LZ data are also satisfied by the relic density requirements.
\subsection{In-Direct Signatures}
\label{subsec:IDD}
In this subsection, we focus on the detection of DM particles via their late-time annihilations into Standard Model particles, which can produce observable gamma rays. These processes may be observed in places where the DM density is still fairly high, yielding detectable signals from DM annihilations. Consequently, indirect detection efforts focus on identifying excess fluxes of such particles that cannot be attributed to known astrophysical processes, particularly in regions expected to host large DM densities, such as the Galactic center~\cite{Strong:2005zx, Maier:2008vw, Thompson:2008rw}. These observations may offer indirect evidence for DM through its annihilation channels. For a Dirac-type DM candidate (as considered in this work), the resulting gamma-ray spectrum typically consists of a broad continuum of final-state radiation (FSR). Due to the continuum nature of the spectrum, it is challenging to identify a distinct mono-energetic gamma-ray line. Consequently, in such scenarios, the predicted diffuse gamma-ray flux is usually compared with experimental observations rather than searching for sharp spectral features. In the plots below, we compute the expected diffuse gamma-ray flux from DM annihilation as a function of DM mass, accounting for the DM density distribution within the Galaxy. The presence of the $Z'$ boson in the model introduces new late-time channels that lead to DM annihilations into SM final states. The Feynman diagrams corresponding to these $Z'$ mediated processes are shown in the Appendix. 
\begin{figure}[htbp]
 \centering
      \begin{subfigure}[b]{0.49\textwidth}
        \centering
        \includegraphics[width=\textwidth]{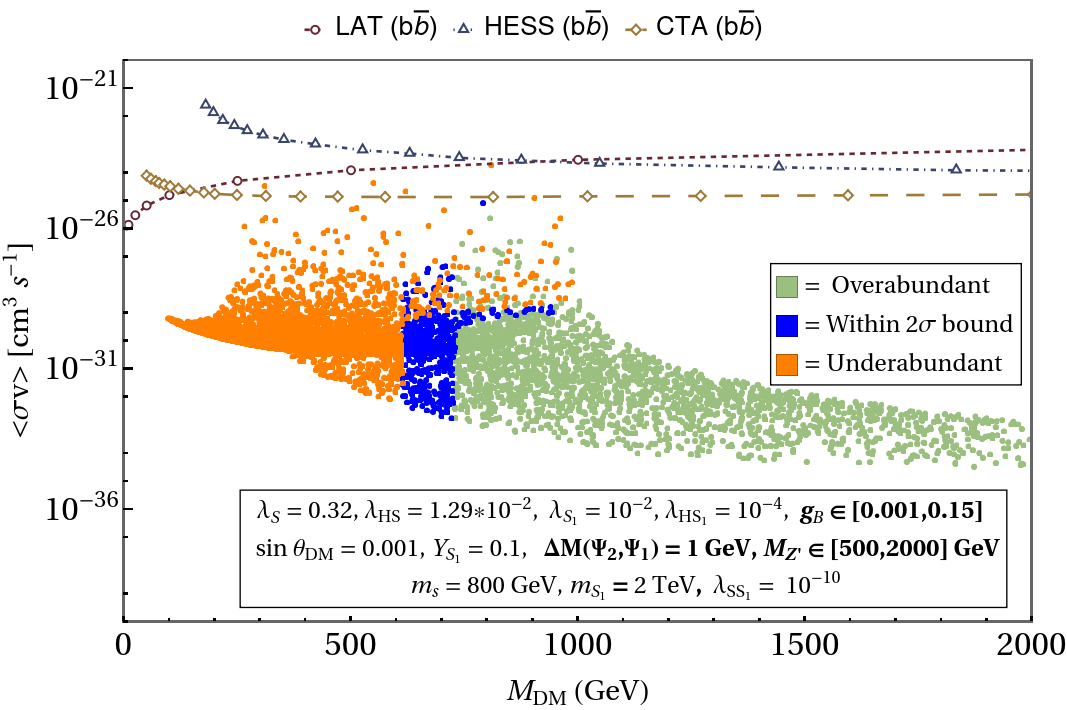}
        \caption{\small $b\bar{b}$ channel with $\Delta M(\Psi_2, \Psi_1)=1$ GeV}
        \label{fig:idd1a}
    \end{subfigure}
    \hfill
    \begin{subfigure}[b]{0.49\textwidth}
        \centering
        \includegraphics[width=\textwidth]{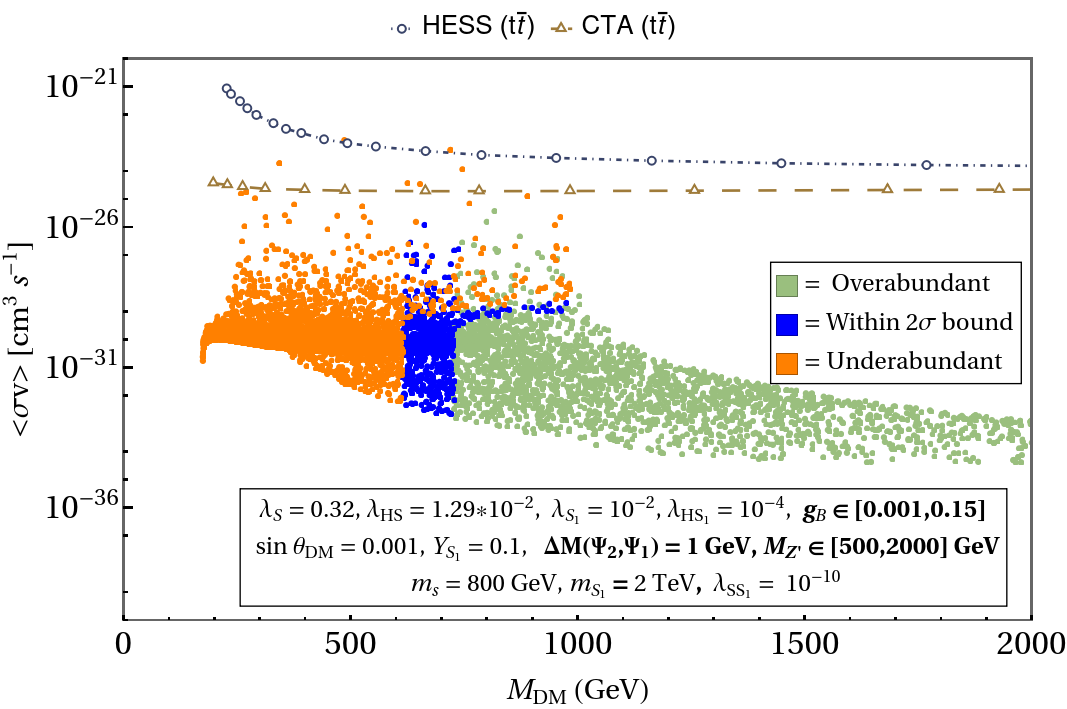}
        \caption{\small $t\bar{t}$ channel with $\Delta{M(\Psi_2, \Psi_1)}=1$ GeV}
        \label{fig:idd1b}
    \end{subfigure}
    
    \vspace{0.5cm} 
    
    \begin{subfigure}[b]{0.49\textwidth}
        \centering
        \includegraphics[width=\textwidth]{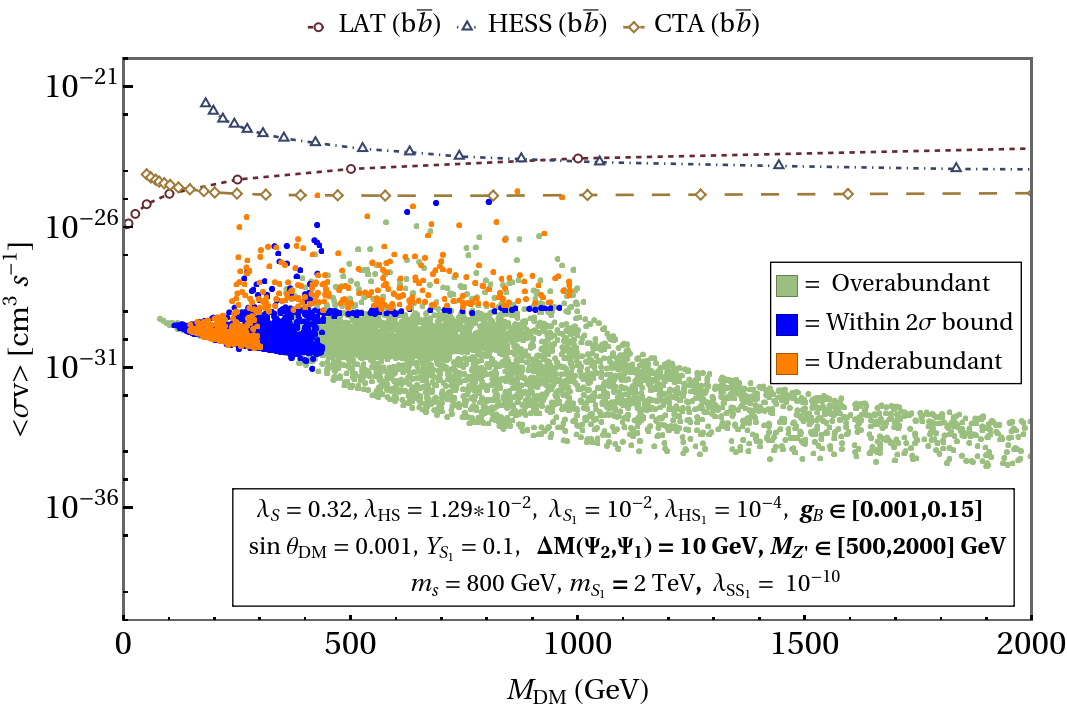}
        \caption{\small $b\bar{b}$ channel with $\Delta{M(\Psi_2, \Psi_1)}=10$ GeV}
        \label{fig:idd1c}
    \end{subfigure}
    \hfill
    \begin{subfigure}[b]{0.49\textwidth}
        \centering
        \includegraphics[width=\textwidth]{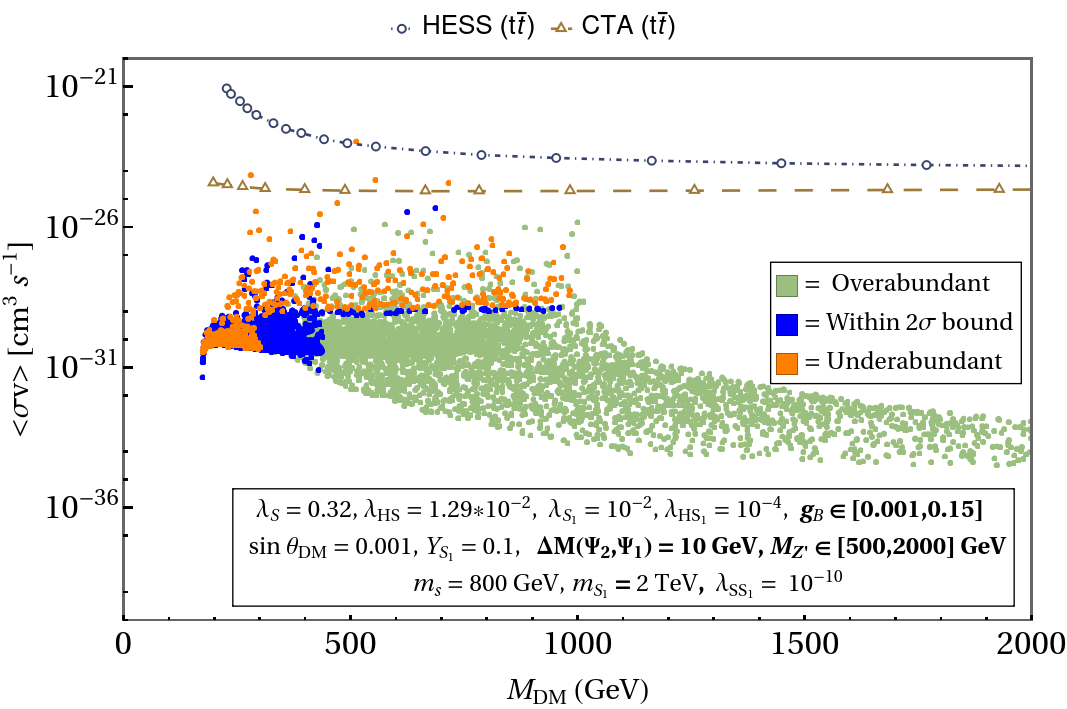}
        \caption{\small $t\bar{t}$ channel with $\Delta{M(\Psi_2, \Psi_1)}=10$ GeV}
        \label{fig:idd1d}
    \end{subfigure}
    \caption{\small Thermally averaged annihilation cross section $\langle\sigma v\rangle$ as a function of the DM mass $M_{\rm DM}$ for two benchmark values $\Delta{M(\Psi_2, \Psi_1)}=1 ~\text{GeV}$ and $10~\text{GeV}$, considering annihilation into $b\bar{b}$ and $t\bar{t}$ final states. The results are shown for varying $M_{Z'}\in[500,2000]$ and $g_B\in[0.001, 0.15]$ corresponding to varying breaking scale $v_B$, while the relic-density allowed (blue), underabundant (orange), and overabundant (green) regions are indicated. Current indirect-detection bounds from \texttt{LAT}, \texttt{HESS}, and projected \texttt{CTA} sensitivities are also displayed for comparison.}
    \label{fig:idd1}
\end{figure}
The analysis specifically considers the dependence on the DM mass and relevant model parameters, such as the Yukawa coupling $Y_{S_1}$ and the mass splitting $\Delta M(S_1, \Psi_1)$, which can modify the annihilation rates through co-annihilation or resonance effects. The predicted gamma-ray fluxes are then compared with current observational limits from \texttt{Fermi-LAT}~\cite{Fermi-LAT:2011vow,Fermi-LAT:2015bhf}, \texttt{HESS}~\cite{HESS:2006vje,HESS:2013rld} and \texttt{CTA}~\cite{CTAConsortium:2017dvg,CTAO:2024wvb}, allowing us to identify regions of parameter space that are consistent with, or excluded by these astrophysical constraints. These signatures can serve as complementary probes to relic density requirements and direct detection constraints for the given DM candidate in our framework.

In Figure~\ref{fig:idd1}, we show the behavior of the velocity-averaged annihilation cross section $\langle\sigma v\rangle$ as a function of the dark-matter mass $M_{\rm DM}$ for two benchmark values of the mass splitting $\Delta M(\Psi_2,\Psi_1)=1$~GeV (Figures~\ref{fig:idd1a} and ~\ref{fig:idd1b}) and $\Delta M(\Psi_2,\Psi_1)=10$~GeV (Figures~\ref{fig:idd1c} and ~\ref{fig:idd1d}). The annihilation final states considered are the $b\bar{b}$ channel (Figures~\ref{fig:idd1a} and ~\ref{fig:idd1c}) and the $t\bar{t}$ channel (Figures~\ref{fig:idd1b} and ~\ref{fig:idd1d}). The input parameters are fixed to $\lambda_{S}=0.32$, $\lambda_{HS}=1.29\times 10^{-2}$, $\lambda_{S_1}=10^{-2}$, $\lambda_{HS_1}=10^{-4}$, $\lambda_{SS_1}=10^{-10}$, $\sin\theta_{\rm DM}=0.001$, $m_{s}=800~\text{GeV}$, $m_{S_1}=2~\text{TeV}$, and $Y_{S_1}=0.1$, while the gauge coupling and mediator mass are varied in the ranges $g_B \in [0.001,0.15]$ and $M_{Z'} \in [500,2000]$~GeV. Current bounds from \texttt{LAT}, \texttt{HESS}, and the projected sensitivity of \texttt{CTA} are also shown for comparison in the dashed line. We find that the blue points satisfying the observed relic density (within $2\sigma$) typically yield annihilation cross sections below the current indirect-detection limits, though sizable regions remain testable in upcoming experiments for both $b\bar{b}$ and $t\bar{t}$ final states. Additionally, the dependence of the final DM relic on the coannihilations via $\Psi_2$ remains evident as we compare the results on top panel plots~(where $\Delta M(\Psi_2,\Psi_1)=1$~GeV) with the bottom panel plots~(where $\Delta M(\Psi_2,\Psi_1)=10$~GeV).

\section{Flavor Physics Analysis}\label{sec:flavor}
In this section, we investigate the impact of the baryonic gauge theory with a mixed-state DM sector on rare $b \to s \ell^+ \ell^-$ transitions. The additional gauge interactions arising from the $U(1)_B$ symmetry, together with the mixing in the dark sector, induce the flavor-changing effects. These contributions can alter the theoretical predictions for semileptonic $B$ meson decays and potentially explain deviations observed in current experimental data. To quantify these effects, we analyze the available measurements from exclusive decay channels, including $B \to (K^{(*)}, \phi)~\ell^+ \ell^-$, and use them to constrain the parameter space of the new physics interactions.

\subsection{Effective Hamiltonian}
The effective Hamiltonian governing the rare semileptonic
$b \to s \ell^+ \ell^-$ transition can be written as
\cite{Ali:1999mm,Altmannshofer:2008dz,Sahoo:2015qha}
\begin{align}\label{Ham-SM}
\mathcal{H}_{\mathrm{eff}}
&= -\frac{4 G_F}{\sqrt{2}}\, V_{tb} V_{ts}^*
\Bigg[
C_7^{\mathrm{eff}}\, \mathcal{O}_7
+ C_7^{\prime}\, \mathcal{O}_7^{\prime}
+ \sum_{i=9,10,S,P}
\left(
(C_i + C_i^{\mathrm{NP}})\, \mathcal{O}_i
+ C_i^{\prime\,\mathrm{NP}}\, \mathcal{O}_i^{\prime}
\right)
\Bigg],
\end{align}
where $G_F$ is the Fermi constant and $V_{ij}$ denote the CKM matrix elements. The local operators $\mathcal{O}_i^{(\prime)}$ encode the short-distance structure of the $b \to s \ell^+ \ell^-$ interaction and are defined (for $\ell = \mu$) as
\begin{eqnarray*}
\mathcal{O}_7^{(\prime)} &=&
\frac{e}{16\pi^2}\, m_b\,
(\bar{s}\sigma_{\mu\nu} P_{R(L)} b)\, F^{\mu\nu}, \\[4pt]
\mathcal{O}_9^{(\prime)} &=&
\frac{e^2}{16\pi^2}\,
(\bar{s}\gamma_\mu P_{L(R)} b)\,
(\bar{\mu}\gamma^\mu \mu),
\qquad
\mathcal{O}_{10}^{(\prime)} =
\frac{e^2}{16\pi^2}\,
(\bar{s}\gamma_\mu P_{L(R)} b)\,
(\bar{\mu}\gamma^\mu \gamma_5 \mu), \\[4pt]
\mathcal{O}_S^{(\prime)} &=&
\frac{e^2}{16\pi^2}\, m_b\,
(\bar{s} P_{R(L)} b)\,
(\bar{\mu}\mu),
\qquad
\mathcal{O}_P^{(\prime)} =
\frac{e^2}{16\pi^2}\, m_b\,
(\bar{s} P_{R(L)} b)\,
(\bar{\mu}\gamma_5 \mu).
\end{eqnarray*}

Here, $P_{L,R} = (1 \mp \gamma_5)/2$ are the chirality projection
operators, and the Wilson coefficients $C_i^{(\prime)}$ parametrize the
corresponding short-distance interactions.
Within the Standard Model, only the unprimed vector and axial-vector
operators contribute, while the primed and (pseudo)scalar operators
arise solely from physics beyond it.
\subsection{The $b \to s \ell^+ \ell^-$ decay observables}

The rare semileptonic decay $B \to K \ell^+ \ell^-$ is mediated by the flavor-changing neutral current transition $b \to s \ell^+ \ell^-$, which is loop-suppressed in the Standard Model. Owing to its relatively simple hadronic structure, this channel provides a theoretically clean environment for probing short-distance dynamics. In the Standard Model, the $q^2$-dependent differential branching ratio can be expressed as~\cite{Bouchard:2013eph}

\begin{equation}
\frac{d\mathcal{BR}}{dq^2}
= \tau_{B_c}
\left( 2 a_\ell + \frac{2}{3} c_\ell \right),
\end{equation}
where the quantities $a_\ell$ and $c_\ell$ encode the dynamical
information about the decay and are defined as
\begin{eqnarray}
a_\ell &=&
\frac{G_F^2 \alpha_{\rm EW}^2 |V_{tb} V_{ts}^*|^2}{2^9 \pi^5 m_{B}^3}
\beta_\ell \sqrt{\lambda}
\Big[
q^2 |F_P|^2
+ \frac{\lambda}{4} (|F_A|^2 + |F_V|^2)
+ 4 m_\ell^2 m_{B}^2 |F_A|^2
\nonumber \\
&&
+ 2 m_\ell (m_{B}^2 - m_{K}^2 + q^2)
{\rm Re}(F_P F_A^*)
\Big], \\
c_\ell &=&
-\frac{G_F^2 \alpha_{\rm EW}^2 |V_{tb} V_{ts}^*|^2}{2^9 \pi^5 m_{B}^3}
\beta_\ell \sqrt{\lambda}
\frac{\lambda \beta_\ell^2}{4}
(|F_A|^2 + |F_V|^2).
\end{eqnarray}
The parameter $\lambda$, and the lepton mass factor $\beta_\ell$ appearing above are given by
\begin{eqnarray}
\lambda &=&
q^4 + m_{B}^4 + m_K^4
- 2 (m_{B}^2 m_{K}^2 + m_{B}^2 q^2 + m_{K}^2 q^2), \\
\beta_\ell &=&
\sqrt{1 - \frac{4 m_\ell^2}{q^2}} .
\end{eqnarray}
The hadronic dynamics are encoded in the parameters
$F_P$, $F_V$, and $F_A$, which can be written in terms of the short-distance Wilson coefficients and the $B \to K$ transition form factors as
\begin{eqnarray}
F_P &=&
- m_\ell C_{10}
\left[
f_+
- \frac{m_{B_c}^2 - m_{D_s}^2}{q^2} (f_0 - f_+)
\right], \\
F_V &=&
C_9^{\rm eff} f_+
+ \frac{2 m_b}{m_{B_c} + m_{D_s}} C_7^{\rm eff} f_T, \\
F_A &=&
C_{10} f_+ .
\end{eqnarray}
For the vector final state, the decay amplitude of
$B \to (K^*, \phi)~\ell^+ \ell^-$ can be derived from the effective Hamiltonian given in Eq.~(\ref{Ham-SM}). The corresponding differential branching ratios read~\cite{Descotes-Genon:2015uva}
\begin{equation}
\frac{d\mathcal{BR}}{dq^2}
=  \frac{1}{4} \tau_B
\left[
3 I_1^c + 6 I_1^s - I_2^c - 2 I_2^s
\right],
\end{equation}
where the angular coefficients $I_i$ depend on $q^2$ and are listed in Appendix~\ref{Ang_coeff}.

In addition to the branching ratio, several angular observables provide valuable probes of the underlying dynamics. Among these, we consider the forward-backward asymmetry $A_{FB}$ and the longitudinal polarization fraction $F_L$, defined as~\cite{Descotes-Genon:2012isb}
\begin{equation}
F_L(q^2)
= \frac{3 I_1^c - I_2^c}{3 I_1^c + 6 I_1^s - I_2^c - 2 I_2^s},
\qquad
A_{FB}(q^2)
= \frac{3 I_6}{3 I_1^c + 6 I_1^s - I_2^c - 2 I_2^s}.
\end{equation}
Another set of notable observables are the six form-factor-independent (FFI) observables~\cite{Descotes-Genon:2012isb}, which are defined as

{\small
\begin{equation}
\begin{aligned}
\langle P_1 \rangle &= \frac{\int_{\text{bin}} dq^2\, I_{3}}
{2 \int_{\text{bin}} dq^2 I_{2s}}, 
\qquad
\langle P_2 \rangle = \frac{\int_{\text{bin}} dq^2\, I_{6}}
{8 \int_{\text{bin}} dq^2 I_{2s}}, \\[0.15cm]
\langle P_3 \rangle &= -\frac{\int_{\text{bin}} dq^2\, I_{9}}
{4 \int_{\text{bin}} dq^2 I_{2s}}, 
\qquad
\langle P_4 \rangle = \frac{\int_{\text{bin}} dq^2 \sqrt{2}\, I_{4}}
{\sqrt{-\int_{\text{bin}} dq^2 I_{2}^c \int_{\text{bin}} dq^2 (2I_{2s}-I_3)}}, \\[0.15cm]
\langle P_5 \rangle &= \frac{\int_{\text{bin}} dq^2 I_{5}}
{\sqrt{-\int_{\text{bin}} dq^2 2I_{2}^c \int_{\text{bin}} dq^2 (2I_{2s}+I_3)}},
\qquad
\langle P_6 \rangle = -\frac{\int_{\text{bin}} dq^2 I_{7}}
{\sqrt{-\int_{\text{bin}} dq^2 2I_{2}^c \int_{\text{bin}} dq^2 (2I_{2s}-I_3)}} .
\end{aligned}
\end{equation}
}

A modified set of optimized clean observables $P_4', P_5', P_6'$ can also be defined, which are related to the original observables via~\cite{Descotes-Genon:2012isb}
\begin{equation}
\langle P^{\prime}_4 \rangle = \frac{\int_{bin}dq^2 I_{4}}{2 \sqrt{-\int_{bin}dq^2 I_{2}^c \int_{bin}dq^2 I_{2}^s}}, \hspace{0.5cm}
 \langle P^{\prime}_5 \rangle = \frac{\int_{bin}dq^2 I_{5}}{2 \sqrt{-\int_{bin}dq^2 I_{2}^c \int_{bin}dq^2 I_{2}^s}}, 
\end{equation}
\begin{equation}
\hspace{0.1cm}
 \langle P^{\prime}_6 \rangle = -\frac{\int_{bin}dq^2 I_{7}}{2 \sqrt{-\int_{bin}dq^2 I_{2}^c \int_{bin}dq^2 I_{2}^s}},\hspace{0.5cm}
 \langle P^{\prime}_8 \rangle = -\frac{\int_{bin}dq^2 I_{8}}{2 \sqrt{-\int_{bin}dq^2 I_{2}^c \int_{bin}dq^2 I_{2}^s}}.
\end{equation}
These clean observables are particularly useful as they reduce dependence on hadronic form factors, making them robust probes for testing the Standard Model predictions and exploring potential new physics effects.
To probe lepton flavor universality (LFU), one can define observables that directly compare the muon and electron final states. In particular, the quantities $\langle Q_4 \rangle$ and $\langle Q_5\rangle$ are given by
\cite{Descotes-Genon:2013vna,Capdevila:2016ivx}
\begin{equation}
\langle Q_{4} \rangle
= \langle P_4^{\prime\,\mu} \rangle - \langle P_4^{\prime\,e} \rangle,
\qquad
\langle Q_5 \rangle
= \langle P_5^{\prime\,\mu} \rangle - \langle P_5^{\prime\,e} \rangle.
\end{equation}
Finally, the ratio of branching fractions between the muon and electron modes for $B \to K^{(*)} \ell^+ \ell^-$ decays are given as
\begin{equation}
R_{K^{(*)}}(q^2)
=
\frac{
\mathcal{BR}(B \to K^{(*)} \mu^+ \mu^-)
}{
\mathcal{BR}(B \to K^{(*)} e^+ e^-)
}.
\end{equation}
The measurement of these observables allows comparisons with the SM contributions and can reveal possible new physics in $b \to s \mu^+ \mu^-$ transition decays. Using the above-discussed observables, we constrain the allowed new physics parameters to the relevant Wilson coefficients, which are analyzed below.
\subsection{Constraint on new physics parameters from $b \to s \mu^+\mu^-$ transitions}
Motivated by the persistent anomalies reported in semileptonic
$b \to s \mu^+ \mu^-$ transitions, we investigate possible contributions
arising from the scalar-assisted baryonic gauge framework.
Our analysis focuses on the set of observables that exhibit
statistically significant deviations from their SM expectations
and are therefore particularly sensitive to potential new physics effects.
Within this setup, the new physics contributions are obtained in the vector and
axial-vector Wilson coefficients $C_{9}^{\mathrm{NP}}$ and
$C_{10}^{\mathrm{NP}}$ governing the short-distance dynamics of
$b \to s \mu^+ \mu^-$ processes. Accordingly, we explore the viable regions of parameter space for the new physics encoded in these coefficients.
For this analysis, the complete set of $b \to s \mu^+ \mu^-$ observables employed in the fit is summarized below.

\begin{itemize}

\item \textbf{$b \to s \mu^+ \mu^-$ observables without lepton flavor universality}

We include a comprehensive set of observables induced by the
$b \to s \mu^+ \mu^-$ transition. In particular, we use the latest
\texttt{LHCb} measurement of the branching ratio
$\mathcal{B}~(B_s \to \mu^+ \mu^-)$ reported at Moriond 2021~\cite{LHCb:2021awg},
which is consistent with earlier combined results from \texttt{ATLAS}, \texttt{CMS},
and \texttt{LHCb}~\cite{ATLAS:2018cur,CMS:2019bbr,LHCb:2017rmj}.
We further incorporate the branching ratios of
$B \to K^{(*)}\mu^+\mu^-$~\cite{LHCb:2014cxe,LHCb:2016ykl} and
$B_s \to \phi \mu^+ \mu^-$~\cite{LHCb:2021zwz},
measured in different $q^2$ bins.
Additionally, the angular observables are also taken into account.
These include the forward-backward asymmetry $A_{FB}$,
the longitudinal polarization fraction $F_L$,
and the optimized observables $P_1$-$P_8^{\prime}$
measured in $B^{0(+)} \to K^{*0(+)} \mu^+ \mu^-$ decays
by \texttt{LHCb}~\cite{LHCb:2015svh,LHCb:2020gog,LHCb:2020lmf}.
We also include the measurement of $F_L$ in
$B_s \to \phi \mu^+ \mu^-$~\cite{LHCb:2021xxq}.

\item \textbf{Lepton flavor universality sensitive observables}

To test lepton flavor universality, we include the \texttt{LHCb} measurements
of the ratios $R_{K_S^0}$ and $R_{K^{*+}}$~\cite{LHCb:2021lvy},
together with the updated results for $R_{K^{(*)}}$~\cite{LHCb:2022vje}.
We also consider the \texttt{Belle} measurements of the optimized differences
$\langle Q_{4} \rangle$ and $\langle Q_{5} \rangle$~\cite{Belle:2016fev},
which provide additional sensitivity to non-universal new physics
effects.
\end{itemize}

In this study, we employ all the observables discussed above and conduct our analysis using the \texttt{flavio} package \cite{Straub:2018kue}.

\begin{figure}[htbp]
\centering
\includegraphics[width=4.5cm,height=4.5cm]{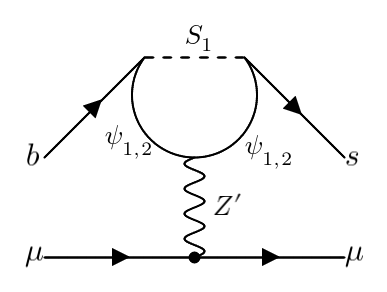}
\vspace{0.01cm}
\includegraphics[width=4.5cm,height=4.5cm]{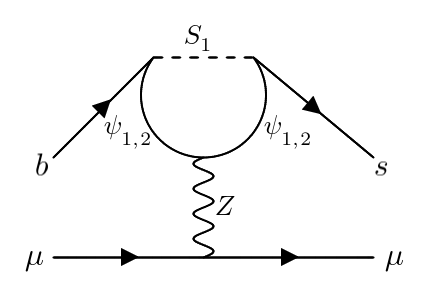}
\vspace{0.01cm}
\includegraphics[width=4.5cm,height=4.5cm]{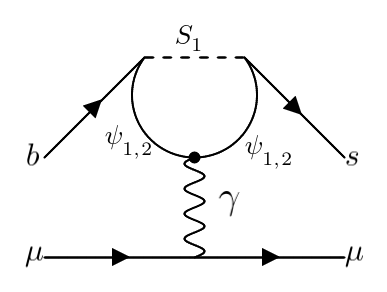}
\vspace{0.01cm}
\includegraphics[width=4.5cm,height=4.5cm]{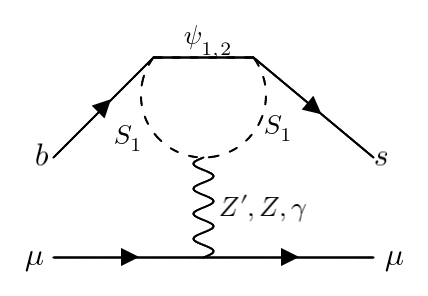}
\hspace{0.5cm}
\includegraphics[width=4.5cm,height=4.5cm]{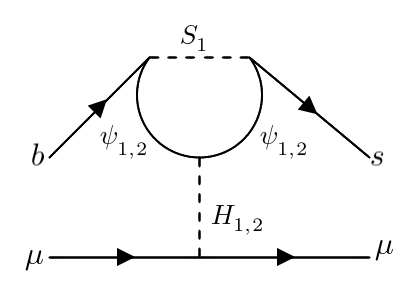}
\vspace{0.01cm}
\includegraphics[width=4.5cm,height=4.5cm]{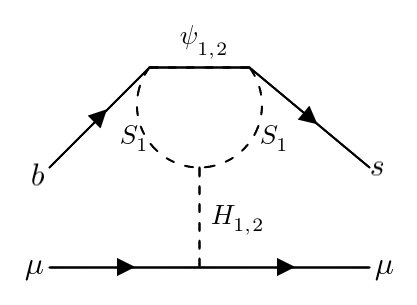}
\caption{Allowed penguin diagrams illustrating the $ b \to s \mu^+\mu^-$ transition.}
\label{penguin1}
\end{figure}
Now in the presence of new physics, the exclusive processes mediated by $b \to s \mu^+ \mu^-$ transitions receive additional contributions beyond the SM. The vertices arise from new interactions, including those associated with the baryonic gauge sector and the mixed-state DM field. These rare transitions receive non-zero contributions at the one-loop level and are shown in Figure~\ref{penguin1}. The effects can eventually be parametrized in terms of modified Wilson coefficients. These couplings include the exchange of the neutral gauge boson $Z'$, $Z$, $\gamma$ as well as the scalar states $H_{1,2}$, with scalar field $S_1$ and fermions $\psi_{1,2}$ propagating inside the loop, as illustrated in Figure~\ref{penguin1}. The penguin diagram in the bottom-left panel is suppressed by the factor $m_b/M_{\psi_{1, 2}}$, thus providing a negligible contribution. The loop functions associated with the Higgs contribution are suppressed by factors of $m_q M_{\psi_{1,2}} / m_{S_1}^2$ and therefore yield numerically negligible effects. Therefore, the Higgs-mediated effects, depicted in the bottom panel, are omitted. On the other hand, the new neutral gauge boson $Z'$ can generate flavor-changing effects at the quark level through loop-induced penguin diagrams. Although its couplings to charged leptons are suppressed, the corresponding $Z'$ contribution is retained for completeness. In addition, the loop-induced penguin diagrams mediated by the SM $Z$ boson and the photon provide relevant contributions to the effective $b \to s \ell^+ \ell^-$ operators. This is due to the presence of an electroweak doublet component and also the presence of the hypercharge in the DM candidate, which induces the couplings to these gauge bosons. Therefore, our analysis includes the $Z$, $Z'$, and $\gamma$ penguin contributions shown in the top panel. On the other hand, the $\gamma$-penguin diagrams generate effective $b \to s \ell^+ \ell^-$ interactions through vector and electromagnetic dipole operators. Consequently, no axial–vector operator is induced with the coefficient $C_{10}^{NP}$, and it remains unaffected by the photon exchange. So the contribution associated with the photon is restricted to the Wilson coefficients $C_9^{NP}$ and $C_7^{NP}$. However, the $C_7^{NP}$ contribution includes a factor of $m_s$. So, we have neglected this contribution in our analysis.

Now, in the presence of $Z^\prime$, $Z$ and $\gamma$ exchanging one loop diagram, the transition amplitudes of semileptonic $b \to s \ell^+ \ell^-$ decay process is given as
\begin{equation}\label{amp}
\begin{aligned}
\mathcal{M}_{Z'-\psi} ={}&
-\frac{1}{(4\pi)^2}(0.015)Y_{S_1}^2
 \frac{\mathcal{R}_{Z'-\psi}(a,b)}{M_{Z'}^{2}}
 \big[\bar{s}(p')\gamma^\mu P_L b(p)\big]
 \big[\bar{\ell}(q_2)\gamma_\mu \ell(q_1)\big] \\
&\phantom{-}
+\frac{1}{(4\pi)^2}(0.008)Y_{S_1}^2
 \frac{\mathcal{R}_{Z'-\psi}(a,b)}{M_{Z'}^{2}}
 \big[\bar{s}(p')\gamma^\mu P_L b(p)\big]
 \big[\bar{\ell}(q_2)\gamma_\mu\gamma_5 \ell(q_1)\big]
\end{aligned}
\end{equation}

\begin{equation}\label{amp}
\begin{aligned}
\mathcal{M}_{Z-\psi} ={}&
-\frac{1}{(4\pi)^2}Y_{S_1}^2 (0.020)
 \frac{\mathcal{R}_{Z'-\psi}(a,b)}{M_{Z'}^{2}}
 \big[\bar{s}(p')\gamma^\mu P_L b(p)\big]
 \big[\bar{\ell}(q_2)\gamma_\mu \ell(q_1)\big] \\
&\phantom{-}
+\frac{1}{(4\pi)^2}Y_{S_1}^2(0.015)
 \frac{\mathcal{R}_{Z'-\psi}(a,b)}{M_{Z'}^{2}}
 \big[\bar{s}(p')\gamma^\mu P_L b(p)\big]
 \big[\bar{\ell}(q_2)\gamma_\mu\gamma_5 \ell(q_1)\big]
\end{aligned}
\end{equation}
\begin{equation}\label{amp}
\begin{aligned}
\mathcal{M}_{\gamma-\psi} ={}&
-\frac{1}{(4\pi)^2}Y_{S_1}^2(0.5)
 \frac{\mathcal{R}_{Z'-\psi}(a,b)}{M_{Z'}^{2}}
 \big[\bar{s}(p')\gamma^\mu P_L b(p)\big]
 \big[\bar{\ell}(q_2)\gamma_\mu \ell(q_1)\big] 
\end{aligned}
\end{equation}
which, in comparison with the generalized effective Hamiltonian, provides additional new Wilson coefficients as
\begin{align}
C_{9}^{ NP, Z'-\psi } &=
-\frac{1}{4 \pi} \frac{\sqrt{2}}{4 G_F M_{Z^\prime}^2} (0.015) 
\frac{1}{\alpha _{em}} \frac{Y_{S_1}^2}{V_{tb}V_{ts}^*} 
\mathcal{R}_{Z'-\psi} (a, b),\\
C_{9}^{ NP, Z-\psi } &=
-\frac{1}{4 \pi} \frac{\sqrt{2}}{4 G_F M_{Z^\prime}^2} (0.020) 
\frac{1}{\alpha _{em}} \frac{Y_{S_1}^2}{V_{tb}V_{ts}^*} 
\mathcal{R}_{Z-\psi} (a, b),\\
C_{9}^{ NP, \gamma-\psi } &=
-\frac{1}{4 \pi} \frac{\sqrt{2}}{4 G_F M_{Z^\prime}^2} (0.5) 
\frac{1}{\alpha _{em}} \frac{Y_{S_1}^2}{V_{tb}V_{ts}^*} 
\mathcal{R}_{\gamma-\psi} (a, b).
\end{align}
and
\begin{align}
C_{10}^{ NP, Z'-\psi } &=
\frac{1}{4 \pi} \frac{\sqrt{2}}{4 G_F M_{Z^\prime}^2} (0.008) 
\frac{1}{\alpha _{em}} \frac{Y_{S_1}^2}{V_{tb}V_{ts}^*} 
\mathcal{R}_{Z'-\psi} (a, b),\\
C_{10}^{ NP, Z-\psi } &=
\frac{1}{4 \pi} \frac{\sqrt{2}}{4 G_F M_{Z^\prime}^2} (0.015) 
\frac{1}{\alpha _{em}} \frac{Y_{S_1}^2}{V_{tb}V_{ts}^*} 
\mathcal{R}_{Z-\psi} (a, b)
\end{align}
where the loop functions are given as
\begin{eqnarray}
\label{eq:R_Zp_psi}
\mathcal{R}_{Z'-\psi}(a,b)
&=& \Bigg[
\sin^2\theta_{\mathrm{DM}}
\bigg\{
\Big(
g_B \cos^2\theta_{\mathrm{DM}}
-\frac{3}{2} g_B \cos 2\theta_{\mathrm{DM}}
+0.017\,\sin^2\theta_{\mathrm{DM}}
\Big)
\nonumber\\
&&
\times
\left(
\frac{1}{8}
-\frac{5a^2-8a+3-2a^2\ln a}{8(a-1)^2}
\right)
+\Big(
0.5 g_B + 0.00028
+ g_B \cos^2\theta_{\mathrm{DM}}
\Big)
\nonumber\\
&&
\times
\left(
\frac{1}{8}
+\frac{-3+3a^2-8a\ln a+2a^2\ln a}{8(a-1)^2}
\right)
\bigg\}
\nonumber\\
+\hspace{-0.18in}
&&\cos^2\theta_{\mathrm{DM}}
\bigg\{
\Big(
- g_B \sin^2\theta_{\mathrm{DM}}
-\frac{3}{2} g_B \cos 2\theta_{\mathrm{DM}}
-0.017\,\sin^2\theta_{\mathrm{DM}}
-0.017 \cos 2\theta_{\mathrm{DM}}
\Big)
\nonumber\\
&&
\times
\left(
\frac{1}{8}
-\frac{5b^2-8b+3-2b^2\ln b}{8(b-1)^2}
\right)
+\Big(
-\frac{3}{2} g_B + 0.00028
\Big)
\nonumber\\
&&
\times
\left(
\frac{1}{8}
+\frac{-3+3b^2-8b\ln b+2b^2\ln b}{8(b-1)^2}
\right)
\bigg\}
\Bigg].
\end{eqnarray}
\begin{eqnarray}
\label{eq:R_Z_psi}
\mathcal{R}_{Z-\psi}(a,b)
&=& \Bigg[
\sin^2\theta_{\mathrm{DM}}
\bigg\{
\Big(
0.005 \cos 2\theta_{\mathrm{DM}}
-0.366\,\sin^2\theta_{\mathrm{DM}}
\Big)
\nonumber\\
&&
\times
\left(
\frac{1}{8}
-\frac{5a^2-8a+3-2a^2\ln a}{8(a-1)^2}
\right)
+\Big(
0.020 g_B -0.014
+0.005 \sin^2\theta_{\mathrm{DM}}
\Big)
\nonumber\\
&&
\times
\left(
\frac{1}{8}
+\frac{-3+3a^2-8a\ln a+2a^2\ln a}{8(a-1)^2}
\right)
\bigg\}
\nonumber\\
&&+
\cos^2\theta_{\mathrm{DM}}
\bigg\{
\Big(
0.371 \cos 2\theta_{\mathrm{DM}}
+0.376 \sin^2\theta_{\mathrm{DM}}
\Big)
\nonumber\\
\times
\hspace{-0.2in}
&&\left(
\frac{1}{8}
-\frac{5b^2-8b+3-2b^2\ln b}{8(b-1)^2}
\right)
+\Big(
0.01 \cos 2\theta_{\mathrm{DM}}
-0.02 g_B
+0.024 \sin^2\theta_{\mathrm{DM}}
\Big)
\nonumber\\
&&
\times
\left(
\frac{1}{8}
+\frac{-3+3b^2-8b\ln b+2b^2\ln b}{8(b-1)^2}
\right)
\bigg\}
\Bigg].
\end{eqnarray}
\begin{eqnarray}
\label{eq:R_gamma_psi}
\mathcal{R}_{\gamma-\psi}(a,b)
&=& \Bigg[
\sin^2\theta_{\mathrm{DM}}
\bigg\{
\Big(
g_B \cos^2\theta_{\mathrm{DM}}
-\frac{3}{2} g_B \cos 2\theta_{\mathrm{DM}}
+0.017\,\sin^2\theta_{\mathrm{DM}}
\Big)
\nonumber\\
&&
\times
\left(
\frac{1}{8}
-\frac{5a^2-8a+3-2a^2\ln a}{8(a-1)^2}
\right)
+\Big(
0.5 g_B + 0.00028
+ g_B \cos^2\theta_{\mathrm{DM}}
\Big)
\nonumber\\
&&
\times
\left(
\frac{1}{8}
+\frac{-3+3a^2-8a\ln a+2a^2\ln a}{8(a-1)^2}
\right)
\bigg\}
\nonumber\\
+\hspace{-0.18in}
&&\cos^2\theta_{\mathrm{DM}}
\bigg\{
\Big(
- g_B \sin^2\theta_{\mathrm{DM}}
-\frac{3}{2} g_B \cos 2\theta_{\mathrm{DM}}
-0.017\,\sin^2\theta_{\mathrm{DM}}
-0.017 \cos 2\theta_{\mathrm{DM}}
\Big)
\nonumber\\
&&
\times
\left(
\frac{1}{8}
-\frac{5b^2-8b+3-2b^2\ln b}{8(b-1)^2}
\right)
+\Big(
-\frac{3}{2} g_B + 0.00028
\Big)
\nonumber\\
&&
\times
\left(
\frac{1}{8}
+\frac{-3+3b^2-8b\ln b+2b^2\ln b}{8(b-1)^2}
\right)
\bigg\}
\Bigg]
\end{eqnarray}
with
\begin{align}
a &= \frac{M_{\psi_1}^2}{m_{S_1}^2}, & \hspace{-3.5cm}
b &= \frac{M_{\psi_2}^2}{m_{S_1}^2}.
\end{align}

Now, \begin{align}
C_9^{\text{NP}} &= C_9^{\text{NP,}\, Z' - \psi}
           + C_9^{\text{NP,}\, Z - \psi}
               + C_9^{\text{NP,}\, \gamma - \psi}, \\
C_{10}^{\text{NP}} &= C_{10}^{\text{NP,}\, Z' - \psi}
                   + C_{10}^{\text{NP,}\, Z - \psi}.
\end{align}
\begin{figure}[H]
\centering
\includegraphics[width=7.5cm,height=5.5cm]{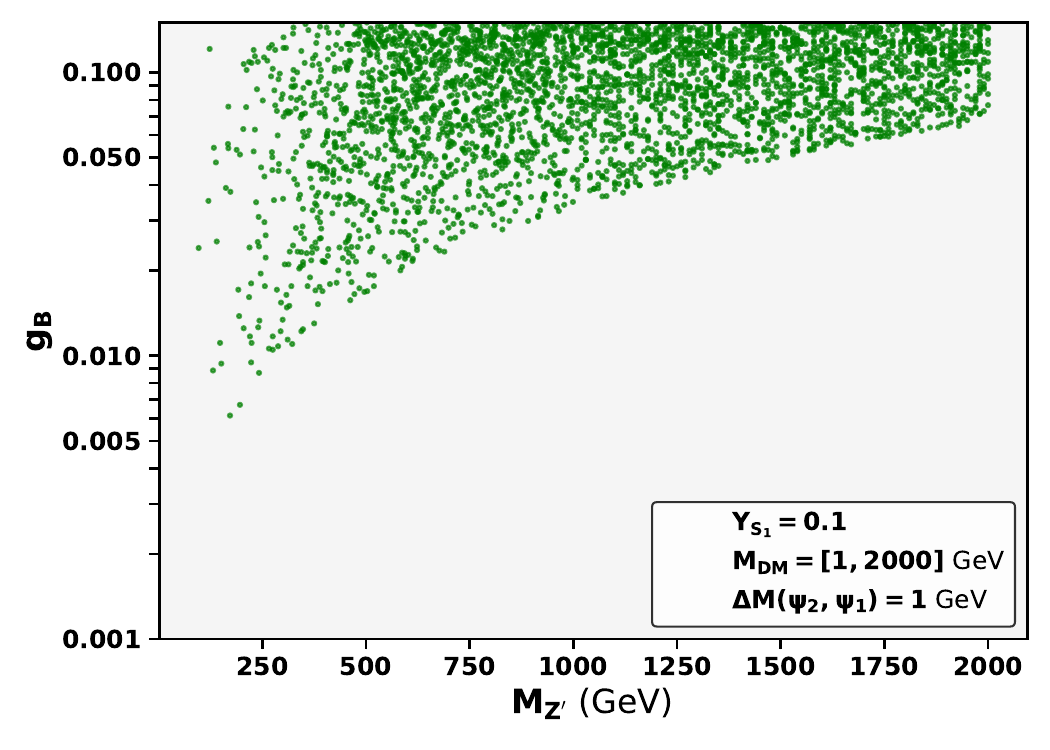}
\includegraphics[width=7.5cm,height=5.5cm]{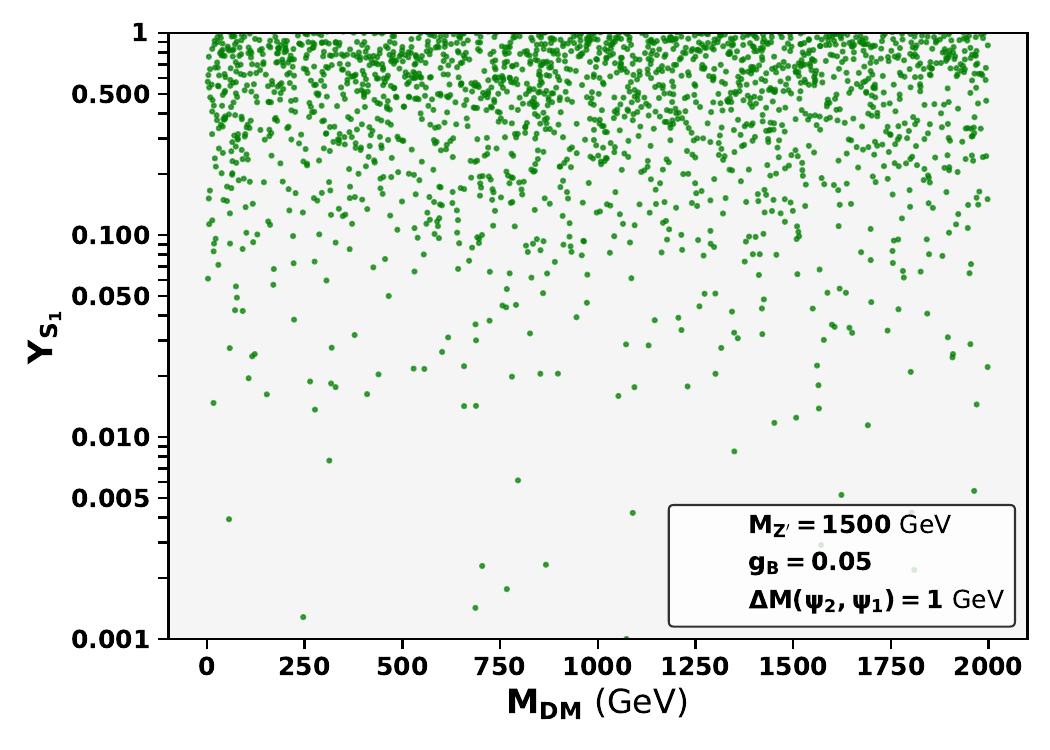}
\includegraphics[width=7.5cm,height=5.5cm]{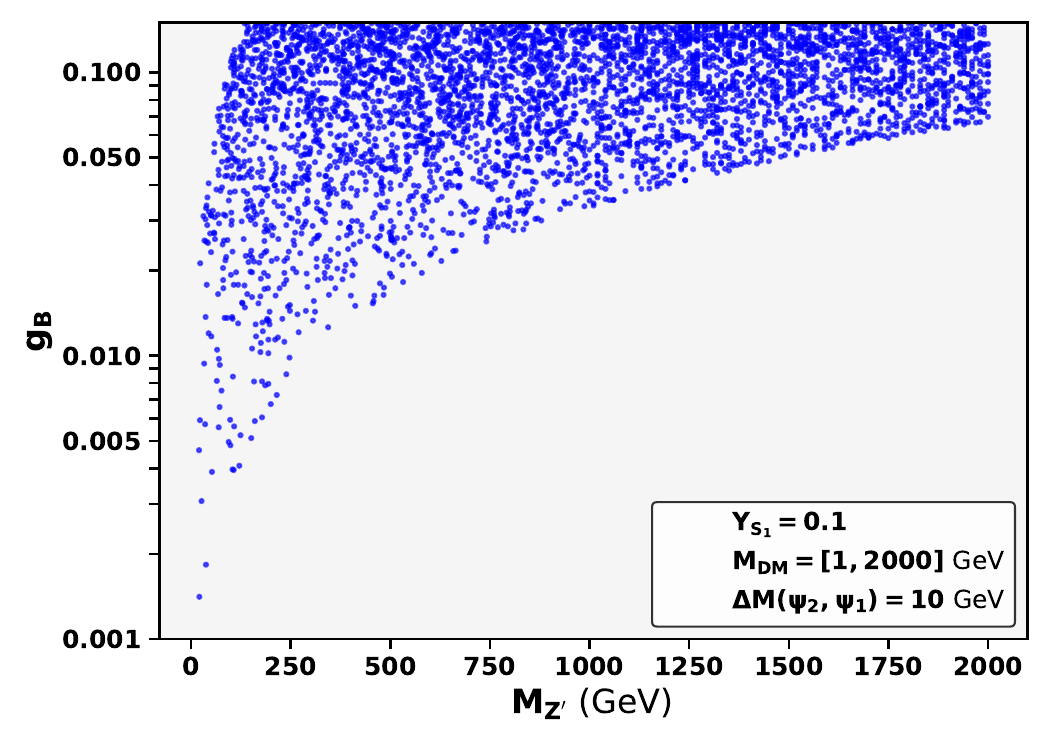}
\includegraphics[width=7.5cm,height=5.5cm]{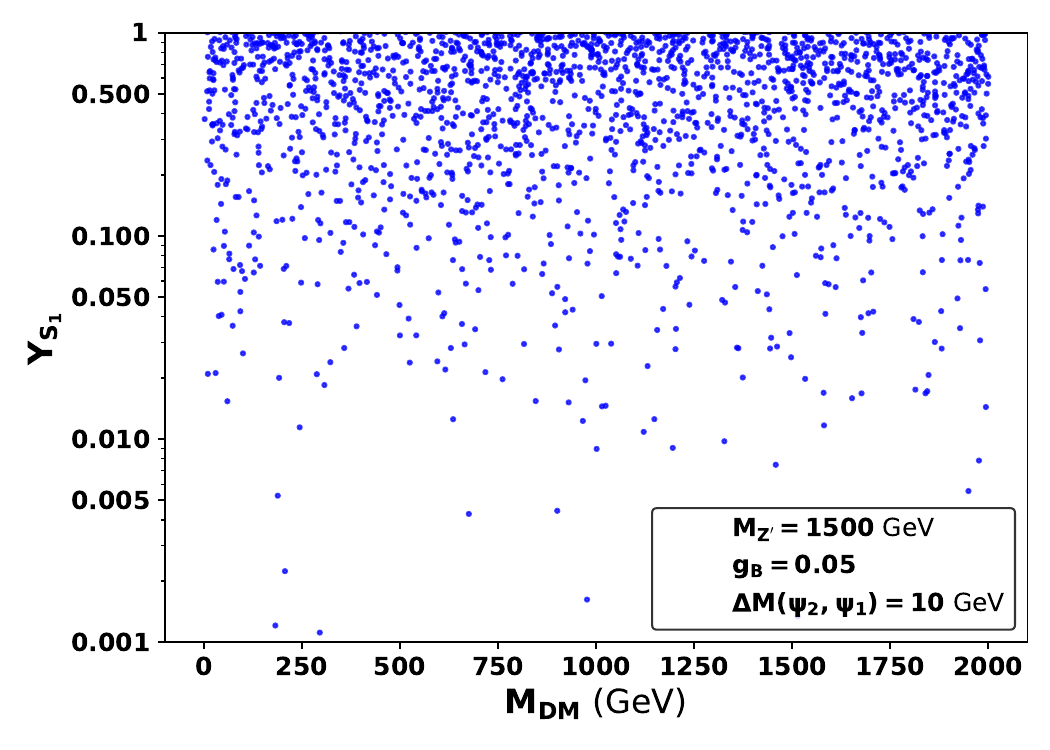}
\caption{Allowed parameter space in the $U(1)_B$ model. {\bf Top-left panel:} $U(1)_B$ gauge coupling $g_B$ versus $Z'$ mass $M_{Z'}$, {\bf Top-right  panel:} Yukawa coupling $Y_{S_1}$ versus DM mass $M_\mathrm{DM}$ for a mass splitting $\Delta M(\Psi_2,\Psi_1) = 1~\mathrm{GeV}$. Bottom panel: Same as top, but for $\Delta M(\Psi_2,\Psi_1)= 10~\mathrm{GeV}$, illustrating the impact of $\Delta M(\Psi_2,\Psi_1)$ on the allowed region.  Legends show benchmark values used in the scans, $M_{Z'} = 1500~\mathrm{GeV}$, $g_B = 0.05$, $Y_{S_1}=0.1$ and $\Delta M(\Psi_2,\Psi_1) = 1$ or $10~\mathrm{GeV}$.}
\label{Flavorconstraint}
\end{figure}
Having established the structure of the new physics contributions to the Wilson coefficients, we now discuss the viable domain arising from the parameters $g_B$, $Y_{S_1}$, $M_{Z'}$, and $M_{\psi_{1,2}}$. In our model, we assume the lighter $\psi_1$ particle to be DM, and denote its mass as $M_{\rm DM}$. Figure~\ref{Flavorconstraint} illustrates the allowed regions of the parameter space after incorporating all relevant phenomenological constraints. In the top-left panel, the correlation between the $U(1)_B$ gauge coupling $g_B$ and the $Z'$ boson mass $M_{Z'}$ is shown, indicating that larger gauge couplings require heavier mediator masses in order to remain consistent with experimental bounds. This behavior reflects the sensitivity of flavor observables and indirect searches to the strength of the new gauge interaction. On the other hand, the top-right panel shows the viable allowed region in the $Y_{S_1}$ - $M_{\mathrm{DM}}$ plane for a small mass splitting $\Delta M(\Psi_2,\Psi_1) = 1~\mathrm{GeV}$. In this nearly degenerate regime, the allowed region is relatively broad, indicating a mild dependence on the DM mass for moderate values of the Yukawa coupling. In contrast, the bottom panel shows that increasing the mass splitting to $\Delta M(\Psi_2,\Psi_1)= 10~\mathrm{GeV}$ leads to a notable modification of the allowed region. This highlights the important role played by the dark sector mass hierarchy in shaping the phenomenology of the model. Overall, these results emphasize that the interplay between the gauge sector, the Yukawa interactions, and the dark-sector mass spectrum imposes nontrivial constraints on the parameter space of the $U(1)_B$ model.

\section{Summarized discussion: Combined Flavor and DM Analysis}
\label{sec:cd}
\begin{figure}[!htbp]
\centering
\includegraphics[width=0.8\textwidth]{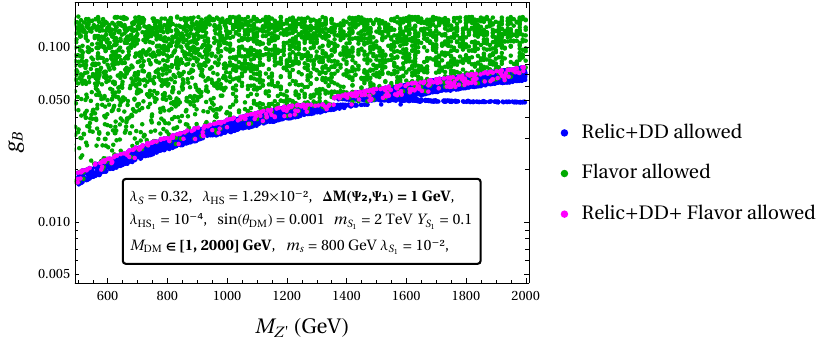}
\includegraphics[width=0.8\textwidth]{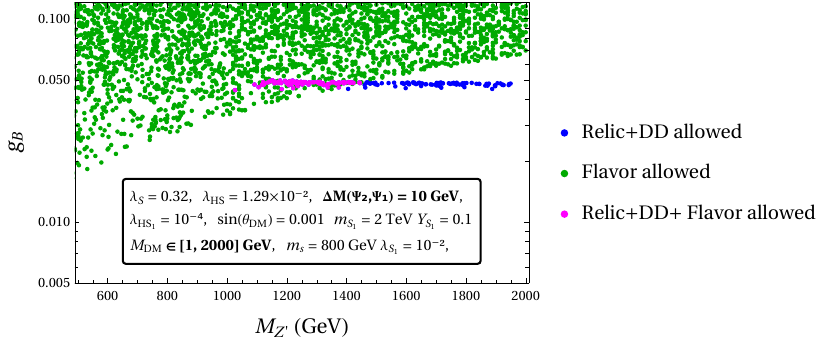}
\caption{Allowed common parameter space in the plane of $M_{Z'}-g_B$ from both DM and flavor phenomenology for {\bf Top Panel:} $\Delta M(\Psi_2,\Psi_1)=1~\text{GeV}$ and {\bf Bottom Panel:} $\Delta M(\Psi_2,\Psi_1)=10~\text{GeV}$, with fixed parameters: $\lambda_S=0.32$, $\lambda_{HS}=1.29\times 10^{-2}$, $\lambda_{S_1}=10^{-2}$, $\lambda_{HS_1}=10^{-4}$, $\sin\theta_{\text{DM}} = 0.001$, $m_s=800~\text{GeV}$, $m_{S_1}=2~\text{TeV}$. The DM mass is varied between $1$ to $2000$ GeV. }
\label{fig:common1}
\end{figure}

\begin{figure}[!htbp]
\centering
\begin{subfigure}{0.8\textwidth}
    \centering
    \includegraphics[width=\textwidth]{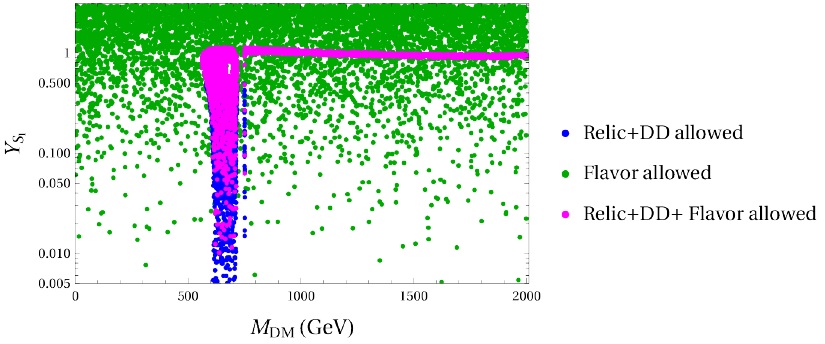}
\end{subfigure}

\vspace{0.4cm}

\begin{subfigure}{0.8\textwidth}
    \centering
    \includegraphics[width=\textwidth]{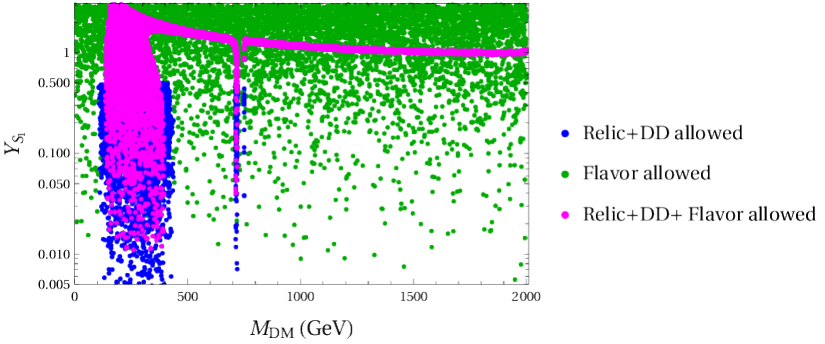}
\end{subfigure}
\caption{Allowed common parameter space in the plane of $M_{\rm DM}-Y_{S_1}$ from both DM and flavor phenomenology for {\bf Top Panel:} $\Delta M(\Psi_2,\Psi_1)=1~\text{GeV}$ and {\bf Bottom Panel:} $\Delta M(\Psi_2,\Psi_1)=10~\text{GeV}$, with fixed parameters:
$\lambda_S=0.32$, $\lambda_{HS}=1.29\times 10^{-2}$, $\lambda_{S_1}=10^{-2}$, $\lambda_{HS_1}=10^{-4}$, $\lambda_{SS_1}=10^{-10}$, $\sin\theta_{\text{DM}} = 0.001$, $m_s=800~\text{GeV}$, $m_{S_1}=2~\text{TeV}$, $M_{Z'} = 1.5~\text{TeV}$, and $g_{B} = 0.05$.}
\label{fig:common2}
\end{figure}

The simultaneous study of flavor physics and DM phenomenology is crucial for establishing a consistent new physics framework. In the present model, the same set of parameters that governs the interactions of the dark sector also enters the physics associated with the \(b \to s \mu^+ \mu^-\) transitions. As a result, constraints from the flavor sector restrict the viable parameter space of the dark sector, and vice versa. Therefore, a combined analysis goes beyond treating flavor physics and DM as independent probes. This approach allows us to identify regions of the parameter space that are simultaneously compatible with the flavor anomalies and DM constraints, thereby providing a robust test of the underlying new physics. To this end, we perform a combined numerical scan of the relevant parameter space consistent with both DM and flavor phenomenology and present our results in Figs.~\ref{fig:common1} and~\ref{fig:common2}.

\begin{table}[!htbp]
\centering
\renewcommand{\arraystretch}{1.0}
\resizebox{\textwidth}{!}{%
\begin{tabular}{|c|c|c|c|c||c|c|c|c|c|}
\hline\hline

\multicolumn{10}{|c|}{
\shortstack{
$\lambda_S = 0.32$,~$\lambda_{HS} = 1.29 \times 10^{-2}$,~$\lambda_{S_1} = 10^{-2}$,~$\lambda_{HS_1} = 10^{-4}$,~$\lambda_{SS_1} = 10^{-10}$\\
$\sin\theta_{\rm DM} = 0.001$,~$m_S = 800~\mathrm{GeV}$,~$m_{S_1} = 2~\mathrm{TeV}$,~$Y_{S_1} = 0.1$
}
} \\
\hline

\multicolumn{5}{|c||}{\textbf{$\Delta M(\Psi_2,\Psi_1) =
1~\mathrm{GeV}$}} &
\multicolumn{5}{c|}{\textbf{$\Delta M(\Psi_2,\Psi_1) =
10~\mathrm{GeV}$}} \\
\hline

$g_B$ & $M_{Z'}~(\mathrm{GeV})$  &  $M_{\rm DM}~(\mathrm{GeV})$ & $\Omega h^2$ &$\sigma_{\rm SIDD/cm^2}$  &
$g_B$ & $M_{Z'}~(\mathrm{GeV})$ &  $M_{\rm DM}~(\mathrm{GeV})$ & $\Omega h^2$ &$\sigma_{\rm SIDD/cm^2}$  \\
\hline

0.0493 & 1145.8 & 535.4 & 0.117 & $9.53\times10^{-47}$ &
0.0600 & 1575.5 &  617.8 & 0.097 & $5.86\times10^{-47}$ \\

0.0481 & 1625.9 &784.7 & 0.095 & $2.19\times10^{-47}$ &
0.0537 & 1395.8 & 610.9 & 0.095 & $6.11\times10^{-47}$ \\

0.0459 & 1374.0 &  649.3 & 0.126 & $3.54\times10^{-47}$ &
0.0660 & 1711.3 & 607.8 & 0.094 & $6.16\times10^{-47}$ \\

0.0487 & 1255.3 &592.3 & 0.11 & $6.36\times10^{-47}$ &
0.0410 & 1148.8 &  658.1 & 0.11 & $4.56\times10^{-47}$ \\

0.0481 & 1722.8 &828.4 & 0.114 & $1.75\times10^{-47}$ &
0.0604 & 1674.1 & 651.9 & 0.11 & $4.79\times10^{-47}$ \\

0.0470 & 1897.19 &918.52 & 0.11 & $1.10\times10^{-47}$ &
0.0399 & 1045.32 & 615.21 & 0.095 & $5.96\times10^{-47}$ \\

0.0485 & 1443.17 & 688.32 & 0.1112 & $3.58\times10^{-47}$ &
0.0229 & 600.49 & 617.21 & 0.095 & $5.92\times10^{-47}$ \\

0.0490 & 1311.16 &620.08 & 0.114 & $5.47\times10^{-47}$ &
0.0288 & 754.83 & 616.60 & 0.095 & $5.92\times10^{-47}$ \\

0.0484 & 1133.79 &531.17 & 0.11 & $9.27\times10^{-47}$ &
0.0462 & 1295.61 & 658.02 & 0.11 & $4.55\times10^{-47}$ \\

0.0481 & 1518.22 &725.75 & 0.115 & $2.84\times10^{-47}$ &
0.0282 & 733.04 &610.56 & 0.095 & $6.08\times10^{-47}$ \\

\hline\hline
\end{tabular}
}
\caption{Randomly selected allowed points in the $(g_B,\,M_{Z'})$ plane for $\Delta M(\Psi_2,\Psi_1)=1~\mathrm{GeV}$~(\textbf{left side}) and $\Delta M(\Psi_2,\Psi_1)=10~\mathrm{GeV}$~(\textbf{right side}). For each allowed point, the corresponding DM mass is also listed in the table. These points satisfy the relic density constraint within $2\sigma$, are consistent with SIDD bounds from \texttt{LZ (2022)}, and are allowed by Flavor physics constraints as shown in Figure~\ref{fig:common1}.}
\label{tab:benchmark1}
\end{table}

In Figure~\ref{fig:common1}, we observe that the overall allowed region~(pink colored) is significantly restricted once both flavor and DM constraints are imposed. Additionally, on comparing the \textbf{Top Panel} with the \textbf{Bottom Panel}, we observe that increasing the mass splitting $\Delta M(\Psi_2,\Psi_1)$ leads to a noticeable reduction of the viable parameter space in the $g_B$-$M_{Z'}$ plane. This is due to the increased co-annihilation efficiency assisted via the next-to-lightest state $\Psi_2$, which drives the final DM relic to be under-abundant for a vast region of parameter space. 

\begin{table}[!htbp]
\centering
\renewcommand{\arraystretch}{1.0}
\resizebox{\textwidth}{!}{%
\begin{tabular}{|c|c|c|c||c|c|c|c|}
\hline\hline

\multicolumn{8}{|c|}{
\shortstack{
$\lambda_S = 0.32$, 
$\lambda_{HS} = 1.29 \times 10^{-2}$,
$\lambda_{S_1} = 10^{-2}$,
$\lambda_{HS_1} = 10^{-4}$,
$\lambda_{SS_1} = 10^{-10}$\\
$\sin\theta_{\rm DM} = 0.001$,
$m_s = 800~\mathrm{GeV}$,
$M_{Z'} = 1.5~\text{TeV}$,
$g_{B} = 0.05$.
}
} \\
\hline

\multicolumn{4}{|c||}{\textbf{$\Delta M(\Psi_2,\Psi_1) =
1~\mathrm{GeV}$}} &
\multicolumn{4}{c|}{\textbf{$\Delta M(\Psi_2,\Psi_1) =
10~\mathrm{GeV}$}} \\
\hline

$Y_{S_1}$ & $M_{\rm DM}~(\mathrm{GeV})$& $\Omega h^2$ &$\sigma_{\rm SIDD/cm^2}$  &
$Y_{S_1}$ & $M_{\rm DM}~(\mathrm{GeV})$ & $\Omega h^2$ &$\sigma_{\rm SIDD/cm^2}$  \\
\hline

0.0797 &674.79 & 0.1101 & $3.51\times10^{-47}$ &
0.4254 &159.45 & 0.112 & $3.44\times10^{-47}$ \\

0.9543 & 1793.22 & 0.0987 & $3.53\times10^{-47}$ &
0.4898 & 259.44 & 0.1035 & $3.46\times10^{-47}$ \\

0.1589 & 677.13 & 0.1124 & $3.49\times10^{-47}$ &
0.0970 &383.61 & 0.1261 & $3.47\times10^{-47}$\\

0.1941 & 632.72 & 0.1021 & $3.47\times10^{-47}$ &
0.1289 &419.87 & 0.121 & $3.49\times10^{-47}$ \\

1.0759 &634.56 & 0.1009 & $3.47\times10^{-47}$ &
1.5315 &532.94 & 0.114 & $3.48\times10^{-47}$ \\

0.9378 &1324.62 & 0.1213 & $3.51\times10^{-47}$ &
1.4103 &630.81 & 0.115 & $3.48\times10^{-47}$ \\

1.0553 & 1056.21 & 0.0943 & $3.51\times10^{-47}$ &
1.2124 & 842.73& 0.123 & $3.49\times10^{-47}$ \\

0.2830 &  680.52 & 0.1170 & $3.48\times10^{-47}$ &
1.0605 & 1266.29 & 0.117 & $3.51\times10^{-47}$ \\

0.1244 &  615.18 & 0.0959 & $3.47\times10^{-47}$ &
1.0062 & 1666.16 & 0.111 & $3.52\times10^{-47}$ \\

0.4452 & 713.68 & 0.1005 & $3.49\times10^{-47}$ &
0.999 & 1994.89 & 0.120 & $3.68\times10^{-47}$  \\
\hline\hline
\end{tabular}
}
\caption{Similar to Table~\ref{tab:benchmark1}, but in the plane of $M_{\rm DM}-Y_{S_1}$ as shown in Figure~\ref{fig:common2}.}
\label{tab:benchmark2}
\end{table}
We then examine the impact of the DM mass splitting on the structure of the allowed parameter space in the $Y_{S_1}$ versus $M_{\rm DM}$ plane (Figure~\ref{fig:common2}). For a small mass splitting $\Delta M(\Psi_2,\Psi_1) = 1~\mathrm{GeV}$ (\textbf{Top Panel}), the dark sector states remain nearly degenerate. As a consequence, the resonant features induced by DM annihilation and co-annihilation processes occur at closely spaced values of the DM mass. This leads to a pronounced clustering of resonance structures in the $(Y_{S_1},\, M_{\rm DM})$ plane, resulting in overlapping or narrowly separated allowed regions. In contrast, for a larger mass splitting $\Delta M(\Psi_2,\Psi_1) = 10~\mathrm{GeV}$ (\textbf{Bottom Panel}), the degeneracy between the dark sector states is lifted. This shifts the resonant conditions to distinct values of $M_{\rm DM}$, thereby reducing the overlap between different resonance regions. Consequently, the allowed parameter space exhibits well-separated resonant bands. This behavior clearly illustrates the sensitivity of the model to the size of the mass splitting. In particular, a modest change in $\Delta M(\Psi_2,\Psi_1)$ from $1~\mathrm{GeV}$ to $10~\mathrm{GeV}$ qualitatively alters the structure of the allowed parameter space, highlighting the strong interplay between flavor and DM constraints. Based on these results, we randomly select a few benchmark points from the overall allowed regions of the parameter space and present them in tabular form in Tables~\ref{tab:benchmark1} and~\ref{tab:benchmark2} for Figs.~\ref{fig:common1} and~\ref{fig:common2}, respectively.


\section{Conclusion}
\label{sec:conc}

In this work, we studied DM and flavor-physics phenomenology in a gauged theory of baryons with the inclusion of an additional colored scalar \(S_1\). We extended the Standard Model by a local \(U(1)_B\) symmetry, a dark sector (consistent with anomaly cancellation), and a color-triplet scalar \(S_1\). The DM candidate is a mixed Dirac fermion arising from the mixing of singlet- and doublet-type exotic fermions, with the lightest state \(\Psi_1\) being dominantly singlet-like. The baryonic gauge symmetry ensures the stability of the DM particle, while the extended particle content allows a simultaneous description of flavor and dark sector observables. The scalar \(S_1\) here opens up novel interactions that can address the anomalies observed in \(b \to s \mu^+ \mu^-\) transitions. In particular, the exchange of \(S_1\) generates the additional contributions to the Wilson coefficients $C_9^{SM}$ and $C_{10}^{SM}$. In addition, \(S_1\) plays an important role in DM phenomenology by opening additional coannihilation channels involving the dark-sector fermions. These processes are especially relevant when the mass splitting between dark-sector states is small, thereby significantly affecting the thermal freeze-out of DM and, in turn, its final relic abundance. Throughout our analysis, we assumed a negligible kinetic mixing between the SM $Z$ boson and the $U(1)_B$ force carrier $Z'$.

A key feature of the model is the Yukawa interaction $Y_{S_1}\, \bar{q}\, S_1\, \Psi_R$, which directly connects the dark sector to the quark sector. The same coupling \(Y_{S_1}\) controls both the size of the flavor-changing neutral current effects and the strength of the DM annihilation and coannihilation processes mediated by \(S_1\). As a result, the constraints from flavor observables and the relic density requirement jointly restrict the allowed parameter space, leading to correlated predictions for DM and flavor physics.

For the dark matter phenomenology, we performed extensive numerical scans to investigate the impact of the Yukawa coupling $Y_{S_1}$ and the mass splittings between the fermionic states $\Psi_1$ and $\Psi_2$, as well as between $\Psi_1$ and the scalar $S_1$. We analyzed their effects on the resulting dark matter relic abundance and on constraints from direct and indirect detection experiments. From our analysis, we found that phenomenologically viable DM emerged for a mass range $m_{\rm DM} \in [100$ - $2000]\,\mathrm{~GeV}$, with the $U(1)_B$ gauge coupling in the interval $g_B \in [0.02,\,0.06]$ and the exotic Yukawa coupling in the range $Y_{S_1} \in [0.005,\,1]$. These results were obtained by varying the fermionic mass splitting within $\Delta M(\Psi_2,\Psi_1) \in [1,\,10]\,\mathrm{GeV}$, as summarized in Tables~\ref{tab:benchmark1} and~\ref{tab:benchmark2}.  Furthermore, we explored the dependence of the DM relic on the mass splittings as shown in Figures~\ref{fig:relic3} and~\ref{fig:relic4}. We observed that for small values of mass splittings, $\Delta M (\Psi_2,\Psi_1)$ \& $\Delta M (S_1, \Psi_1)$, i.e. when these $\Delta M \in [1$-$10]\,\mathrm{GeV}$, the obtained DM relic is within the $2\sigma$ \texttt{PLANCK} bound for a broad DM mass window in the range $M_{\rm DM} \simeq 100$-$900\,\mathrm{GeV}$, while simultaneously satisfying latest constraints from direct and indirect detection experiments for the same parameter space as shown in Figs.~\ref{fig:dd1},~\ref{fig:dd2} and \ref{fig:idd1}. Thus, the framework allows a significant parameter space and mass range for a cosmologically viable Dirac DM candidate that is allowed by the latest DD limits from \texttt{LZ}, particularly when coannihilation effects via next-to-lightest dark state and via $S_1$ are taken into account. 

On the other hand, the analysis reveals a clear hierarchy between the constraints arising from flavor observables and those from the dark sector. The flavor observables mediated by $b \to s \mu^+\mu^-$ transition decays allow a comparatively large region of parameter space. In contrast, the DM relic abundance and direct detection impose stronger bounds. Consequently, the common allowed parameter space consistent with both sectors is effectively governed by the dark contributions, with the flavor observables remaining close to their SM expectations throughout the region. Thus, the correlated structure of the model leads to several testable predictions. Additionally, the scalar \(S_1\) can be probed at colliders through jet-plus-missing-energy signatures, while direct detection experiments test the \(Z'\)-mediated interactions of DM. Indirect searches, such as diffuse gamma-ray observations at the Cherenkov Telescope Array (\texttt{CTA}), provide complementary probes of the parameter space.

Finally, we note a few possible extensions of this work. The breaking of the \(U(1)_B\) symmetry could proceed through a first-order phase transition, potentially giving rise to an observable stochastic gravitational-wave signal at upcoming experiments like \texttt{LISA} and \texttt{ET}. In addition, by carefully tuning the quantum numbers of exotic fermions, the framework can accommodate Majorana DM candidates too, which may lead to qualitatively different phenomenology and distinct indirect detection signatures, including loop-induced gamma-ray lines. Overall, the presented framework provides a simple, predictive setting in which DM and flavor physics phenomenology are closely linked, and its predictions within a given parameter space are falsifiable with upcoming \texttt{CTA} and \texttt{XENONnT} data.

\section*{Acknowledgment}
Taramati acknowledges the financial support from the Department of Science and Technology, INSPIRE (DST/INSPIRE Fellowship/IF200289), New Delhi. MKM would like to acknowledge the financial support from the IoE PDRF at the University of Hyderabad. UP acknowledges SINP, Kolkata, as his current source of research funding.  SP would like to acknowledge the funding support from SERB, Government of India, under the MATRICS project with grant no. MTR/2023/000687.\\

\noindent \begin{LARGE}\textbf{Appendix}\end{LARGE}
\appendix
\label{APP:app}

\section{Relevant Feynman Diagrams due to extra scalar $S_1$}\label{app:C}
Here, we display the relevant Feynman diagrams contributing
to DM annihilation and co-annihilation processes in the presence of the additional scalar $S_1$. The diagrams include $s$- and $t$-channel processes mediated by scalar ($h, s, S_1$) and gauge ($W^\pm, Z, Z'$) bosons, leading to quark, scalar, and gauge boson final states. These channels provide the dominant contributions to the DM relic abundance.
\subsection{Feynman Diagrams Contributing to Relic Density}
\label{app:Cc}
\begin{figure}[H]
    \centering
    \begin{subfigure}[b]{0.48\textwidth}
        \centering
        \begin{tikzpicture}[line width=0.5 pt, scale=1.35]
            \draw[solid] (1.7,1.0)--(3.2,1.0);
            \draw[solid] (1.7,-0.5)--(3.2,-0.5);
            \draw[dashed](3.2,1.0)--(3.2,-0.5);
            \draw[solid] (3.2,1.0)--(4.7,1.0);
            \draw[solid] (3.2,-0.5)--(4.7,-0.5);
            \node at (1.5,1.0) {${\Psi_i}$};
            \node at (1.5,-0.5) {$\overline{\Psi}_j$};
            \node [right] at (3.08,0.25) {${S_1}$};
            \node at (4.9,1.0) {$d_k$};
            \node at (4.9,-0.5) {$d_k$};
        \end{tikzpicture}
        \caption{}
        \label{fig:subfig1}
    \end{subfigure}
    \hfill
    \begin{subfigure}[b]{0.48\textwidth}
        \centering
        \begin{tikzpicture}[line width=0.5 pt, scale=1.35]
            \draw[solid] (-3.5,1.0)--(-2.0,1.0);
            \draw[solid] (-3.5,-0.5)--(-2.0,-0.5);
            \draw[solid](-2.0,1.0)--(-2.0,-0.5);
            \draw[dashed] (-2.0,1.0)--(0.0,1.0);
            \draw[dashed] (-2.0,-0.5)--(0.0,-0.5);
            \node at (-3.7,1.0) {${\Psi_i}$};
            \node at (-3.7,-0.5) {$\overline{\Psi}_j$};
            \node [right] at (-2.05,0.25) {${d_k}$};
            \node at (0.28,1.0) {$S_1$};
            \node at (0.28,-0.5) {$\overline{S}_1$};
        \end{tikzpicture}
        \caption{}
        \label{fig:subfig2}
    \end{subfigure}

    \vspace{0.3cm}
    \begin{subfigure}[b]{0.48\textwidth}
        \centering
        \begin{tikzpicture}[line width=0.5 pt, scale=0.85]
            \draw[solid] (-3.0,1.0)--(-1.5,0.0);
            \draw[solid] (-3.0,-1.0)--(-1.5,0.0);
            \draw[snake] (-1.5,0.0)--(0.8,0.0);
            \draw[dashed] (0.8,0.0)--(2.5,1.0);
            \draw[dashed] (0.8,0.0)--(2.5,-1.0);
            \node at (-3.3,1.0) {${\Psi_i}$};
            \node at (-3.3,-1.0) {$\overline{\Psi}_j$};
            \node [above] at (0.0,0.05) {$Z$};
            \node at (2.9,1.0) {$S_1$};
            \node at (2.9,-1.0) {$\overline{S}_1$};
        \end{tikzpicture}
        \caption{}
        \label{fig:subfig3}
    \end{subfigure}
    \hfill
    \begin{subfigure}[b]{0.48\textwidth}
        \centering
        \begin{tikzpicture}[line width=0.5 pt, scale=0.85]
            \draw[solid] (5.0,1.0)--(6.5,0.0);
            \draw[solid] (5.0,-1.0)--(6.5,0.0);
            \draw[dashed] (6.5,0.0)--(8.5,0.0);
            \draw[dashed] (8.5,0.0)--(10.1,1.0);
            \draw[dashed] (8.5,0.0)--(10.1,-1.0);
            \node at (4.7,1.0) {${\Psi_i}$};
            \node at (4.7,-1.0) {$\overline{\Psi}_j$};
            \node [above] at (7.4,0.05) {$h/s$};
            \node at (10.4,1.0) {$S_1$};
            \node at (10.6,-1.0) {$\overline{S}_1$};
        \end{tikzpicture}
        \caption{}
        \label{fig:subfig4}
    \end{subfigure}

    \vspace{0.3cm}
    \begin{subfigure}[b]{0.48\textwidth}
        \centering
        \begin{tikzpicture}[line width=0.5 pt, scale=0.85]
            \draw[solid] (-3.0,1.0)--(-1.5,0.0);
            \draw[solid] (-3.0,-1.0)--(-1.5,0.0);
            \draw[snake] (-1.5,0.0)--(0.8,0.0);
            \draw[dashed] (0.8,0.0)--(2.5,1.0);
            \draw[dashed] (0.8,0.0)--(2.5,-1.0);
            \node at (-3.3,1.0) {${\Psi_i}$};
            \node at (-3.3,-1.0) {$\overline{\Psi}_j$};
            \node [above] at (-0.1,0.05) {$Z'$};
            \node at (2.7,1.0) {$S_1$};
            \node at (2.7,-1.0) {$\overline{S}_1$};
        \end{tikzpicture}
        \caption{}
        \label{fig:subfig5}
    \end{subfigure}
    \caption{Feynman diagrams for annihilation ($i=j$) and co-annihilation ($i\neq j$) channels of dark fermion states into quarks and the exotic scalar $S_1$, mediated
    by scalar ($h,s$) and gauge ($Z, Z'$) bosons in the $s$- and $t$-channels.}
    \label{fig:annihilation_diagrams}
\end{figure}
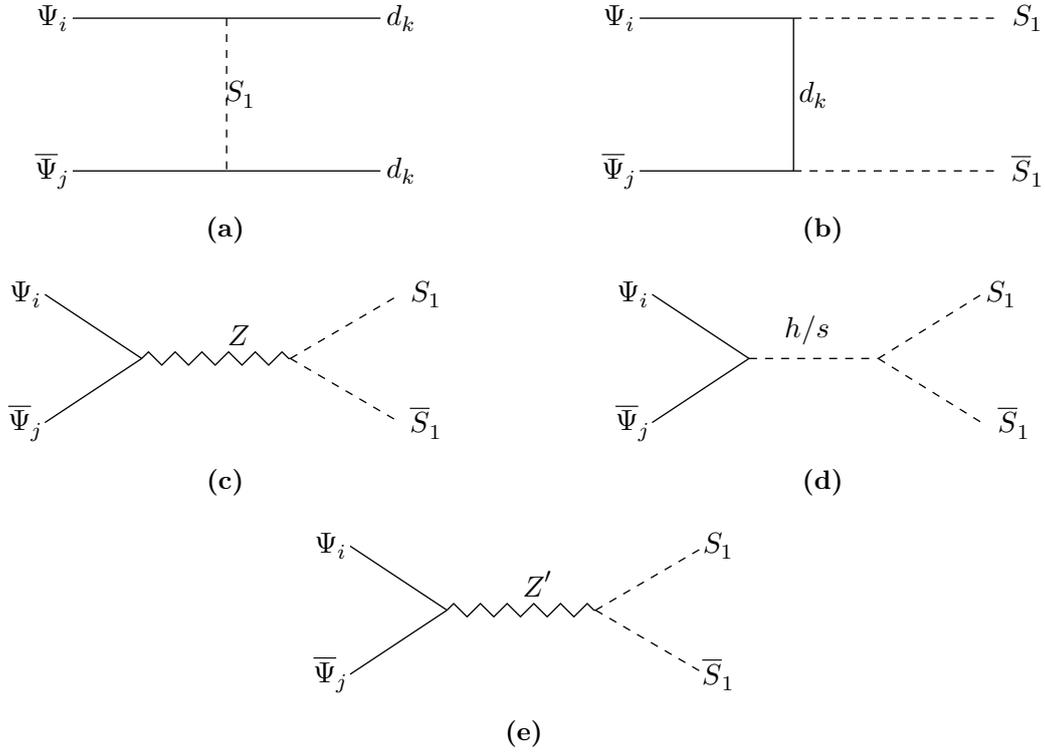


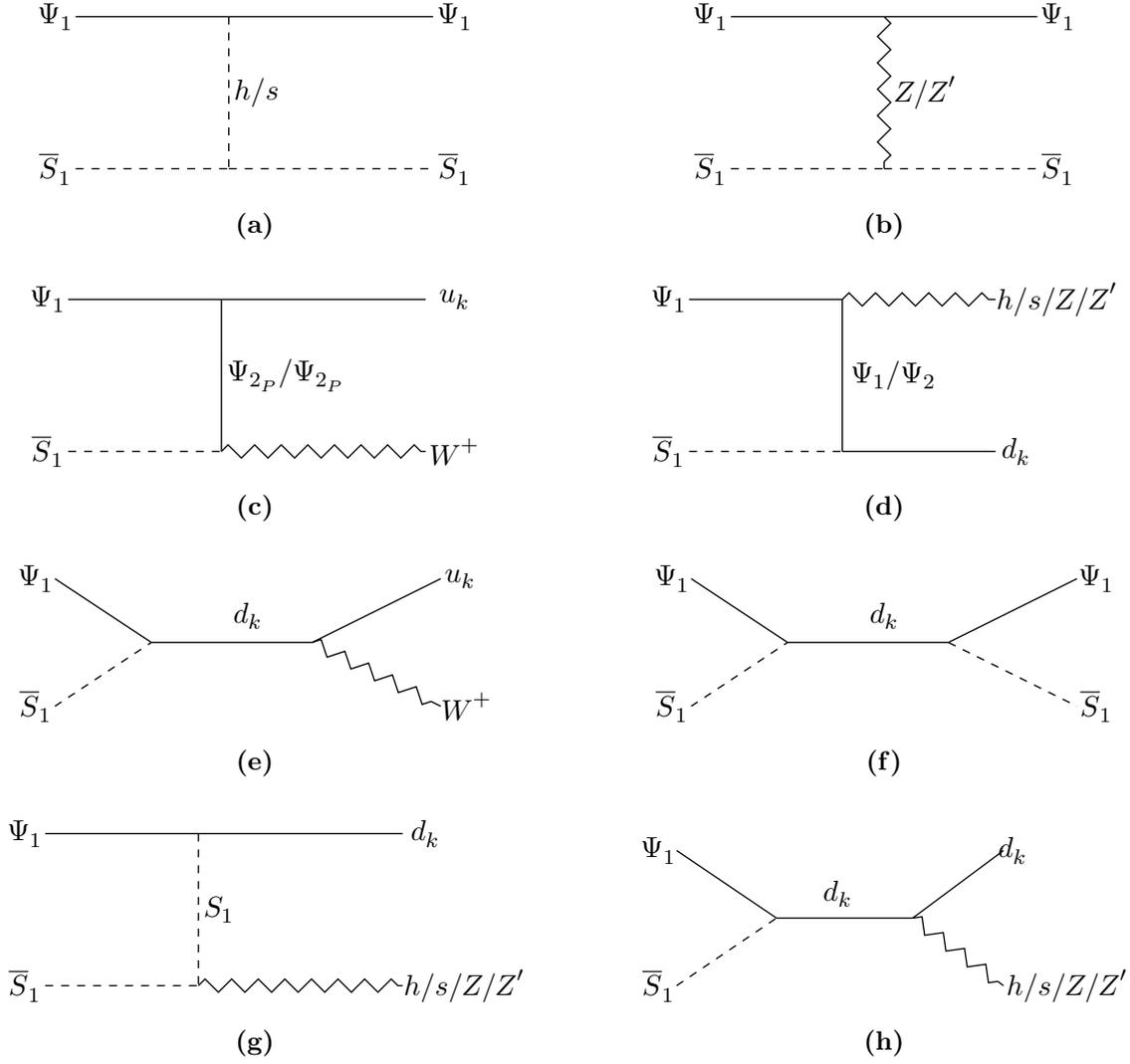
\begin{figure}[H]
\centering
\begin{subfigure}[b]{0.45\textwidth}
\centering
\begin{tikzpicture}[line width=0.5 pt, scale=1.35]
    \draw[solid] (-3.5,1.0)--(-2.0,1.0);
    \draw[dashed] (-3.5,-0.5)--(-2.0,-0.5);
    \draw[dashed](-2.0,1.0)--(-2.0,-0.5);
    \draw[solid] (-2.0,1.0)--(0.0,1.0);
    \draw[dashed] (-2.0,-0.5)--(0.0,-0.5);
    \node at (-3.7,1.0) {${\Psi_1}$};
    \node at (-3.7,-0.5) {$\overline{S}_1$};
    \node [right] at (-2.05,0.25) {${h/s}$};
    \node at (0.21,1.0) {${\Psi_1}$};
    \node at (0.21,-0.5) {$\overline{S}_1$};
\end{tikzpicture}
\caption{}
 \label{fig:4a}
\end{subfigure}
\hfill
\begin{subfigure}[b]{0.45\textwidth}
\centering
\begin{tikzpicture}[line width=0.5 pt, scale=1.35]
    \draw[solid] (1.7,1.0)--(3.2,1.0);
    \draw[dashed] (1.7,-0.5)--(3.2,-0.5);
    \draw[snake](3.2,1.0)--(3.2,-0.5);
    \draw[solid] (3.2,1.0)--(4.7,1.0);
    \draw[dashed] (3.2,-0.5)--(4.7,-0.5);
    \node at (1.5,1.0) {${\Psi_1}$};
    \node at (1.5,-0.5) {$\overline{S}_1$};
    \node [right] at (3.19,0.25) {$Z/Z'$};
    \node at (4.9,1.0) {$\Psi_1$};
    \node at (4.9,-0.5) {$\overline{S}_1$};
\end{tikzpicture}
\caption{}
\label{fig:4b}
\end{subfigure}

\vspace{1em}

\begin{subfigure}[b]{0.45\textwidth}
\centering
\begin{tikzpicture}[line width=0.5 pt, scale=1.35]
    \draw[solid] (-3.5,1.0)--(-2.0,1.0);
    \draw[dashed] (-3.5,-0.5)--(-2.0,-0.5);
    \draw[solid](-2.0,1.0)--(-2.0,-0.5);
    \draw[solid] (-2.0,1.0)--(0.0,1.0);
    \draw[snake] (-2.0,-0.5)--(0.0,-0.5);
    \node at (-3.7,1.0) {${\Psi_1}$};
    \node at (-3.7,-0.5) {$\overline{S}_1$};
    \node [right] at (-2.05,0.25) {$\Psi_{2_P}/\Psi_{2_P}$};
    \node at (0.28,1.0) {$u_k$};
    \node at (0.28,-0.5) {$W^+$};
\end{tikzpicture}
\caption{}
\label{fig:4c}
\end{subfigure}
\hfill
\begin{subfigure}[b]{0.45\textwidth}
\centering
\begin{tikzpicture}[line width=0.5 pt, scale=1.35]
    \draw[solid] (1.7,1.0)--(3.2,1.0);
    \draw[dashed] (1.7,-0.5)--(3.2,-0.5);
    \draw[solid](3.2,1.0)--(3.2,-0.5);
    \draw[snake] (3.2,1.0)--(4.7,1.0);
    \draw[solid] (3.2,-0.5)--(4.7,-0.5);
    \node at (1.5,1.0) {${\Psi_1}$};
    \node at (1.5,-0.5) {$\overline{S}_1$};
    \node [right] at (3.18,0.25) {${\Psi_{1}/\Psi_2}$};
    \node at (5.3,1.0) {$h/s/Z/Z'$};
    \node at (4.9,-0.5) {$d_k$};
\end{tikzpicture}
\caption{}
\label{fig:4d}
\end{subfigure}

\vspace{1em}

\begin{subfigure}[b]{0.45\textwidth}
\centering
\begin{tikzpicture}[line width=0.5 pt, scale=0.85]
    \draw[solid] (-3.0,1.0)--(-1.5,0.0);
    \draw[dashed] (-3.0,-1.0)--(-1.5,0.0);
    \draw[solid] (-1.5,0.0)--(1.0,0.0);
    \draw[solid] (1.0,0.0)--(3.0,1.0);
    \draw[snake] (1.0,0.0)--(3.0,-1.0);
    \node at (-3.3,1.0) {${\Psi_1}$};
    \node at (-3.3,-1.0) {$\overline{S}_1$};
    \node [above] at (0.0,0.05) {$d_k$};
    \node at (3.3,1.0) {$u_k$};
    \node at (3.4,-1.0) {$W^+$};
\end{tikzpicture}
\caption{}
\label{fig:4e}
\end{subfigure}
\hfill
\begin{subfigure}[b]{0.45\textwidth}
\centering
\begin{tikzpicture}[line width=0.5 pt, scale=0.85]
    \draw[solid] (-3.0,1.0)--(-1.5,0.0);
    \draw[dashed] (-3.0,-1.0)--(-1.5,0.0);
    \draw[solid] (-1.5,0.0)--(1.0,0.0);
    \draw[solid] (1.0,0.0)--(3.0,1.0);
    \draw[dashed] (1.0,0.0)--(3.0,-1.0);
    \node at (-3.3,1.0) {${\Psi_1}$};
    \node at (-3.3,-1.0) {$\overline{S}_1$};
    \node [above] at (0.0,0.05) {$d_k$};
    \node at (3.3,1.0) {${\Psi_1}$};
    \node at (3.3,-1.0) {$\overline{S}_1$};
\end{tikzpicture}
\caption{}
\label{fig:4f}
\end{subfigure}

\vspace{1em}
\begin{subfigure}[b]{0.45\textwidth}
\centering
\begin{tikzpicture}[line width=0.5 pt, scale=1.35]
	\draw[solid] (-3.5,1.0)--(-2.0,1.0);
    \draw[dashed] (-3.5,-0.5)--(-2.0,-0.5);
    \draw[dashed](-2.0,1.0)--(-2.0,-0.5);
    \draw[solid] (-2.0,1.0)--(0.0,1.0);
    \draw[snake] (-2.0,-0.5)--(0.0,-0.5);
    \node at (-3.7,1.0) {$\Psi_{1}$};
    \node at (-3.7,-0.5) {$\overline{S}_1$};
    \node [right] at (-2.05,0.25) {$S_1$};
    \node at (0.22,1.0) {$d_k$};
    \node at (0.6,-0.5) {$h/s/Z/Z'$};
\end{tikzpicture}
\caption{}
\label{fig:subfig5}
\end{subfigure}%
\hfill
\begin{subfigure}[b]{0.45\textwidth}
\centering
\begin{tikzpicture}[line width=0.5 pt, scale=1.2]
    \draw[solid] (-2.9,1.0)--(-1.8,0.25);
    \draw[dashed] (-2.9,-0.5)--(-1.8,0.25);
    \draw[solid](-1.8,0.25)--(-0.3,0.25);
    \draw[solid] (-0.3,0.25)--(0.7,1.0);
    \draw[snake] (-0.3,0.25)--(0.7,-0.5);
    \node at (-3.10,1.0) {${\Psi_{1}}$};
    \node at (-3.10,-0.5){$\overline{S}_1$};
    \node [right] at (-1.4,0.52) {$d_k$};
    \node at (0.8,1.0) {$d_k$};
    \node at (1.4,-0.5) {$h/s/Z/Z'$};
\end{tikzpicture}
\caption{}
\label{fig:subfig6}
\end{subfigure}
\caption{Dominant co-annihilation channels induced by the additional scalar $S_1$ involving the dark fermion $\Psi_1$, mediated by scalar ($h,s$) and gauge ($W^\pm, Z, Z'$) bosons, and leading to quark, scalar, and gauge boson final states. The index $k$ labels the quark generation.}
\label{fig:coann_diagrams}
\end{figure}

\subsection{Relevant Feynman Diagrams with diffused gamma ray production}
\label{app:Cb}
\begin{figure}[H]
\centering
    \begin{tikzpicture}[line width=0.5 pt, scale=0.85]
          \draw[solid] (-3.0,1.0)--(-1.5,0.0);
        \draw[solid] (-3.0,-1.0)--(-1.5,0.0);
         \draw[snake] (-1.5,0.0)--(0.8,0.0);
        \draw[solid] (0.8,0.0)--(2.5,1.0);
        \draw[snake] (1.59,0.5)--(2.5,0.0);
         \draw[solid] (0.8,0.0)--(2.5,-1.0);
         \node at (-3.4,1.0) {$\text{DM}$};
         \node at (-3.4,-1.0) {$\overline{\text{DM}}$};
         \node [above] at (0.0,0.05) {$Z'$};
        \node at (2.7,1.0) {$f$};
        \node at (2.7,0.0) {$\gamma$};
        \node at (2.7,-1.0) {$\overline{f}$};
         \draw[solid] (5.0,1.0)--(6.5,0.0);
        \draw[solid] (5.0,-1.0)--(6.5,0.0);
         \draw[snake] (6.5,0.0)--(8.5,0.0);
         \draw[solid] (8.5,0.0)--(10.4,1.0);
         \draw[solid] (8.5,0.0)--(10.4,-1.0);
          \draw[snake] (9.5,-0.5)--(10.4,0.0);
         \node at (4.6,1.0) {$\text{DM}$};
         \node at (4.6,-1.0) {$\overline{\text{DM}}$};
         \node [above] at (7.4,0.05) {$Z'$};
         \node at (10.5,1.0) {$f$};
        \node at (10.5,-1.0) {$\overline{f}$};
        \node at (10.5,0.0) {$\gamma$};
     \end{tikzpicture}
\caption{\small Feynman diagram illustrating DM annihilation into a fermion-antifermion pair accompanied by a photon, arising via final state radiation (FSR) at tree level. The fermion \(f\) can be either a Standard Model or an exotic fermion, and the process is mediated by the additional gauge boson \(Z'\) associated with the extended gauge symmetry.}
\label{fig:FSR}
\end{figure}
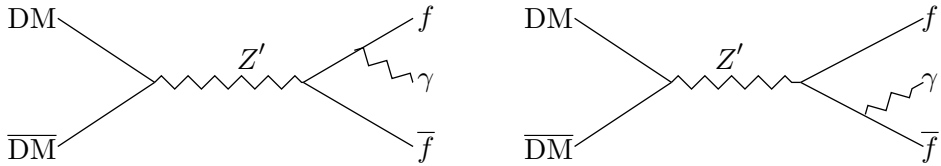
\section{Details of the form factors and the angular coefficietns for $B \to (P, V) \ell^+ \ell^-$ processes}
We provide the relevant form factors of $B \to K \ell^+ \ell^-$ in \ref{BtoKFF} and the angular coefficients of $B \to V \ell^+ \ell^-$ processes in \ref{Ang_coeff} in term of the transverse amplitudes below. 
\subsection{ Form factors for $B \to (K^{(*)}, \phi)\, \ell^+ \ell^- (\ell = e, \mu, \tau)$} \label{BtoKFF}
The hadronic matrix elements for the exclusive $B \to K$ transition in terms of form factors are given by \cite{Bouchard:2013eph}
\begin{align}
J_\mu&= \langle K | \bar{s}\gamma^{\mu} b | B \rangle \nonumber \\
&= f_{+}(q^2)\Big[p_{B}^{\mu}+p_{K}^{\mu}
-\frac{M_{B}^2-M_{K}^2}{q^2}\,q^{\mu}\Big]
+ f_{0}(q^2)\,
\frac{M_{B}^2-M_{K}^2}{q^2}\,q^{\mu}\,, \\[0.2cm]
J_\mu^{T} 
&= \langle K | \bar{s}\sigma^{\mu\nu} q_{\nu} b | B \rangle \nonumber \\
&= \frac{i\,f_{T}(q^2)}{M_{B}+M_{K}}
\Big[q^2\left(p_{B}^{\mu}+p_{K}^{\mu}\right)
-\left(M_{B}^2-M_{K}^2\right)q^{\mu}\Big] \, .
\end{align}
where $q =p_{B} - p_{K}$ with the mass of $B$ and $K$ mesons are $M_B$ and $M_K$, respectively. The form factors given above the expression satisfy the following relations:
\bea
f_+(0)= f_0, \hspace*{.5cm} f_0(q^2) = f_+ (q^2) + \frac{q^2}{M_{B}^2-M_{K}^2}f_-(q^2).
\eea

Similarly, for the $B \to V$ transition  where ($V = K^{\ast}, \phi)$, the hadronic matrix elements can be given in terms of the form factors as~\cite{Ebert:2010dv}
\begin{eqnarray}
&&\langle V|\bar{s}\gamma^{\mu}b|B \rangle = \frac{2\,i\,V(q^2)}{M_{B}+M_{V}}\,\epsilon^{\mu\nu\rho\sigma}\epsilon^{\ast}_{\nu}\,
p_{B_{\rho}}\,p_{{V}_{\sigma}}\,, \nonumber \\
&&\langle  V|\bar{s}\gamma^{\mu}\gamma_5\,b|B\rangle = 2\,M_{V}\,A_0(q^2)\frac{\epsilon^{\ast}\cdot q}{q^2}\,q^{\mu} + 
(M_{B} + M_{V})\,A_1(q^2)\Big(\epsilon^{{\ast}^{\mu}}-\frac{\epsilon^{\ast}\cdot q}{q^2}\,q^{\mu}\Big)\, \nonumber \\
&&\hspace*{4cm}-
A_2(q^2)\frac{\epsilon^{\ast}\cdot q}{M_{B} + M_{V}}\,\bigg[p_{B}^{\mu}+p_{V}^{\mu}-\frac{M_{B}^2-M_{V}^2}
{q^2}\,q^{\mu}\bigg]\,, \nonumber \\
&&\langle V|\bar{s}\,i\,\sigma^{\mu\nu}\,q_{\nu}\,b|B \rangle = 2\,T_1(q^2)\,\epsilon^{\mu\nu\rho\sigma}\epsilon^{\ast}_{\nu}\,p_{B_{\rho}}\,
p_{{V}_{\sigma}}\,, \nonumber \\
&&\langle V|\bar{s}\,i\,\sigma^{\mu\nu}\,\gamma_5\,q_{\nu}b|B \rangle= T_2(q^2)\,\bigg[(M_{B}^2-M_{V}^2)\epsilon^{{\ast}^{\mu}} - 
(\epsilon^{\ast}\cdot q)(p_{B}^{\mu}+p_{V}^{\mu})\bigg] \nonumber \\
&&\hspace*{4cm}+ 
T_3(q^2)\,(\epsilon^{\ast}\cdot q)\bigg[q^{\mu}-\frac{q^2}
{M_{B}^2-M_{V}^2}(p_{B}^{\mu}+p_{V}^{\mu})\bigg]\,,
\end{eqnarray}
where $q^{\mu}=(p_B ^ \mu -p_{V}^ \mu)$ is the four momentum transfer and $\epsilon_{\mu}$ is polarization vector of the $V$ meson. 

\subsection{Angular coefficients}\label{Ang_coeff}
The $q^2$ dependent angular coefficients required for $B \to V\, \ell^+ \ell^- (\ell = \mu, e, \tau)$ processes are given as follows:
\begin{eqnarray}
I_{1}^{c} &=& \bigg(|A_{L0}|^2 + |A_{R0}|^2\bigg) + 8\frac{m_{l}^2}{q^2} Re\bigg[A_{L0}A_{R0}^{*}\bigg] + 4\frac{m_{l}^2}{q^2}|A_{t}|^2, \nonumber \\
I_{2}^{c} &=& -\beta_{l}^2 \bigg(|A_{L0}|^2 + |A_{R0}|^2\bigg), \nonumber \\
 I_{1}^{s} &=& \frac{3}{4} \bigg[|A_{L\perp}|^2 + |A_{L\parallel}|^2 + |A_{R\perp}|^2 + |A_{R\parallel}|^2\bigg] \bigg(1-\frac{4m_{l}^2}{3q^2}\bigg) +
 \frac{4m_{l}^2}{q^2} Re\bigg[A_{L\perp} A_{R\perp}^{*} + A_{L\parallel} A_{R\parallel}^{*}\bigg], \nonumber \\
 I_{2}^{s} &=& \frac{1}{4} \beta_{l}^2 \bigg[|A_{L\perp}|^2 + |A_{L\parallel}|^2 + |A_{R\perp}|^2 + |A_{R\parallel}|^2\bigg],   \nonumber \\
 I_{3} &=& \frac{1}{2} \beta_{l}^2 \bigg[|A_{L\perp}|^2 - |A_{L\parallel}|^2 + |A_{R\perp}|^2 - |A_{R\parallel}|^2\bigg], \nonumber \\
 I_{4} &=& \frac{1}{\sqrt{2}} \beta_{l}^2 \bigg[Re\bigg(A_{L0}A_{L\parallel}^{*}\bigg) + Re\bigg(A_{R0}A_{R\parallel}^{*}\bigg)\bigg],  \nonumber \\
 I_{5} &=& \sqrt{2} \beta_l \bigg[Re\bigg(A_{L0}A_{L\perp}^{*}\bigg) - Re\bigg(A_{R0}A_{R\perp}^{*}\bigg)\bigg], \nonumber \\
 I_{6} &=& 2 \beta_l \bigg[Re\bigg(A_{L\parallel}A_{L\perp}^{*}\bigg) - Re\bigg(A_{R\parallel}A_{R\perp}^{*}\bigg)\bigg], \nonumber \\
 I_{7} &=& \sqrt{2} \beta_l \bigg[Im\bigg(A_{L0}A_{L\parallel}^{*}\bigg) - Im\bigg(A_{R0}A_{R\parallel}^{*}\bigg)\bigg], \nonumber \\
 I_{8} &=& \frac{1}{\sqrt{2}} \beta_{l}^2 \bigg[Im\bigg(A_{L0}A_{L\perp}^{*}\bigg) + Im\bigg(A_{R0}A_{R\perp}^{*}\bigg)\bigg], \nonumber \\
 I_{9} &=& \beta_{l}^2 \bigg[Im\bigg(A_{L\parallel}A_{L\perp}^{*}\bigg) + Im\bigg(A_{R\parallel}A_{R\perp}^{*}\bigg)\bigg]\,,
\end{eqnarray}
where $\beta _ \ell = \sqrt{1-4m_ \ell ^2/q^2}$. 
According to Ref. \cite{Altmannshofer:2008dz}, the transversity amplitude in terms of form factors and Wilson coefficients is given as
\begin{eqnarray}
 A_{L0} &=& N \frac{1}{2m_{V}\sqrt{q^2}}
 \bigg\{(C_{9}^{eff} - C_{10}) \bigg[(M_{B}^2 - M_{V}^2 - q^2)(M_{B} + M_{V})A_1 
 - \frac{\lambda}{M_{B} + M_{V}}A_2\bigg] \nonumber \\  
 && +\, 2\,m_b\, C_{7}^{eff}\, \bigg[(M_{B}^2 + 3M_{V}^2 - q^2)T_2 
 - \frac{\lambda}{M_{B}^2 - M_{V}^2}T_3 \bigg] \bigg\}\,, \nonumber \\
 A_{L\perp} &=& - N\sqrt{2} \bigg[(C_{9}^{eff} - C_{10})
 \frac{\sqrt{\lambda}}{M_{B} + M_{V}} V 
 + \frac{\sqrt{\lambda}\,2\,m_b\,C_{7}^{eff}}{q^2} T_1 \bigg]\,, \nonumber \\
 A_{L\parallel} &=& N\sqrt{2} \bigg[(C_{9}^{eff} - C_{10})
 (M_{B} + M_{V}) A_1 
 + \frac{2\,m_b\,C_{7}^{eff}(M_{B}^2 - M_{V}^2)}{q^2} T_2 \bigg]\,, \nonumber \\
 A_{Lt} &=& N (C_{9}^{eff} - C_{10}) \frac{\sqrt{\lambda}}{\sqrt{q^2}}A_0\,,
\end{eqnarray}

\[
\lambda = M_{B}^4 + M_{V}^4 + q^4 
- 2\left(M_{B}^2 M_{V}^2 + M_{V}^2 q^2 + q^2 M_{B}^2\right)
\]
 and 
$N$, the normalization factor, which is defined as 
\begin{equation}
 N = \bigg[\frac{G_{F}^2\alpha_{em}^2}{3\cdot 2^{10}\pi^5\,M_{B}^3}
 |V_{tb}V_{ts}^{*}|^2 q^2 \sqrt{\lambda}
 \bigg(1-\frac{4m_{l}^2}{q^2}\bigg)^{1/2}\bigg]^{1/2}\,.
\end{equation}

The right chiral component $A_{Ri}$ of the transversity amplitudes can be obtained by replacing $A_{Li}$ by $A_{Li}|_{C_{10} \to -C_{10}} (i= 0, \parallel, \perp, t)$.

\bibliographystyle{JHEP}
\bibliography{dm}

\end{document}